\documentclass[aps,10pt,
prd,preprintnumbers,
nobibnotes,nofootinbib,floatfix,
amsmath,amssymb,
longbibliography,superscriptaddress
]{revtex4-2}
\usepackage[english]{babel}
\usepackage[utf8]{inputenc}
\usepackage{siunitx}
\usepackage{amsfonts}
\usepackage{subcaption}
\usepackage{braket}
\usepackage{mathtools}
\usepackage{cancel}
\usepackage{slashed}
\usepackage{pifont}
\usepackage{soul}
\usepackage{comment}
\usepackage[colorlinks=true,linkcolor=blue]{hyperref}%
\usepackage[htt]{hyphenat}
\usepackage{bbm}
\usepackage{bm}
\usepackage{graphicx}
\usepackage{dcolumn}
\usepackage{bm}

\newcommand{\PDF}{f(x)}
\newcommand{\PDFv}{\mathbf{f}}

\newcommand{\x}{\mathbf{x}}
\newcommand{\m}{\mathbf{m}}
\newcommand{\xm}{\langle x^m \rangle}

\newcommand{\bb}{\mathbf{b}}
\newcommand{\y}{\mathbf{y}}
\newcommand{\K}{K}
\newcommand{\FK}{\mathcal{F}}
\newcommand{\FKN}{\mathcal{F}_\mathrm{NL}}
\newcommand{\Cy}{C_\y}
\newcommand{\Cyk}{C_{\y\mathbf{k}}}
\newcommand{\nmax}{n_\mathrm{max}}

\newcommand{\FANTO}{\texttt{FANTO10\_n15 }}

\newcommand{\MATERN}[1]{Mat\`ern$_{\nu=#1}$}
\newcommand{\GIBBS}[1]{$\ell_{d=#1}(x)$}

\newcommand{\biascorrnote}{\textbf{Bias correction applied.}}
\newcommand{\muvec}{\boldsymbol{\mu}}
\newcommand{\xmom}{\langle x\rangle}
\newcommand{\NNPDF}{\texttt{NNPDF30\_lo\_as\_0130\_nf\_3 }}

\begin{document}

\title{Tackling the noisy truncated moment problem with Gaussian Processes:\\An application to parton distribution functions}

\author{Rohith Karur}
\affiliation{Department of Physics, University of California, Berkeley, CA 94720, U.S.A}
\affiliation{Nuclear Science Division, Lawrence Berkeley National Laboratory, Berkeley, CA 94720, USA}

\date{\today}

\begin{abstract}
In this work, we study the noisy truncated Hausdorff moment problem in the context of recovering parton distribution functions (PDFs) from their moments. We start by reviewing basics of the Hausdorff moment problem and analyze how the relation between the amount of moment data provided and the ability for $f(x)$ to be reconstructed varies across different classes of parton distribution functions; we then turn to using Bayesian methods centered around Gaussian Processes (GPs) to solve the moment problems. Standard aspects of Bayesian regression such as covariance kernel choice, analytical characterization of the posterior, hyperparameter sampling with Monte Carlo methods, and validation metrics are detailed. In addition to this, powerful novel tools, such as a method to correct for mean bias in posterior distributions, and a simple class of covariance kernels whose $x$ correlation length is learned during sampling are introduced. 

Finally, we test our framework on 5 datasets from valence, gluon, and sea quark phenomenological pion and nucleon PDF datasets, selected to reflect diversity in PDF behavior. We find that our GP framework robustly reconstructs PDFs given a moderate number of moments. Additionally, the novel tools developed in this work resolve persistent issues that would be difficult to address via conventional means. The presented framework can be readily applied to moments obtained via lattice Quantum Chromodynamics (LQCD) to reconstruct PDFs from theoretical first principles.


\end{abstract}

\maketitle

\tableofcontents

\section{\label{sec:level1}Introduction}

Quantum Chromodynamics (QCD), the fundamental theory of the strong nuclear force, predicts that hadrons in our universe are composed of quarks and gluons. The roles of these constituent \textit{partons} in hadron structure can in part be quantified through density functions known as \textit{parton distribution functions} (PDFs). PDFs are density functions of a quantity $x\in(0,1)$, denoted ``Bj\"orken $x$", which denotes the fraction of a hadron's longitudinal momentum carried by a given parton. Experimentally, hadronic structure can be probed through a variety of processes, including Deep Inelastic Scattering (DIS), Drell-Yan (DY) processes, and prompt photon production; the partonic dependence of measured cross sections can be split into different \textit{structure functions} $F_i(x,Q^2)$, which are sensitive to both $x$ as well as the energy scale $Q^2$ at which hadrons are probed. Structure functions can be factorized into a high-energy ($Q^2$ dependent) component and a low-energy ($Q^2$ independent) component containing PDF information. Experimental cross section data is sensitive to the $x$ value \footnote{In neutral current leptonic DIS experiments for example, the incoming and outgoing lepton energies and angle of deflection $\theta$ are measured, from which cross sections can be determined} of a parton it is detecting, and upon extracting a discrete set of 
PDF values as a function of $x$, the PDF is commonly fit to the parametric form:

\begin{equation}
    f(x) = x^{\alpha}(1-x)^\beta\times p(x),\; x\in [0,1]
    \label{eq:para_form}
\end{equation}

$f$ can refer to all possible partons: valence quarks, gluons, and sea quarks. Valence quarks ($q_V$) are quarks that define the quantum numbers of the parent hadron (like charge and spin), gluons ($g$) are gauge bosons that mediate the strong force and thereby confine a hadron's quarks, and sea quarks are quarks that arise from the process $g\to q \bar{q}$. The form of Eq. \ref{eq:para_form} stems from several requirements for PDFs; they are required to be zero at $x=1$, as the failure to meet this requirement breaks the parton model. Regge theory \cite{Ball:2016spl,Kuti:1971ph} predicts $x^\alpha$-like power divergence behavior for low $x$, which restricts $\alpha \leq 0 $. Partons also must obey momentum sum rules (the total integrated momentum sum over all partons must equal one: $\sum_i \int_0^1 dx\;xf_i(x)=1$), and for this to hold, the function $x f(x)$ must be integrable, and this necessitates $\alpha \in (-2,0]$.\footnote{For gluons, low $x$ behavior is dominated by the $g\to gg$ process \cite{Ellis:1996mzs}.} For valence quarks, an extra number conservation constraint is required, which is tantamount to a normalization constraint; in this case, $\alpha$ is tightened to  $\alpha \in (-1,0]$. Finally, the function $p(x)$ can refer to an arbitrarily complex polynomial, or a different parametrization (i.e. neural network,...etc).

Both $x\ll 1 $ and $x\sim1$ regions of PDFs are difficult to experimentally probe, and while the range and quality of data will only improve with upcoming experiments (EIC, HL-LHC, ...etc), first-principles theoretical input from QCD is invaluable: the short scale components of structure functions (``Wilson coefficients") can be computed perturbatively, while the long range component must be computed non-perturbatively, i.e. with lattice QCD. Lattice QCD is a computational tool that is the only known general method for evaluating quantities pertaining to QCD from first principles.  It is formulated on a Euclidean 4-dimensional lattice; the transformation from Minkowski to Euclidean time allows the Boltzmann factor in the QCD path integral to be expressed as a real number: $e^{iS} \to e^{-S}$, where $S$ is the QCD action. This allows us to importance sample QCD gauge configurations with Monte Carlo methods, from which correlation functions of interest can be constructed.

When expressing the hadronic vertices present in scattering processes relevant to experiment in terms of time ordered current products: $T[J(z)J(0)]$, terms lying on the light-cone ($z^2 \sim 0$) dominate the expression. As such, PDFs themselves are defined on the light-cone, i.e. we can write a PDF (in the light-cone gauge) as:

\begin{equation}
    f(x) = \int \frac{d\xi}{2\pi} e^{i\xi x} \langle P | \bar \psi(0) \cancel n \psi(\xi n)|P\rangle,
    \label{eq:PDFnonlocal}
\end{equation}

where $\xi n$ is the (light-like) distance between the two parton operators $\psi$. Being constructed on a Euclidean lattice, light-like dispersion relations are inaccessible to LQCD. To circumvent this, we can use the Operator Product Expansion (OPE) to recast the expressions that define PDFs (Eq. \ref{eq:PDFnonlocal}) to ones containing a sum of local operators, the matrix elements of which can be accessed on the lattice. From this reformulation, one can obtain the moments of PDFs: $f: f\to \langle x^{n}\rangle = \int_0^1 dx\; x^{n} f(x)$\footnote{In lattice QCD literature, it is standard to write moments with the \textit{Mellin moment} definition: $f: f\to \langle x^{n}\rangle = \int_0^1 dx\; x^{n-1} f(x)$. We chose not to use this notation and instead use the definition of moments as used in applied math.} through matrix elements of twist-2 operators.\footnote{Twist-2 operators are operators where the quantity mass dimension - spin = 2} An added complication of the lattice is that the O(4) symmetry of the continuum is broken to the H(4) hyper-cubic symmetry; operators with zero spatial momenta that would normally provide optimal signal to noise now straddle different irreps of H(4), leading to power divergent mixing as the lattice spacing $a$ is taken to zero. For this reason, operators with nonzero spatial momenta were used, leading to an increased noise-to-signal ratio, meaning that only the first few moments of PDFs were realistically accessible \cite{Cichy:2018mum} up until recently. Recent studies \cite{Shindler:2023xpd,Shindler:2024grr} proposed the use of gradient flow techniques \cite{Luscher:2010iy,Luscher:2013cpa} to resolve these issues and thereby calculate moments of any order on the lattice, and initial studies of this concept have already been performed on the pion \cite{Francis:2025pgf, Francis:2025rya}
 with success, with subsequent studies for pions at the physical point and nucleons soon expected. Given these developments, the volume and quality of moment data from the lattice will only improve. With the moments of $f(x)$ at hand, however, the question of reconstructing $f(x)$ itself still remains.

The mathematical term for this moment problem is the \textit{truncated Hausdorff moment problem}. Appearing in a number of scientific disciplines, this inverse problem aims to recover a density function $\PDF$ with compact support given a finite number of moments. Tackling the finite moment problem has been looked at in a few contexts, the majority of which consider the problem with vanishing experimental error; however, these studies neglect the impact of uncertainty present in most simulation/experimental datasets, including lattice QCD results. In this study, we address the truncated Hausdorff moment problem in the presence of noisy data, and in particular demonstrate the efficacy of a Bayesian framework revolving around \textit{Gaussian Process Regression} (GPR) in PDF reconstruction.

The layout of this paper is as follows: first we give an outline of the Hausdorff finite moment problem, with some simple discussions of convergence with applications to PDFs. We then present our GPR analysis framework, covering both established Bayesian tools as well as introducing powerful novel elements for performing PDF reconstruction and validation. Finally, the GPR framework is applied to real PDF datasets of pion and nucleon valence quarks and gluons obtained from phenomenological studies to highlight its effectiveness across PDF classes. Phenomenological datasets serve as ``truth models", where we can reverse engineer moment values from an underlying PDF $f(x)$ to output moment data and can perform associated closure studies. 

\subsection{Background on the Hausdorff Moment Problem}

The Hausdorff finite moment problem asks to recover a density function $f(x)$ with compact support given a finite number of moments. Because there exist complete sets of orthonormal polynomials (with no reweighting factor) on finite intervals, such as the Legendre polynomial set, knowing an infinite number of moments of a function with compact support amounts to knowing that function uniquely. Once we truncate this number to a finite number of moments, this is no longer the case. There are still, however, quantitative claims we can make regarding constraints to a finite-moment-reconstructed function. For this, let us first introduce a ``truncated" scenario where we are provided the first $n+1$ ascending moments, with the first moment being the norm constraint.

Suppose we are given a sequence of numbers: $\boldsymbol{\mu}_n \coloneqq(\mu_1,\mu_2,...)$. In Felix Hausdorff's seminal 1923 paper \cite{hausdorff_momentprobleme_1923}, a difference relation is derived and presented that identifies this sequence as a \textit{moment sequence}, i.e. $\mu_n=\langle x^n \rangle = \int_0^1 x^n f(x) dx$, if and only if ($\forall n,k\geq 0$):

\begin{equation}
    (-1)^k (\Delta^k \mu)_n \geq 0
    \label{eq:hausdorff}
\end{equation}

Here, the difference operator is defined as $(\Delta \mu)_n = \mu_{n+1}-\mu_n$. Eq \ref{eq:hausdorff} enforces that a moment sequence must be monotonically decreasing, and if it is, then, for an infinite sequence, a solution exists (and is unique). From this, we already have a constraint on what our truncated moment sequence can be. Ref \cite{karlin_geometry_nodate} continued off of work in Refs. \cite{hausdorff_momentprobleme_1923,shohat_problem_1943} to develop a geometric formulation of the infinite moment problem that could readily be applied to the truncated problem.

Suppose we have a sequence of $n$ terms $\muvec_n \coloneqq(\mu_1,...,\mu_n)\in \mathcal{R}^n$. The \textit{moment space} $\mathcal{D}^n$ is a subspace of $\mathcal{R}^n$ that contains all possible $\boldsymbol{\mu}_n$ that correspond to the first $n$ moments of \textit{a} density function. Ref \cite{karlin_geometry_nodate} showed that this subspace is a convex body. Whether $\muvec_n$ lies on the boundary, the interior, or the exterior of $\mathcal{D}^n$ can be completely specified by  $\muvec$'s Hankel matrix determinants, details of which can be found in Refs. \cite{karlin_geometry_nodate,athanassoulis_truncated_2002}.  If $\muvec_n$ lies within the interior of $\mathcal{D}^n$, the truncated moment problem is indeterminate, and $\muvec_n$ corresponds to the first $n$ moments of an infinite numbers of target function. If $\muvec_n$ lies on the boundary of $\mathcal{D}^n$, then a target function can be uniquely determined by the first $n$ moments given in  $\muvec_n$. Ref. \cite{karlin_geometry_nodate} further showed that if $\muvec_n$ lies in the interior of $\mathcal{D}^n$, the uncertainty of the $(n+1)^\mathrm{th}$ moment is bounded by $2^{-2n}$, $\forall n>2$.

Consider an orthogonal set of functions with compact support on $[0,1]$. Henceforth, we will refer to two ``spaces": \textit{$x$ space}, which refers to the space where our target function is reconstructed, where $x\in(0,1]$ is our independent variable, and \textit{data space}, or \textit{moment space}, where our moments live and where our dependent variable is moment number $n$. Two immediately obvious basis function sets that can be used to recreate functions on a compact interval are the Chebyshev ($T_n(2x-1)$) and Legendre ($P_n(2x-1)$) polynomials, where $x$ coordinates have been transformed from $[-1,1]$ to $[0,1]$ (we will refer to them in this shifted state from now on). The Chebyshev basis is orthogonal under the weight $\frac{1}{\sqrt{(2x-1)^2-1}}$ whereas the Legendre polynomials are orthogonal under a unit weight. We can pick the Legendre polynomials to continue our analysis, as the uniform dot product will be readily applicable to our case. We will express our $f(x)$ function in the Legendre basis, where we define $\sqrt{k+\frac{1}{2}}\cdot P_k(2x-1)\coloneqq|P_k\rangle $ to normalize our basis functions, and define our orthogonality convention as $\langle P_k|P_l\rangle = \int_0^1 (k+\frac{1}{2})P_k(2x-1)P_l(2x-1)dx = \delta_{k,l}$. From here, we can define our function in the Legendre basis, $f(x)$, and our function projected onto the first $n+1$ Legendre polynomials, $f_n(x)$:

\begin{equation}
    \PDF \coloneq \sum_{k=0}^\infty a_k |P_k\rangle,\;\; \; f_n(x) \coloneqq \sum_{k=0}^n a_k |P_k\rangle
    \label{eq:leg_basis}
\end{equation}

If we have the first $n$ moments of a function, using the Legendre decomposition, we can place several bounds on what the error from the true function $|\Delta_n (x)| \coloneqq |f(x)-f_n(x)|$ is. In some sense, these bounds (in addition to those performed with Chebyshev polynomials) describe the best possible reconstruction we can get with a truncated sequence with zero uncertainty. Note that this is a point-wise error, and is equivalent to the infinity norm $||f(x)-f_n(x)||_\infty$.This has been done numerous times, and for the purpose of this study, we will refer to two studies, Refs. \cite{wang_convergence_2012,wang_new_2018}, and a pedagogical introduction with Chebyshev polynomials, Ref. \cite{trefethen_approximation_2019}. Ref. \cite{wang_convergence_2012} proved that bounds on $|\Delta_n|$ depend explicitly upon two classes of $f$ (slightly modified to fit our example): 

\textit{If  $\{f,f',...,f^{(k-1)}\}$ is absolutely differentiable on $[0,1]$ up to $k$ times and the $k^\mathrm{th}$ derivative is bounded $||f^{(k)}||_T \coloneqq \int_0^1\frac{|f^{(k)}(x)|}{\sqrt{1-(2x-1)^2}}dx = V_k<\infty$, then for $n > k+1$ we have:}

\begin{equation}
    |\Delta_n(x)|\leq \frac{V_k}{(k-1)}\frac{(n-(\frac{2k-1}{2}))!}{(n-\frac{1}{2})!}\sqrt{\frac{\pi}{2(n-k)}}
    \label{eq:diff_bound_v0}
\end{equation}

\textit{Consider a Bernstein ellipse $\mathcal{E}_\rho$ in the complex plane with the sum of semi-major/minor axes $\rho>1$ (with complex coordinate $z$), a circumference $\ell(\mathcal{E}_\rho)$, and foci at $\pm1$. If $f$ is analytic inside and on a Bernstein ellipse $\mathcal{E}_\rho$, then for $n\geq 0$ and $\max_{z\in\mathcal{E}_\rho}|f(z)|=M$}

\begin{equation}
    |\Delta_n((x+1)/2)|\leq \frac{(2n\rho+3\rho -2n-1) \ell(\mathcal{E_\rho})M}{\pi\rho^{n+1}(\rho-1)^2(1-\rho^{-2})}
    \label{eq:ana_bound}
\end{equation}

Ref \cite{wang_new_2018} derived an even sharper bound for finitely differentiable $f$, but here as well our error scales as $\sim n^{-(k-\frac{1}{2})}$, and for analytic $f$, our error scales to leading order as $\sim \rho^{-n}$.

\subsubsection{Connection to parton physics}  
\label{subsec:math_theory_disc}
The most import feature of PDFs with respect to convergence analysis is their power divergence $x^\alpha$, with $\alpha$ such that either $xf(x)$ is integrable for the gluon/sea quark PDF and both $f(x)$ and $xf(x)$ are integrable for valence quark PDFs. It is useful to understand the dependence of $f(x)$'s convergence on this power divergence in the zero error case. If we try to reconstruct $f(x)$ either by approximating it with spline interpolants or by assuming it is analytic at all points except at $x=0$, we will have to make some approximations to account for the singularity at $f(0)$. To attempt this, let us consider only the domain $x\in[\epsilon,1]$, where $0<\epsilon\ll 1$. 

\paragraph{Finitely differentiable case:}
Let us approximate $f(x)$ with cubic spline interpolants along a uniformly split grid in the region $x\in(\epsilon,1)$. In this case, we know that $f(x)$ will be 2 times absolutely differentiable ($k=3$), and so $V_k \leq \alpha(\alpha-1)(\alpha-2)\epsilon^{\alpha-3}$. As a function of moment added, our error will scale like $n^{-2.5}\epsilon^{\alpha}$.

\paragraph{Analytic case:}To consider the analytic approximation in the domain $x\in [\epsilon,1]$, we can  stretch our function in this domain out to $[-1,1]$ to get $f(x) = (\frac{1+\epsilon}{2}+\frac{1-\epsilon}{2}x)^\alpha$. This function has a singularity at $x = \frac{1+\epsilon}{\epsilon-1}$. Invoking (8.4) from \cite{trefethen_approximation_2019} we calculate $\rho = \frac{1+\epsilon}{1-\epsilon} + \sqrt{\Big(\frac{1+\epsilon}{1-\epsilon}\Big)^2 -1}$ which is in our case a number quite close to the semi-major axis length $\xi = \frac{1+\rho^2}{2\rho}$. The circumference can be calculated with the series expression for focus length $1$: $\ell = 2 \xi \pi \Big[ 1 - \sum^\infty_{n\;\mathrm{odd}}\frac{1}{n \xi^{n+1}}\Big(\frac{n!!}{(n+1)!!} \Big)^2 \Big]$. Our error scaling (accounting for $\alpha$ as in the finitely differentiable case) is $\mathcal{O}\big(\rho^{-n}\epsilon^{\alpha}\big)$.

The bounds we have here depend on both the degree of divergence in $x^\alpha$ as well as the amount of that divergence that we wish to reconstruct $f(x)$ (signified by $\epsilon$). Valence quark PDFs are restricted to $\alpha \in (-1,0] $, while gluon/sea quark PDFs have $\alpha \in (-2,0]$. The convergence rates from the two bounds above indicate that gluon/sea quark PDFs will take significantly more data to converge.

\paragraph{Unknown moment extrapolation + adding noise:}Given a sequence of moments, \cite{karlin_geometry_nodate} proved that provided $n>2$ exactly determined consecutively ascending moments that lie within a moment space of dimension $n$, the extrapolation to the $(n+1)^\mathrm{th}$ moment has a uniform uncertainty of at most $2^{-2n}$. When we introduce an error associated with each moment $\mu_i$ however, the possible modes of extrapolation explode combinatorially. For this reason, analytically estimating how noise may factor into our calculations is very complicated, and it is not clear what effect a single moment's uncertainty will have on the convergence bounds discussed directly above; this indicates that there is no analytical substitute for an empirical reconstruction of $f$. In this study we will see how our GPR empirical framework can robustly deal with PDF data across a spectrum of $x^\alpha$ divergence degrees and moment uncertainty levels.

\section{A Bayesian Reformulation}

While the reconstruction of PDFs in the zero error case can be determined with a large amount of moments through many diverse methods, this determination is often not robust upon the introduction of uncertainty in data. Ref. \cite{athanassoulis_truncated_2002}, for example, uses kernel density methods for reconstruction, and circumvents the ill-posed problem by filtering moment values to lie within appropriate moment spaces. This type of solution, however, depends on the data having zero error, and for quoted examples, a lot of moment data is needed to unambiguously converge to a given density distribution. This is not ideal, as in the vast majority of scenarios (the lattice being no exception, see Refs.  \cite{Francis:2025pgf,Francis:2025rya,Pang:2024kza}), the moments with which we are tasked with recovering a density function from are a) few in number, and b) have associated uncertainty. Satisfying the Hausdorff moment condition subject to the quadratic constraint placed by the existence of noisy data can often make the complexity of solving the said problem spiral out of control. Rather than solving such problems rigorously from scratch, it is often far easier to devise a solution via regression and verify it. One approach can be to use neural networks: much like the fitting done by NNPDF \cite{NNPDF:2014otw}, lots of data can be used to train a neural network to surmise what distribution can generate the noisy data that is obtained from simulation/experiment. The uncertainty of said reconstruction is determined by the volume and diversity of the training data used, and data will have to be carefully curated to prevent either overfitting or underfitting. It is with this consideration in mind that we turn to Bayesian methods. Bayesian methods offer a promising framework to carry out such calculations, where minimal prior assumptions imply posteriors conditioned on data that are robust to both changes in dataset size as well as error. The Bayesian framework we utilize revolves around \textit{gaussian processes} (GPs). The central tenant of the GP hypothesis is the claim that our PDF $\PDF$ can be completely specified by a mean value $m(x)$ and a covariance kernel $K(x,x')$ such that if we select a finite number of $x$ coordinates to form the vectors $\x$, $m(\x) = \m$, $K(\x,\x') = \K$, then our corresponding vector $\PDFv$ will be normally distributed: $\PDFv \sim \mathcal{N}(\m,K)$. 

\subsection{Analytical Posterior Formulation}
\label{subsec:analytical_posterior_formulation}
Suppose we select a prior mean and covariance \textit{kernel} $\m$ and $K$. Then our PDF is distributed according to:\footnote{We largely use variable conventions as presented in Ref. \cite{candido_bayesian_2024}}

\begin{equation}
    p(\PDFv|\m,\K) = \frac{1}{\sqrt{\det(2\pi\K)}} \exp\Big(-\frac{1}{2}(\PDFv-\m)^TK^{-1}(\PDFv-\m)\Big)
    \label{eq:plain}
\end{equation}

Now we need to account for having observed some moments of $\PDF$, $\y$, where $y_m = \xm$, with some associated error encapsulated in the covariance matrix $\Cy$. $\y$ is related to $\PDFv$ through an integral operator $\FK$: $\FK\PDFv=\y$. $\FK$ for the finite moment problem is defined as:

\begin{equation}
    \FK_{i,j} = \int dx \;x^i p_j(x)
    \label{eq:FKmom}
\end{equation}

Here, $p_j$ is the interpolating polynomial on the $j^\mathrm{th}$ spatial grid interval. Many splines can be used: cubic splines ensure continuity at each $\x$ grid-point for the first and second derivatives, B-splines can be constructed out of basis functions that are hyper-local, and further variants of B-splines like M-splines and I-splines ensure positivity and monotonicity, respectively. Since we obviously have a non-identity $C_\y$, we amend our previous equation to  $\FK \PDFv = \y + \epsilon$, where $\epsilon \sim \mathcal{N}(0,\Cy)$. Because our moment error and prior estimate of $\PDF$ are uncorrelated, we can modify \ref{eq:plain} to arrive at the probability of observing $\PDFv$ given $\m$, $\K$, and $\y$:

\begin{align}
    p(\PDFv|\m,\K,\y) &= \frac{1}{\sqrt{\det(2\pi\K)}} \times  \frac{1}{\sqrt{\det(2\pi\Cy)}} \nonumber \\
    & \quad \times \exp\left(-\frac{1}{2}(\PDFv-\m)^T K^{-1}(\PDFv-\m) \right. \nonumber \\
    & \quad \left. - \frac{1}{2} (\FK \PDFv-\y)^T \Cy^{-1}(\FK \PDFv-\y)  \right)
    \label{eq:master}
\end{align}

\ref{eq:master} is also a Gaussian, and we can easily rearrange terms to view what the posterior Gaussian distribution is:

\begin{align}
    p(\PDFv|\m,\K,\y) &= \frac{1}{\sqrt{\det(2\pi\K)}} \times \frac{1}{\sqrt{\det(2\pi\Cy)}} \nonumber \\
    & \quad \times \exp\left(-\frac{1}{2}\PDFv^T (K^{-1} +\FK^T \Cy^{-1} \FK) \PDFv \right. \nonumber \\
    & \quad \left. + (\m^T \K^{-1} + \y^T \Cy^{-1} \FK) \PDFv \right. \nonumber \\
    & \quad \left. -\frac{1}{2} \y^T \Cy^{-1} \y \right. \nonumber \\
    & \quad \left. -\frac{1}{2} \m^T \K^{-1} \m \right)
    \label{eq:posterior}
\end{align}

Let us view the exponential term in Gaussian form (obtained by ``completing the square")\footnote{We can view it in the form: $e^{-\frac{1}{2}\x^T A\x + \bb^T\x + c} = e^{-\frac{1}{2}(\x+A^{-1}\bb)^T A(\x+A^{-1}\bb) + \mathrm{const.}}$}. From here we can immediately see how our posterior inverse covariance matrix is $\tilde{\K}^{-1} = {\K}^{-1} + \FK^T \Cy^{-1} \FK$. Using the Woodbury matrix identity, and defining $\Cyk = \FK\K\FK^T+\Cy$, we can solve for both the posterior covariance $\tilde\K$ and posterior mean $\tilde\m$ that will completely characterize the posterior PDF $\tilde\PDFv$.

\begin{equation}
    \tilde{\K} = \K - \K\FK^T\Cyk^{+}\FK\
    \label{eq:post_cov}
\end{equation}

\begin{equation}
    \tilde \m = \m + \K\FK^T\Cyk^{+} (\y-\FK\m)
    \label{eq:post_mean}
\end{equation}

\subsection{Choice of priors + hyper-parameters}

 We can choose our priors given that we want a number of degrees of freedom such that we maximize both the simplicity and specificity of our model. We therefore introduce a limited number of \textit{hyper-parameters} to define the prior mean and covariance matrix. If we encode all of our hyper-parameters into a vector $\Theta$ then our posterior probability distribution will read:

\begin{equation}
    p(\PDFv,\Theta|\y) = p(\PDFv|\Theta,\y)p(\Theta|\y)
    \label{eq:full_post}
\end{equation}

In other words, we want to find the probability distribution function of our reconstructed PDF and hyper-parameters, both conditioned on our moment data. The $p(\PDFv|\Theta,\y)$ term is the density function that can be completely analytically determined (through GPR). The $p(\Theta|\y)$ term will be discussed later. In both of these terms, the determination of appropriate hyper-parameters is imperative.
As we shall see, both empirically and semi analytically, the mean prior is weakly important, while the covariance kernel plays a major role in determining posterior results. Below we explore construction of several different types of covariance matrices with a few hyper-parameters.

\subsubsection{Covariance Kernels}
\label{subsec:kernels_ootb}
Covariance kernels can be defined in a plethora of ways, under the general requirement of them being \textit{symmetric} and \textit{positive semi-definite}. Below are a few kernels, which we will term ``out of the box" (OOTB) for their simple forms and pervasive use across several fields of study.

\paragraph{Squared Exponential (SE) (Hermite):}

\begin{equation}
    K(x,x')=\sigma^2 \exp\Bigg(\frac{-(x-x')^2}{2\ell^2}\Bigg)
    \label{eq:SE}
\end{equation}

This kernel is commonly used \cite{Rasmussen2006Gaussian} as a one shot in giving quick reliable posterior results. The two hyper parameters used here will be used in analogous contexts in later kernels: $\sigma$ signifies the kernel variance, while $\ell$ signifies the \textit{$x$ correlation length}. This kernel only depends on the distance between two points $x,x'$, and is thus a popular choice to model time series. For the purposes of reconstructing our density function, we can see how this may not be desired: the functions we desire to reconstruct is only supported on $(0,1)$, and other than a few constraints we will impose, we want a much more flexible model than this one. Before we turn away from this, it is worth specifying the eigendecomposition of this kernel. Mehler's formula \cite{kibble_extension_1945} can be used to express the outer product of Hermite polynomials as a form of the squared exponential kernel, with parameter $\rho\in(0,1]$:

\begin{align}
    K(x, x') &= \sum_{k=0}^{\infty} \frac{\rho^k}{k!} H_k(x) H_k(x') \nonumber \\
    &= \frac{1}{\sqrt{1 - \rho^2}} \exp\left( \frac{-\rho^2(x^2 + x'^2) - 2\rho xx'}{2(1 - \rho^2)} \right)
\end{align}

We can see that $\ell^2=\frac{1-\rho^2}{\rho^2}$ when matching onto \ref{eq:SE}. Mercer's theorem \cite{Rasmussen2006Gaussian} allows us to re-express a kernel in terms of its eigen-functions/values and a probability density weight that defines the set of eigen-functions' orthogonality relations. In our case we can express the SE kernel as being composed of a complete orthonormal set of (Hermite) polynomials over $x$ with a Gaussian orthogonality weight over the domain $x\in \mathbb{R}$. This means that the set of eigen-functions over the domain $x\in(0,1]$ is dense, so $f$ can be predicted to an arbitrary degree of precision given increasing quantities of data. In an application to the moment problem, the primary caveat is that we do not have any information (other than perhaps an endpoint or two) regarding spatial point-wise constraints, only integral constraints. It is in any case a good idea to begin with a kernel that is a universal approximator as opposed to one that is not; we will see how the SE kernel and its variants are powerful nevertheless.

\paragraph{Mat\`ern and Rational Quadratic (RQ):}

It is useful to mention the Mat\`ern and Rational Quadratic kernels, defined in terms of the spatial difference unit $r = |x-x'|$, with a variance term added in:

\begin{equation} \mathrm{Mat\grave{e}rn: \;}
    K(r)=\sigma^2\frac{2^{1-\nu}}{\Gamma(\nu)}\Big(\frac{\sqrt{2\nu}r}{l}\Big)^{\nu}K_\nu\Big( \frac{\sqrt{2\nu}r}{l}\Big),
    \label{eq:Matern}
\end{equation}

\begin{equation}
    \mathrm{RQ:\;}K(r)=\sigma^2 \Big(1 + \frac{r^2}{2\alpha l^2}\Big)^{-\alpha}
    \label{eq:RatQ}
\end{equation}

Both of the functions can be proven to be p.s.d through B\"ochner's theorem \cite{Rasmussen2006Gaussian} which states that a kernel is p.s.d. if its Fourier's representation is fully supported on the interval in which regression is being performed. Both of these kernels yield the SE kernel in specific cases: for the Mat\`ern kernel, when $\nu\to\infty$, and for the RQ kernel when $\alpha \to 1$. The Mat\`ern kernel can model a wide class of data with much more flexibility than the SE kernel. The covariance kernel has simple expressions for half-integer values of $\nu$. The $\nu$ parameter directly influences the $x$ correlation length of the prior/posterior; for $\nu=\frac{1}{2}$ we have very sort $x$ correlation lengths making this kernel ideal for simulating processes like Brownian motion or stock options \cite{Rasmussen2006Gaussian}. As we increase $\nu$, we simulate processes with increasingly longer correlation lengths until we arrive at the SE kernel at $\nu \to \infty$. A good rule of thumb to keep in mind when using the Mat\`ern kernel is that (eq 2.7 in Ref. \cite{stein_interpolation_1999}) the path sampled with the Mat\`ern kernel at a smoothness parameter $\nu$ is $m$ times mean-square-differentiable iff $\nu > m$ \cite{Porcu2023-ga}.

\paragraph{Logarithmic SE (LSE)}

The SE kernel is stationary, i.e. a function of only $|x-x'|$. This assumption may be rather restricting for our purposes, and it is possible to design simple nonstationary kernels that can capture physical priors that stationary ones cannot: in the context of hadron physics, PDFs are understood to behave differently for large $x$ than they are for $x\ll 1$. To account for this, Ref. \cite{dutrieux_simple_2024} first suggested making the non-linear transform $x\to\log(x)$ to create a logarithmic SE kernel (LSE). The kernel would then look like:

\begin{equation}
    K(x,x')=\sigma^2\exp\Bigg(\frac{-(\log x-\log x')^2}{2\ell^2}\Bigg)
    \label{eq:LSE}
\end{equation}
This enforces the constraint of PDF behavior at $x\ll 1$ being fundamentally decorrelated from behavior at $x\sim 1$, and is quasi-stationary in the sense that the kernel can be specified in terms of the ratio of spatial coordinates $q=x/x'$. We can also think about this transform as ``stretching" the probability measure out for all $x\in(0,1]$ while suppressing $x>1$. While the Hermite eigen-functions constituting \ref{eq:SE} are orthogonal on $x\in\mathbb{R}$ under a Gaussian weight, the LSE kernel \ref{eq:LSE} is composed of variants of the Hermite polynomials, $H_n(\log|x|)$, which are orthogonal on $x\in\mathbb{R}^+$ under a log normal weight. The eigen-bases for both SE and LSE kernels are dense on $x\in[0,1]$ under Gaussian and log normal density functions, respectively. It may well be advantageous that both kernels are not orthogonal on $x\in[0,1]$, as an orthonormal basis would induce rapid unstable fluctuations at $x=1$, which are not reflective of our physical expectation of relatively smooth behavior of $f(x)\to0$ as $x\to 1$ . Additionally, the log normal probability weight which governs the kernel expansion for the LSE kernel, in the limit that $\ell\to\infty$, is equivalent to $\frac{1}{x}$; this may be indicative of the LSE kernel's particularly effectiveness at reconstructing the power-divergence-like behavior in $f(x)$ from PDF data, although further studies of this are warranted.

\paragraph{Non-Stationary (Gibbs) Kernels}
\label{para:Gibbs}
The kernels discussed above are but a tiny assortment of kernels that can be used to model data. These kernels can also maintain their positive semi-definite-ness upon multiplication and addition, which can improve their flexibility. A particularly useful extension modifies the correlation $\ell(x)$ itself to be a function of $x$ to correct for any systematic limitation that a pre-defined kernel may have. To gain control of our $x$ correlation function, Ref. \cite{Gibbs1997} first suggested the alteration of the basis functions used to create the SE kernel such that we are now free to make $\ell$ an arbitrary function of $x$. This ``Gibbs" kernel reads:

\begin{equation}
    K(x,x')=\sigma^2\sqrt\frac{2\ell(x)\ell(x')}{\ell(x)^2+\ell(x')^2}\exp\Bigg(\frac{-(x-x')^2}{\ell(x)^2+\ell(x')^2}\Bigg)
    \label{eq:GibbsSE}
\end{equation}

This Gibbs kernel can in principle be applied to any existing kernel to adjust for any imposed spatial dependent $x$ correlation lengths. We will see later how this adjustment can be used to effectively reconstruct a PDF given a finite number of moments.

\paragraph{Modifying kernels}
Given existing kernels, we can apply multiplicative, additive, and nonlinear transformations and still maintain properties of covariance matrices \cite{Rasmussen2006Gaussian}. We may even combine different kernels as a function of spatial regime as described in Refs. \cite{Rasmussen2006Gaussian,medrano_gaussian_2025}.

\subsubsection{Mean priors}
We may parameterize our mean prior in similar ways to our kernels. As discussed in Ref. \cite{dutrieux_simple_2024}, we can set our mean to be flat and of a magnitude set by the variance of our covariance kernel: $\m = \sigma$, thereby weakly enforcing positivity of our PDF. For even stronger priors we can use parametric forms.

\subsubsection{Hyperparameter priors}
\label{subsubsec:hyperparameter_priors}
Let us now turn to the $p(\Theta|\y)$ term in Eq. \ref{eq:full_post}. This is the distribution from which we draw our hyperparameter prior samples. Bayes theorem tells us that it is proportional to: $p(\Theta|\y)\propto p(\y|\Theta)p(\Theta)$. At the prior level, both the mean and covariance matrix are functions of the hyperparameter $\Theta$  vector: $\m(\Theta)$, $K(\Theta)$, and according to this hypothesis; $\y$ is then centered around $\FK \m(\Theta)$ and has an error informed by both the inherent data covariance $\Cy$ and the covariance kernel $\FK K(\Theta) \FK^T$. Therefore we can write down the appropriate distribution using quantities defined above Sec. \ref{subsec:analytical_posterior_formulation}:

\begin{equation}
    p(\y|\Theta) = \frac{e^{-\frac{1}{2}(\y-\FK\m)^T\Cyk^+ (\y-\FK\m)}}{\sqrt{\det(2\pi\Cyk)}}
    \label{eq:hyp_samp}
\end{equation}

Tackling \ref{eq:hyp_samp} is only possible non-analytically due to the variable denominator term.  We can optimize this function and find the value of $\Theta$ that is a maximum a posteriori estimator of $p(\Theta|y)$. The most error-proof way, as detailed in Ref. \cite{candido_bayesian_2024} of characterizing $p(\y|\Theta)$ without underestimating the variance in hyper-parameter selection is through sampling \ref{eq:hyp_samp} through joint sampling of all elements of $\Theta$ with a Markov Chain Monte Carlo (MCMC)-like unbiased sampling algorithm. With this in place, we can effectively characterize $p(\PDFv,\Theta|\y)$, and thus accurately quantify the posterior error on our $\PDFv$.

\paragraph{MC Sampling}

The most popular way of sampling a distribution in an unbiased manner is through MCMC methods, where our hyperparameter state space  (in our case $\Theta$) is thoroughly sampled via stochastic updates of a given $\Theta$ vector consistent with the sequence $\Theta_1\to\Theta_2\to...\to\Theta_n$ being a Markov process. Arguably the most successful MCMC framework is the Hamiltonian Monte Carlo algorithm (class) \cite{Duane:1987de}, where a  momentum space conjugate to the state space is introduced, returning state sequences that are sampled through Hamiltonian dynamics. HMC is much more efficient than vanilla Metropolis MC, and is used for a wide array of scientific fields, including gauge field generation in lattice QCD. The arguably biggest problem plaguing HMC is the difficulty in tuning Monte Carlo parameters within HMC to optimize efficiency.\footnote{For example, minimizing the autocorrelation length for a certain quantity we want to sample} This is an area of active research, and several modifications to HMC have been proposed to further increase sampling efficiency and ameliorate tuning. The current state of the art HMC variant is the No U-turn Sampler (NUTS) \cite{hoffman_no-u-turn_2011}, whose dynamics are designed to avoid sampling redundant areas of phase space (that HMC might repeatedly sample, thereby wasting compute time), while being amenable to techniques to tune MC parameters automatically. For our study, we will use NUTS as a part of the \texttt{Blackjax} \cite{cabezas2024blackjax} library to sample $p(\Theta|\y)$.

\subsection{Prior Induced Bias/Variance }
\label{subsec:prior_bias}

In order to understand the sources of error in our Bayesian reconstruction scheme, we follow an analysis that can be found in both Refs. \cite{valentine_gaussian_2020,candido_bayesian_2024}. To begin, we notate the difference between the true PDF $\PDFv_T$ (selected from an existing dataset in $x$ space) and a reconstructed PDF $\tilde{\PDFv}$, which is one we obtain through GPR (Bayesian framework or whatever). Let us consider a ``resolution kernel" $R(x,x')$, that can act as our original PDF $\PDFv_T$ to yield a reconstructed PDF. For our GPR framework (i.e. our characterization of $p(\PDFv|\y,\Theta)$), this kernel is defined as \cite{valentine_gaussian_2020}:

\begin{equation}
    R(x,x') = K(x,x')\FK^T\Cyk^+\FK 
    \label{eq:Res}
\end{equation}

This object measures the ``distance" between the reconstructed function and the true function: consider this kernel in the limit that experimental error vanishes, i.e. $\Cy\to0$ (with the $\_^0$ subscript denoting quantities with zero data uncertainty):

\begin{align}
    R^0(x,x') &= \lim_{\Cy \to 0} R(x,x') \notag \\
              &= K(x,x')\,\FK^T \left(\FK\, K(x,x')\, \FK^T\right)^{-1} \FK
    \label{eq:Res0}
\end{align}

We can then express the \textit{bias}, i.e. the posterior mean deviation from the true function, and the \textit{variance} of our posterior prediction as: \footnote{note that an intermediate step is $\tilde{\m} -\m = R^0(x,x')(\PDFv-\m)$. Ref. \cite{candido_bayesian_2024} points out that if our mean prior is zero, we immediately arrive at $\tilde{\m} = R^0(x,x')\PDFv$, however, we additionally notice that if our mean prior is a function of a form that can be completely specified by one (or some) of the rows of $\FK$, then $R^0(x,x')\m = \m $ and the identity $\tilde{\m} = R^0(x,x')\PDFv$ holds nevertheless}

\begin{equation}
    \mathrm{bias:\;}\mathcal{B^0} = \Delta\m = \tilde\m - \PDFv^0_T = (R^0(x,x')-I)\PDFv^0_T
\end{equation}

\begin{equation}
    \mathrm{variance:\;}\mathcal{V^0} = \FK \tilde{K} \FK^T = \FK (I-R^0(x,x'))K \FK^T
\end{equation}

$R^0$ looks somewhat like a smoothing kernel: information r.e. the original function is lost upon $\FK$ acting on $\PDFv_T$, much in the same was as a frequency filter. If we had rows of $\FK$ forming a complete orthogonal basis, or if $\FK$ is able to completely specify $\PDFv_T$, then $R^0 = I$, and our bias and variance of our posterior estimates go to zero. Moreover, if $\K$ is not of sufficient rank to capture the features of our PDF, we will have an additional source of information loss.

 In practice, we always have errors associated with our measurements, and in this case we must use the full resolution kernel \ref{eq:Res}. Here, the expressions for the bias become significantly more complicated. Quoting from Ref. \cite{candido_bayesian_2024}, $a^T   \coloneq \K \FK^T \Cyk^+$, we have:

\begin{equation}
    \mathcal{B} = \FK(R-I)\PDFv^0_T+\FK a^T\epsilon
    \label{eq:post_bias}
\end{equation}

\begin{equation}
    \mathcal{V} = \FK (I-R)K (I-R)^T\FK^T + \FK a^T \Cy a\FK^T
    \label{eq:post_var}
\end{equation}

A crucial realization is that the bias expression contains an \textit{unwanted} term depending on $\Cy$. This will skew the posterior mean from its true location, and make the error much more sensitive to the choice of hyper parameters. It is by all means possible to achieve near-zero bias in this case, although this will depend on the covariance kernel that is chosen given a covariance matrix. Such a selection is nontrivial, as it depends not only on the covariance kernel but also on quality and the quantity (i.e. number of moments) of data provided. 

\subsubsection{Bias correction}
\label{subsubsec:bias_corr}
We can circumvent the problem of error induced mean bias if we are provided with an estimate of what the (unbiased) mean posterior is (we will term this \textit{bias correction}). Naively we might assume that we can simply remove the $\Cy$ term from our posterior, thereby setting $\Cyk\to\FK K \FK^T$, after which we can proceed with our sampling + GPR pipeline. Unfortunately, the matrix $\FK K \FK^T$ is often ill-conditioned, making efficient sampling difficult. We instead introduce a simple extrapolation technique consisting of repeating our sampling + GPR framework with progressively deflated values of $\Cy$: $\Cy \to \lim_{\lambda\to\infty} \frac{\Cy}{\lambda}$, resulting in convergence to the zero-noise unbiased mean, $\Cyk\to\FK K \FK^T$, and $R(x,x') \to R^0(x,x')$. As we will see, $\lambda$ does not have to be too high for us to empirically observe a convergence to the unbiased mean, and we can thus avoid having to deal with an ill conditioned problem. Following this, adjustments to the posterior calculations performed with the true $\Cy$ can be made to eliminate data induced bias.

\subsection{Implementing constraints}

Constraints fall into two broad categories: equality constraints and inequality constraints. Equalities (with any associated uncertainty) can be rather easily implemented within the $\FK$ matrix itself as an extra row. For example, enforcing that $f(1)=0$ is achieved by adding a row to $\FK$ where the final entry is set to one, and appending a zero onto the existing $\y$ vector. The direct implementation of inequality constraints is a bit more subtle. A naive method would be to sample from a truncated Gaussian: we simply reject samples which fail an inequality constraint; depending on the strength of the constraint, this method may be computationally inefficient. A more computationally efficient method would be to project the reconstructed PDF onto a basis that enforces any inequality constraint explicitly. As an example, take the restriction that $\PDF \geq0$. We can enforce this by replacing the cubic spline basis of $\FK$ with M-splines \cite{Ramsay:1988mspline}. M-splines are non-negative variants of B-Splines and a projection onto them will result in a non-negative function. The non-negativity property of M-splines can be easily extended to any inequality constraint where $\PDF$ is bounded by a constant value. Another example could be the enforcement of monotonicity,\footnote{There is no theoretical need to enforce monotonicity of PDFs, however, it may be employed if moment data is too unconstrained.} in which case M-splines can be further replaced by I-splines \cite{Ramsay:1988mspline}. While these properties of M and I-splines seem attractive, they will incur a cost of numerically conducting the integration within $\FK$ that was previously analytic with cubic splines. No studies compare the choice of interpolating polynomials in constructing $\FK$, but there may perhaps be no gain in using non-cubic spline interpolants if they are not absolutely needed. In our empirical study we will stick entirely to cubic spline interpolants.

\subsection{Techniques to adapt to non-linear integral transforms}
\label{subsec:nonlin}
We have discussed the consequences of the GP hypothesis in the context of $\FK$ being a linear (integral) operator. If $\FK$ is non-linear, we no longer have an analytical GPR posterior prediction; there are however a few things we can do to overcome this. The last resort solution would be to treat each entry in $\PDFv$ as a hyper-parameter which would be sampled in tandem with all other hyper-parameters. This can be applied to any operator in practice, but we will not consider this case due to a lack of implementation strategy to solve this problem efficiently. We will instead discuss two more practical methods. 

\subsubsection{Linearization}
\label{subsec:nonlin:linear}
This technique, mentioned in Ref. \cite{debbio_bayesian_2022}, is applicable to any operator, but  requires the  assumption that we pick a suitable mean prior $\m_0$ such that our mean posterior will not be ``far off" from. After this, we can approximate $\FK_\mathrm{NL}$ as:

\begin{equation}
    \FKN = \FKN (\m_0)+ \nabla_\PDFv \FKN|_{\m_0} (\m-\m_0) + \mathcal{O}(\Delta\m^2)
    \label{eq:FK_nonlin}
\end{equation}

With this approximation, we need only make a few redefinitions to our prior data, mean, and covariance matrices above, and we can carry out our analysis just like we did in the linear case:

\begin{equation}
    \y \to \y - \FKN(\m_0) + \nabla\FKN|_{\m_0}(\m_0)
\end{equation}

\begin{equation}
    \FK \to \nabla\FKN|_{\m_0}
\end{equation}

This ``linearized" method is a simple way of casting a nonlinear problem as a linear one. An obvious drawback of this method is that it relies on the prior mean value being close to the posterior, and the magnitude of the error $\mathcal{O}(\Delta\m^2)$ and its effects on bias and variance of the posterior are uncontrolled. Another major drawback of this method is that we cannot perform bias correction on it, as deflating $\Cy$ will no longer result in convergence to the mean values of our data $\y$.

\subsubsection{Exact sampling}
\label{subsec:nonlin:exact}
 Suppose we have a nonlinear operator $\FKN$ which acts on  a PDF $\PDFv$. If there exists a parameter $\Pi$ within $\FKN$ which we can set constant and thereby retrieve a linear operator: $\FKN|_{\Pi =\mathrm{const.}} \coloneq \FK$, we can exactly sample $\Pi$ and maintain analyticity of our GPR. 
 
As an example, let us take this operator acting on $f(x)$:

\begin{equation}
    \FK^R_{m,j}f(x) = \frac{\int_{x \in [x_j, x_{j+1}]} dx \;x^{m} p_j(x)f(x)}{\sum_{j}\int_{x \in [x_j, x_{j+1}]} dx \;x p_j(x)f(x)}
    \label{eq:FKratmom}
\end{equation}

This operator returns ratios of moments $\xm/\langle x \rangle$, and is of particular relevance to a new gradient flow-based method on the lattice \cite{Francis:2025pgf,Francis:2025rya} which calculates $\langle x^n \rangle/ \xmom$ up till $n=5$.\footnote{Ratios of moments $\langle x^n \rangle/ \langle x^m \rangle$ are the simplest moment-related quantity that can be obtained from gradient flow
\cite{Shindler:2023xpd}. When $\xmom$ of PDFs is calculated with gradient flow, additional renormalization of fermions needs to be taken into account.} According to the prescription above, our parameter $\Pi$ is the first moment: $\Pi=\xmom$. We thus add $\xmom$ as an additional hyperparameter to sample \ref{eq:hyp_samp} by replacing $\FK^R\to\FK/\xmom$. This effectively ``off-loads" any nonlinearities of \ref{eq:FKratmom} to the sampling step of our framework while maintaining the analyticity of the GPR step.

The exact sampling method eliminates some of the persistent bias issues with the linearization method; regarding bias correction, if the sampled value of $\xmom$ is sufficiently peaked, i.e. there is an unambiguous peak of sampled values of $\xmom$ that, when combined with the moment ratio sequence produces a most probable moment sequence distribution then bias correction can be performed successfully. We will leave the analysis of moment ratios to a subsequent study.

\subsection{Validating Results}
\label{subsec:validation}
Validation is the final step of our analysis pipeline, and aims to provide a quantitative description of how well our Bayesian framework reconstructs our PDF. Given some sparse data $(\y,\Cy)$, we can measure how well our reconstructed PDF moments (and variants) match our experimental data, i.e. a measure of the ``loss" $L=||\y_\mathrm{recon}-\y||$. Even a perfect moment reconstruction (zero loss) does not guarantee a perfect match to the underlying model. If we know the provided model and are thus provided with ``truth" PDFs, we can create a ``fractional error" statistic between reconstruction and truth and see, i.e. how well respective percentile bands match between the reconstruction and the truth PDFs, and thereby perform closure tests. 

If for a given prior choice our reconstruction loss and validation error is not low (egregiously high), then our reconstruction is unsatisfactory, and we must turn to a new prior choice. However, if it is low, then we must ask the additional question of how much our choice of hyper-parameters influenced the posterior reconstruction. This can be further developed into a Bayesian analogue of a goodness of fit test which captures not only how well our model reconstructs given data, but how well it does not over/under-fit data. Some ways of how to do this are discussed in \cite{candido_bayesian_2024,valentine_gaussian_2020}. We will focus on just one validation measure, which casts our problem through the lens of \textit{information gain} in our GPR step. We want to quantify how the GPR flow from a prior to a posterior distribution (of $p(\PDF|\Theta,\y)$) has constrained our reconstruction. To do this, we need to use a metric that calculates the divergence between two density functions. Following from Ref. \cite{valentine_gaussian_2020}, we use the KL divergence 
\cite{kullback_leibler} as metric applied to two functions $f,\;\tilde f$:

\begin{equation}
    D_{KL}(\tilde f | f) = \int\tilde{f}(x) \log\frac{\tilde f(x)} {f(x)}dx
    \label{eq:KL}
\end{equation}

If these density functions are Gaussian, their analytical form can be written in closed form (Eq. 23 from \cite{valentine_gaussian_2020}), and we will use it specifically to visualize information gain in PDF reconstruction as a function of $x$.

Here the base of the logarithm sets the unit of ``information gain", and for the  natural base, this unit is a `nat', and differences in values of this metric for different GPR models can offer valuable input on which models describe data better. For a given posterior reconstruction, this measure also sees the highest increase when the prior is as unconstrained as possible, and thus has maximum variance, which would come about from a distribution with a relatively small amount of hyper-parameters, whereas if a large amount of hyper-parameters constrain our prior, our posterior is likely to be peaked around the prior peak and the information gain will not be as dramatic. In other words, a reconstruction that results in the most information gain from as few parameters as possible is the best one. An entire Bayesian goodness of fit test can be seen as being composed of two parts: a) measuring this KL divergence between the prior and posterior, thereby quantifying the degree to which over-under-fitting occurs and b) measuring reconstruction fidelity/loss.

\subsection{Contrasting values of $D_{KL}$ for successful reconstructions}
\label{subsec:contrast_DKL}
A reconstruction with a \textit{high} information gain and \textit{low} loss corresponds to a successful and robust Gaussian Process. Let us talk about the case with \textit{low} information gain and \textit{low} loss. In this case, our priors are highly constrained: consequently the Gaussian process played no meaningful role in our reconstruction, and rather, our sampling of $p(\Theta|\y)$ played the primary role in characterizing the posterior. In terms of hyper-parameters, this occurs when our mean prior is itself a parametrized hypothesis of the final function, and has with hyper-parameters $\Theta_\mathrm{para}$ (as opposed to a uniform function). In this case, our characterization of $p(\Theta|\y)$  will result in well sampled values of $\Theta_\mathrm{para}$, but will also result in the our covariance kernel variance $\sigma$ being sharply peaked at $0$, with all other GP related hyper-parameters effectively becoming nuisance parameters. Therefore, in this case, our reconstruction is akin to a non-Gaussian parametric fit, and is a fundamentally different type of reconstruction than a Gaussian Process fit, a table of which is presented in \ref{tab:DKL_diff}. Our framework can seamlessly process both types of reconstruction, and we will view instances of both in our applications section. The matter of whether to use a parametric fit as opposed to
GPR is beyond the scope of this paper. Needless to say that different metrics will have to be used to compare the quality of fits to each other.

\begin{table}[]
\resizebox{\textwidth}{!}{%
\begin{tabular}{lll}
                                              & High $D_\mathrm{KL}$ & Low $D_\mathrm{KL}$                            \\ \cline{2-3} 
\multicolumn{1}{l|}{Primary Reconstruction Mechanism} & Gaussian Process Fit & (non-Gaussian) Parametric Fit                  \\
\multicolumn{1}{l|}{Primary Error Source}     & GP Covariance Kernel & Distribution of sampled $\Theta_\mathrm{para}$
\end{tabular}%
}
\caption{}
\label{tab:DKL_diff}
\end{table}

\section{General Guidelines for Application of GPR Framework on PDF Datasets}

The developed GPR framework can in principle be applied to a reconstruction problem given any arbitrary density function. However, depending on the problem field, the prior assumptions that govern the tuning of $p(\Theta)$ and the covariance kernel vary significantly, so for our purpose, we will limit our study to reconstructing PDFs from moments, and that too, from models directly obtained from experimental data.

For the purposes of our problem, PDFs behave almost exactly like normal density functions. The only hard requirement we place on them is the requirement that $f(1)=0$ for each parton. An important difference between PDFs and density functions is the requirement of integrability, the details of which are presented in the introduction. A looser requirement we place is the decorrelation between mid-to-high $x$ and low $x$ regions in $\PDF$. The positivity condition $\PDF \geq 0$, is loosely enforced through allowing $\sigma > 0$ when defining $ p(\Theta)$. Although positivity is a general property of probability density functions, only loose as opposed to strict enforcement is implemented  for a couple of reasons. First, we will see how a strict enforcement will empirically not be needed. Second, PDFs are not required to be positive definite at all energy scales \cite{Candido:2023ujx}, and especially for gluon/sea PDFs with high uncertainties, we want our framework to model this occasional negativity if it is reflected in moment data. 

\subsubsection{Grid/Interpolation details}
\label{subsec:grid_details}
As discussed in Sec. \ref{subsec:analytical_posterior_formulation}, our FK table operator $\FK$ acts on a discretized function $\PDFv$ that is a function of a discretized $x$ grid, $\x$. The ``integration" of the operator is achieved by interpolating $\PDFv$ with splines (cubic in our case).  Polynomial interpolation allows us freedom in choosing grid spacings while maintaining as robust a reconstruction as possible. This will be useful in our various use cases, as the power divergences in the $x\ll 1$ region may warrant logarithmic spacing while for the high $x$ region, a linear spacing may suffice.  For this reason, we employ a hybrid grid scheme, where we specify a low, mid, and max $x$ point $(x_\mathrm{min},x_\mathrm{midpoint},x_\mathrm{max})$, where for $x\leq x_\mathrm{midpoint}$, we employ logarithmic spacing and for $x\geq x_\mathrm{midpoint}$ we employ a linear spacing. \textit{The grid scheme(s) to generate truth PDF(s) (and their moments) is kept independent of the grid schemes that we will use in reconstruction.} The different grid schemes are detailed below.
\begin{itemize}
    \item Truth grid scheme: This hybrid grid scheme is used to generate the truth PDF sets with \texttt{LHAPDF6}, and is kept constant for all PDF sets. The $x$-grid points are given in the LHAPDF config file for each dataset and generally follow a logarithmic spacing at low $x$ and linear spacing at high $x$. For valence PDFs it is validated that this grid results in valence sum rules being satisfied.
    \item Grid scheme 1: 100 linearly spaced points in the region $x\in[1\times 10^{-1},1]$ and 5 logarithmically spaced points in $x\in[1\times 10^{-5},5\times 10^{-1}]$
    \item Grid scheme 2: 50 linearly spaced points in the region $x\in[5\times 10^{-1},1]$ and 50 logarithmically spaced points in $x\in[1\times 10^{-5},5\times 10^{-1}]$
\end{itemize}

Grid scheme 1 is effectively a linear grid, while grid scheme 2 is effectively logarithmic.

\subsubsection{Enforcing priors} Recalling from our introductory discussion in Sec. \ref{subsec:math_theory_disc}, there is a power divergence at low $x$. This can be directly encoded into our covariance kernel: $K(x,x')\to x^{\alpha}K(x,x')(x')^{\alpha}$. $\alpha$ itself turns into a hyperparameter to be sampled. In order to satisfy valence number conservation, we require that valence PDFs are integrable ($\alpha\in(-1,0]$), and for gluon and sea quark PDFs, momentum sum rules must be satisfied ($\alpha\in(-2,0]$). Thus, when we add a power divergence prior, the prior bound for $\alpha$ will be modified appropriately. The $f(1)=0$ condition can be directly enforced via insertions of extra rows into $\FK$ and $(\y,\Cy)$ each. The characterization of the kernel will be done in a manner laid out throughout the  text. For our mean prior, we will largely consider the uniform case $m = \sigma$; this maintains weak positivity of our reconstruction when $\sigma$ is sampled from a positive distribution coupled with the fact that moment sequences satisfy the Hausdorff moment condition; otherwise it enforces very unconstrained priors. Another scenario we will briefly investigate is the parametric prior $m = x^\alpha (1-x)^\beta$.  This enforces a rather strict prior, and the uniform mean prior choice will almost always lead to a significantly higher information gain from prior to posterior. As laid out in Sec. \ref{subsec:contrast_DKL}, our framework can seamlessly handle this case where our PDF may not necessarily be a GP. With the $f(1) = 0$ requirement, we will find that restricting the range in which $\sigma$ is sampled in the uniform mean case will have an impact on gluon and sea PDF forms due to their very divergent behavior for $x\ll 1$.

\subsection{Sampling}
\label{subsec:sampling}
In our application section, we aim to completely characterize the probability density (as opposed to simply finding the MAP estimator) given in Eq. \ref{eq:full_post}, so to do this we must first sample hyper-parameter values according to Eq. \ref{eq:hyp_samp}. With increasing data quantity and precision, the $\Cyk^+$ term may become increasingly ill conditioned. Tuning our sampler is a nontrivial task, even with NUTS; we use a dual averaging technique \cite{nesterov_primal-dual_2009} that picks an optimal step size and inverse mass matrix to run the MC dynamical evolution and generate samples. Our tuning criteria aims for an acceptance rate of $>50\%$. To start off, we run 8 chains through an initial tuning step, and determine how many return a tuned MC step-size of $0.1$ or greater. Any less than this and we reject the initial configuration, as sampling will be too slow. If none of the 8 chains survives, we increase the number of chains and repeat the tuning process. We find that sampling becomes increasingly difficult for larger datasets where the fractional error remains fairly constant, sometimes gradually, and sometimes quite abruptly (i.e. all $n$ chains are accepted for $\nmax = 10$ and none are accepted for  $\nmax = 11$). A possible strategy around this is to use the sampled distribution from the immediately preceding $\nmax$ value, although this may introduce a non-negligible bias. Yet another solution is to try a change of kernel.

There is still the matter of characterizing $p(\Theta)$, i.e. the prior density function from which hyper-parameters are drawn. We will limit our study to choosing each of the parameters in $\Theta$ to follow uniform distributions on finite bounds. The reason we choose uniform distributions is due to a) simplicity of implementation and b) an effort to keep our framework as model agnostic as possible. Even with uniform distributions, we will see that in some cases, changing the bounds makes a noticeable impact on our reconstruction. In our entire framework, these bounds on uniformly sampled $p(\Theta)$ are the only hand-tuned parameters.

\subsection{Our datasets}  
\label{subsec:datasets}
The parton content of hadrons can be broken up into three categories: valence quarks, sea quarks, and gluons. Valence quark PDFs have the most constrained forms, whereas sea/gluon PDF uncertainties often are far less constrained, especially at low $x$. In phenomenological determinations, the degree of PDF constraint further depends on the total phase space ($x$ range) that experiments used in said determination were able to survey: the nucleon's partonic content has been extensively studied, for a wide $x$ and $Q^2$ range. On the other hand, the pion's parton content is more challenging to measure, and has not been studied nearly as extensively as the nucleon (only a handful of important experiments have been conducted  \cite{E615:1989bda,NA10:1987hho,WA70:1987bai}). To this end, we will study PDF phenomenological extractions performed by the NNPDF collaboration for the LHC Run II \cite{NNPDF:2014otw} for the nucleon and by the CTEQ-TEA collaboration using the \texttt{Fant\^omas} fitting framework \cite{Courtoy:2023bme,Kotz:2025lio,Kotz:2023pbu} for the pion. The NNPDF dataset, \NNPDF, includes DIS, DY production, vector boson production, ... etc. experiments in its analysis (a few examples are detailed in Refs. \cite{H1:2009pze,Moreno:1990sf,ATLAS:2011qdp}) and surveys a $x$ range of approximately $4\times 10^{-5}\lesssim x\lesssim 9\times 10^{-1}$ (full kinematic details including volume of data are given in Fig. 1 of Ref. \cite{NNPDF:2014otw}). Our \texttt{Fant\^omas} dataset, which we term \FANTO, contains much less statistics, with its experiments spanning a kinematic range $8\times 10^{-4}\lesssim x \lesssim 9.8 \times 10^{-1}$.\footnote{The maximum $x$ value was inferred from Fig. 2 of Ref. \cite{Pasquini:2023aaf}.}
\NNPDF datasets fit $x$ sensitive data to the parametric form given in Eq \ref{eq:para_form}, with the auxiliary $p(x)$ determined via neural network, for the \FANTO dataset, the model is much less reliant on Eq. \ref{eq:para_form}, and relies on universal approximator polynomials to fit data, additionally factoring in several sources of uncertainty to more faithfully represent PDF uncertainty given data. We do not consider lattice data in this study due to lacking a truth PDF; such an analysis will be left for a subsequent study. All of our PDF datasets were obtained at a renormalization scale of $\mu=2\; \mathrm{GeV}$ and processed with the \texttt{LHAPDF6} library \cite{Buckley:2014ana} and further processed with the \texttt{gvar} library \cite{lepage_gvar_2026}.

\begin{figure*}[h!]
    \centering
    \begin{subfigure}[t]{0.32\textwidth}
        \centering
        \includegraphics[width=\textwidth]{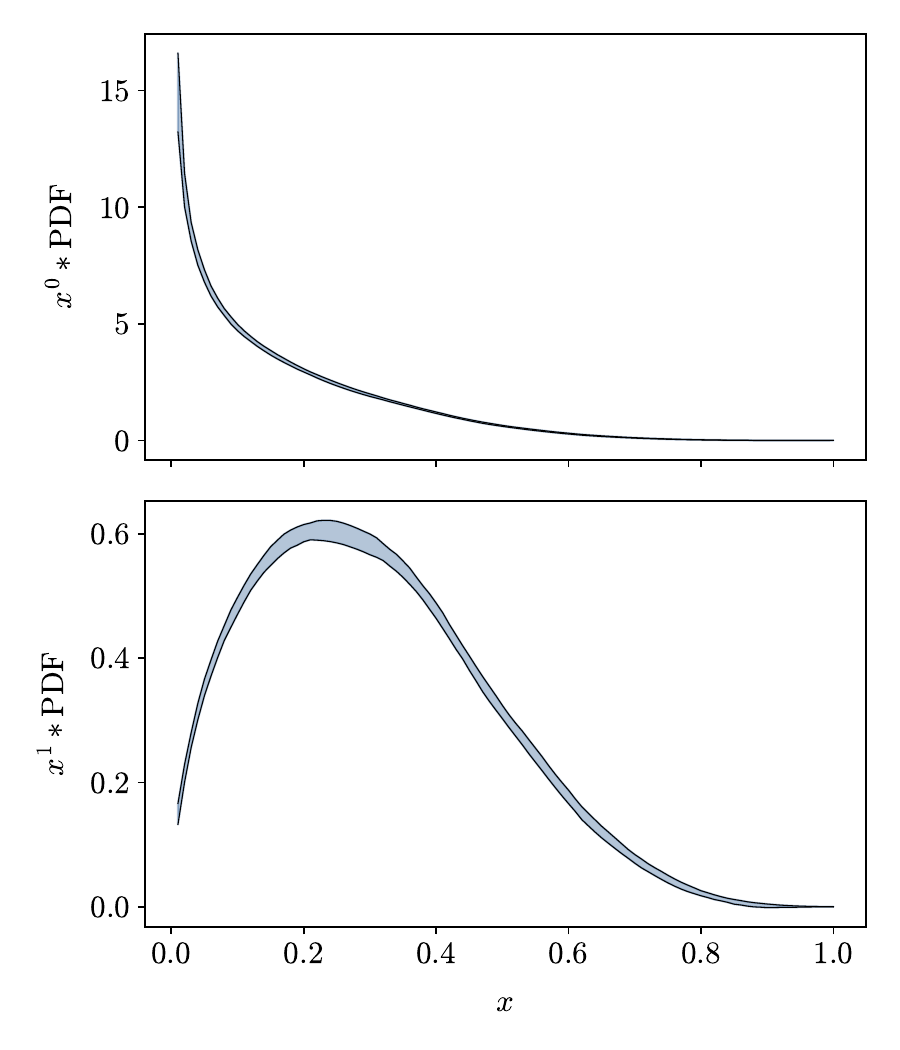}
        \caption{$u$ quark (valence) PDF}
    \end{subfigure}
    \begin{subfigure}[t]{0.32\textwidth}
        \centering
        \includegraphics[width=\textwidth]{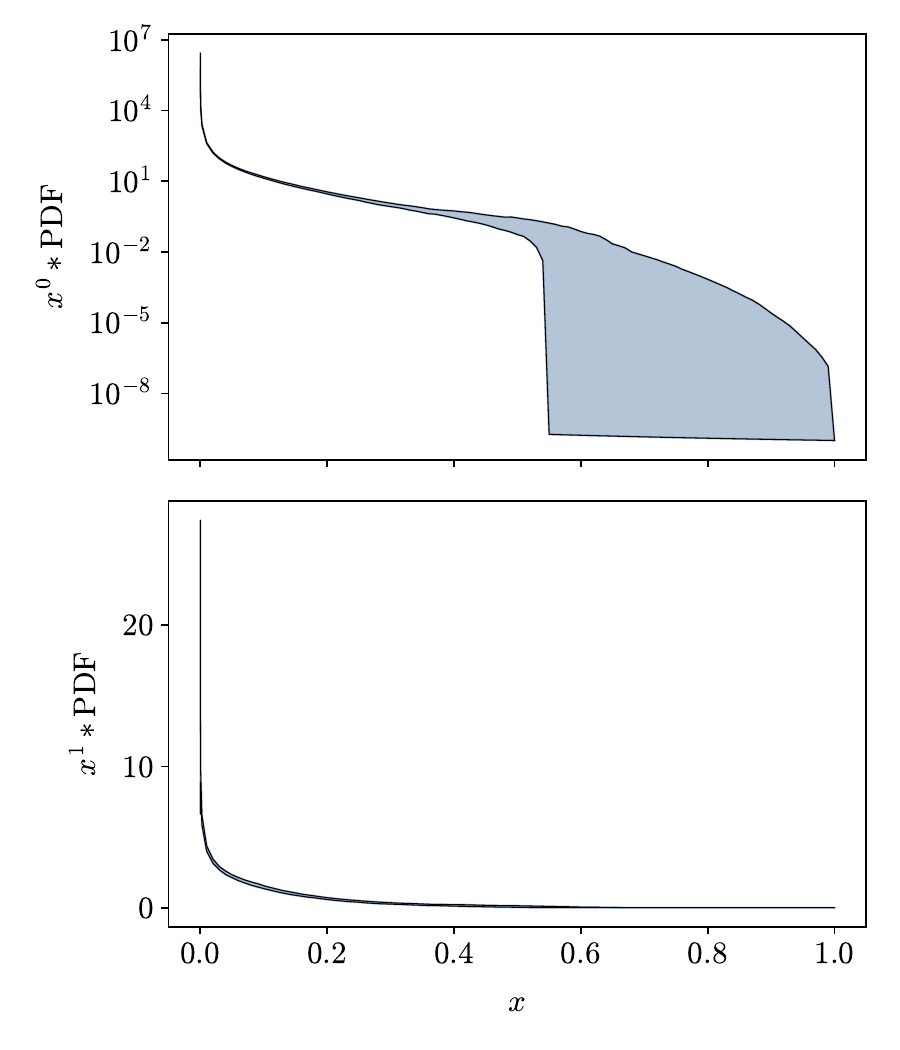}
        \caption{gluon PDF}
    \end{subfigure}
    \begin{subfigure}[t]{0.32\textwidth}
        \centering
        \includegraphics[width=\textwidth]{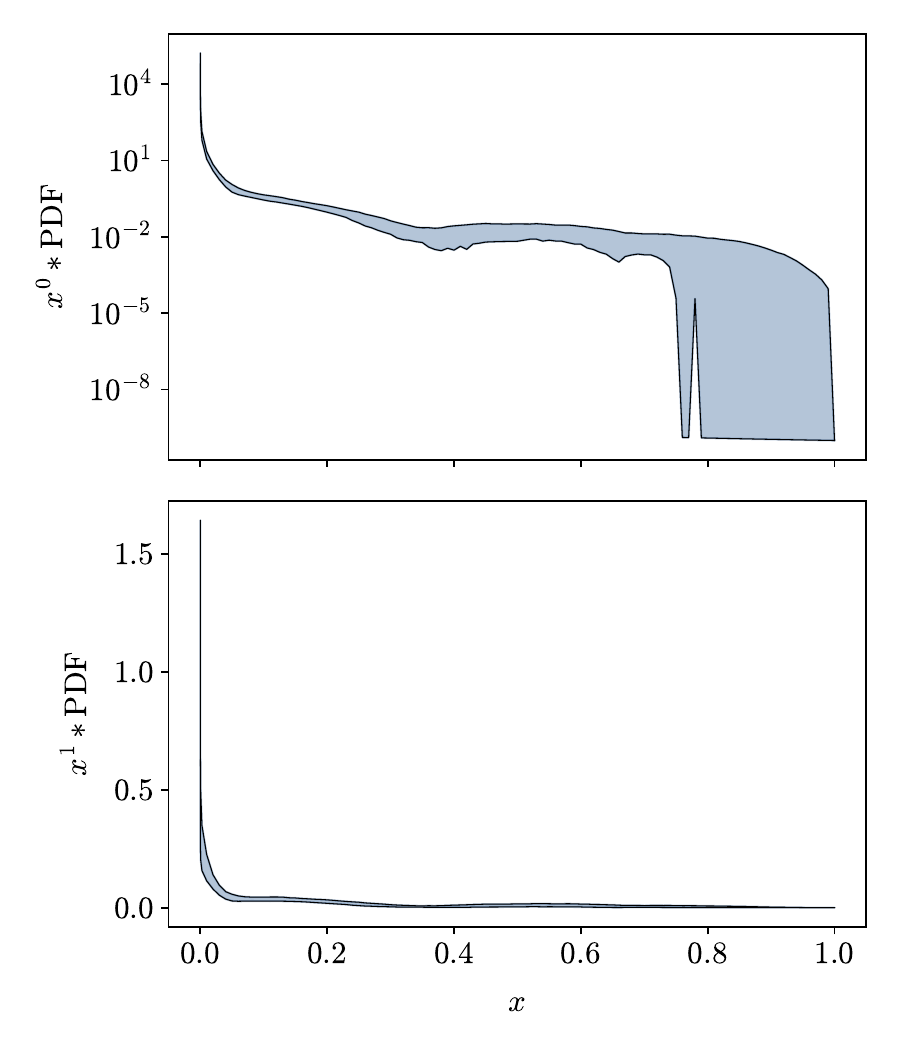}
        \caption{$s$ quark (sea) PDF}
    \end{subfigure}

    \caption{$f(x)$ and $x f(x)$ for the \NNPDF PDF dataset}
    \label{fig:NNPDF_truth}
\end{figure*}

\begin{figure*}[h!]
    \centering
    \begin{subfigure}[t]{0.32\textwidth}
        \centering
        \includegraphics[width=\textwidth]{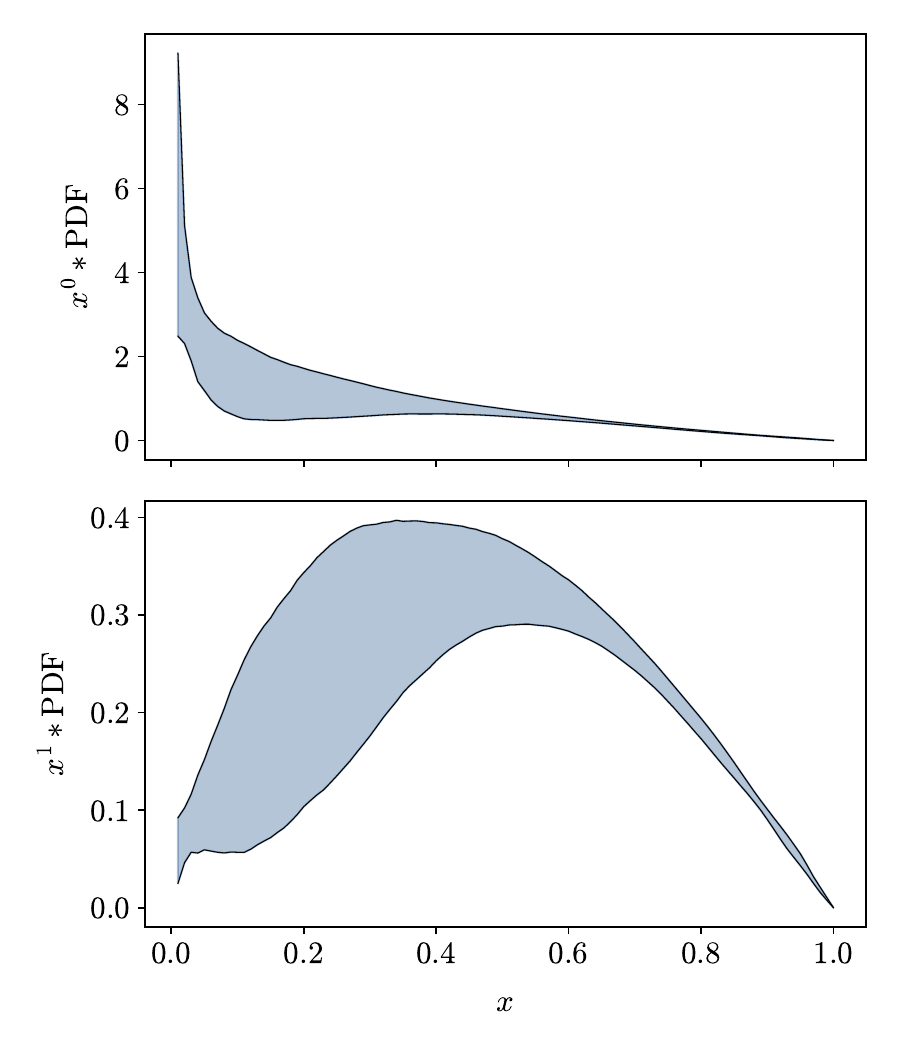}
    \caption{$u$ quark (valence) PDF}
    \end{subfigure}
    \begin{subfigure}[t]{0.32\textwidth}
        \centering
        \includegraphics[width=\textwidth]{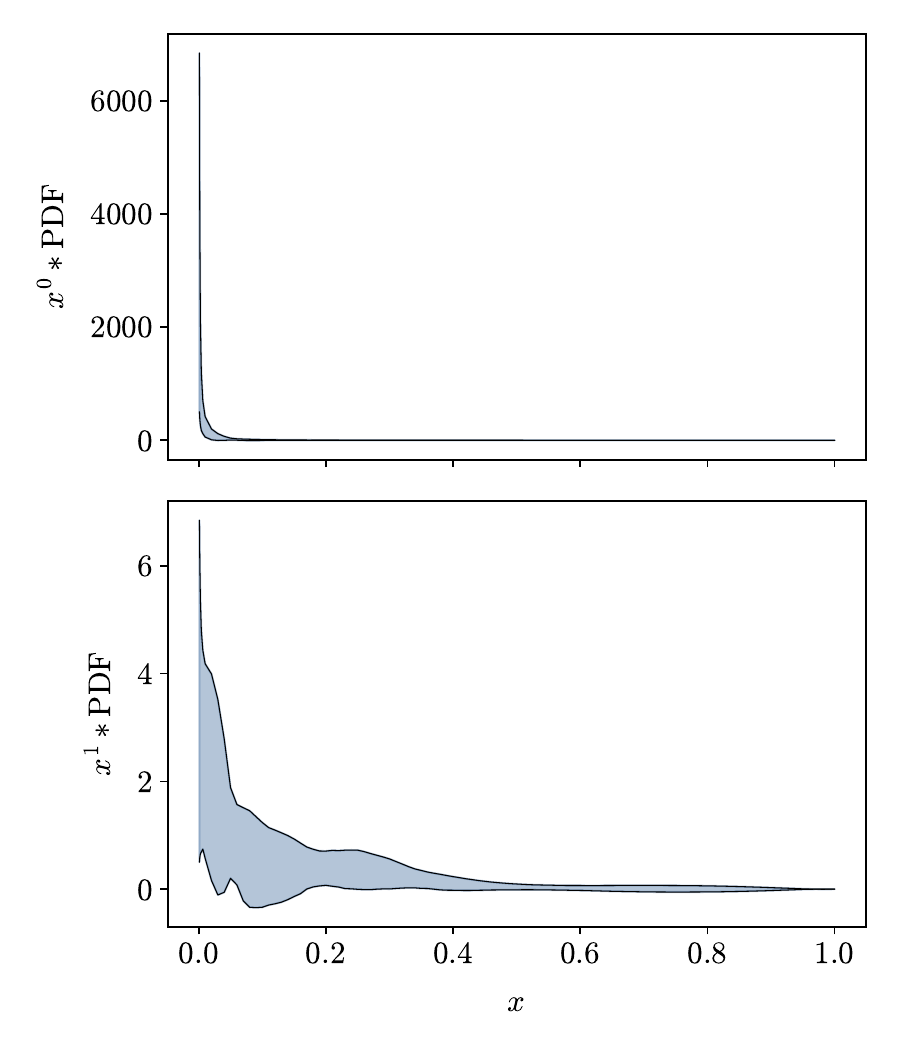}
        \caption{gluon PDF}
    \end{subfigure}

    \caption{$f(x)$ and $x f(x)$ for \FANTO PDF dataset}
    \label{fig:FANTO_truth}
\end{figure*}

\begin{figure*}[h!]
    \centering
    \begin{subfigure}[t]{0.3\textwidth}  
        \centering
        \includegraphics[width=\textwidth]{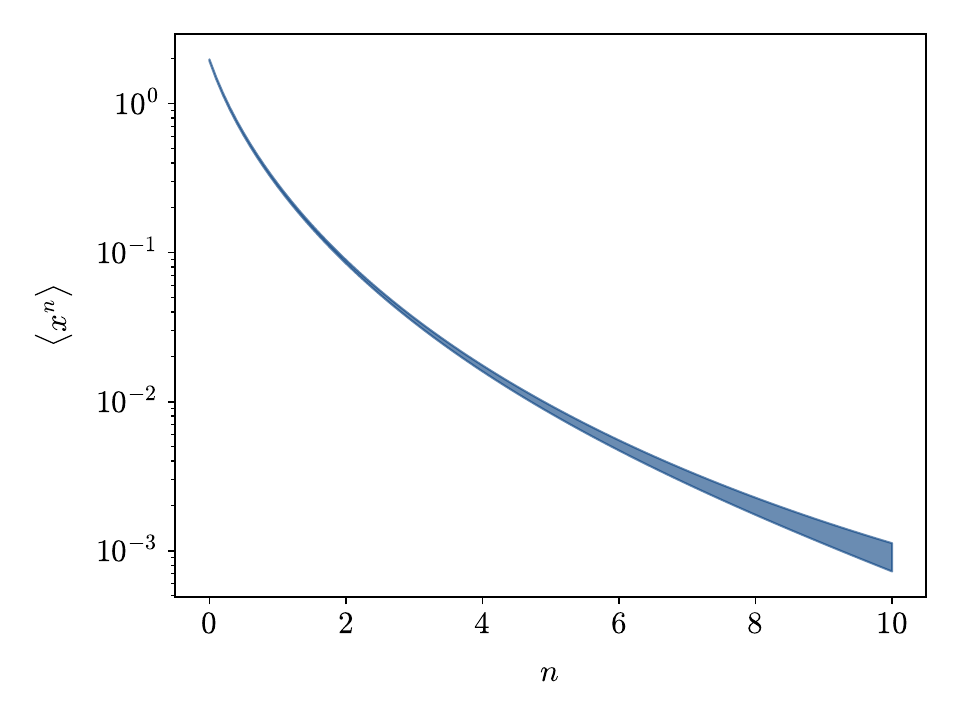}
        \caption{$u$ quark (valence) PDF}
    \end{subfigure}
    \begin{subfigure}[t]{0.3\textwidth}  
        \centering
        \includegraphics[width=\textwidth]{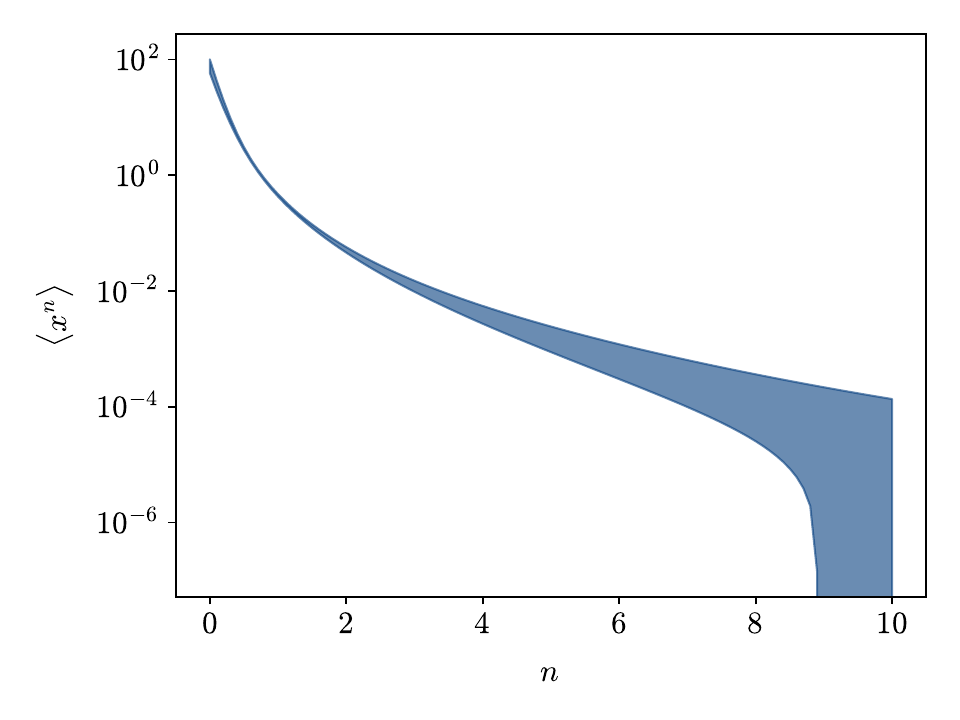}
        \caption{gluon PDF}
    \end{subfigure}
    \begin{subfigure}[t]{0.3\textwidth}  
        \centering
        \includegraphics[width=\textwidth]{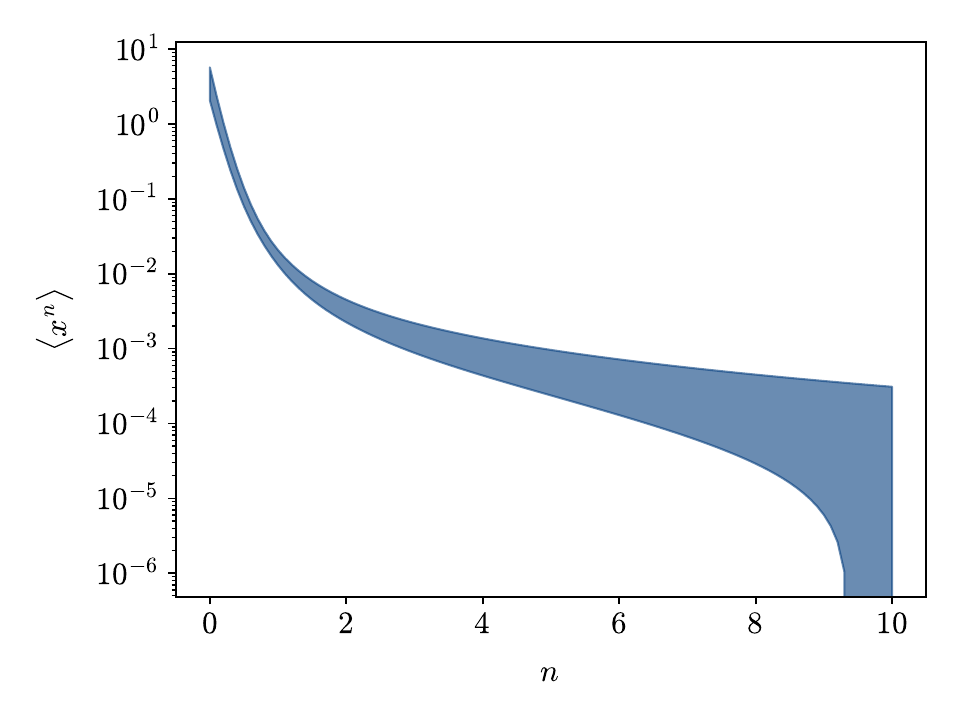}
        \caption{$s$ quark (sea) PDF}
    \end{subfigure}
    
    \caption{$\langle x ^n\rangle$ plotted against $n$ (interpolated between integer $n$) for the \NNPDF dataset}
    \label{fig:NNPDF_mom_truth}
\end{figure*}

\begin{figure*}[h!]
    \centering
    \begin{subfigure}[t]{0.32\textwidth}
        \centering
        \includegraphics[width=\textwidth]{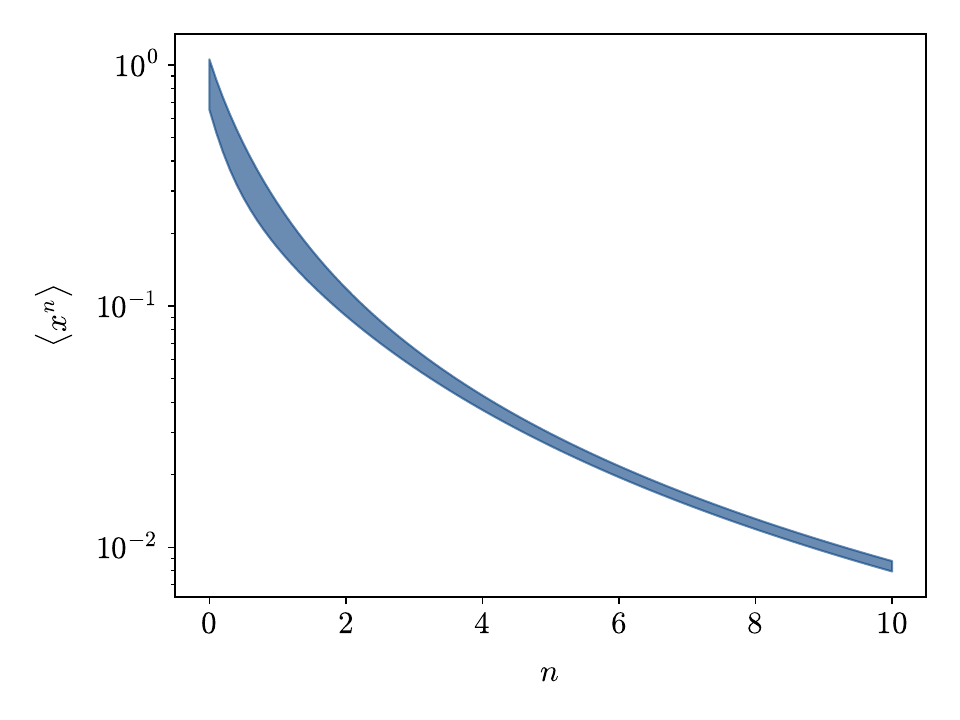}
    \caption{valence PDF}
    \end{subfigure}
    \begin{subfigure}[t]{0.32\textwidth}
        \centering
        \includegraphics[width=\textwidth]{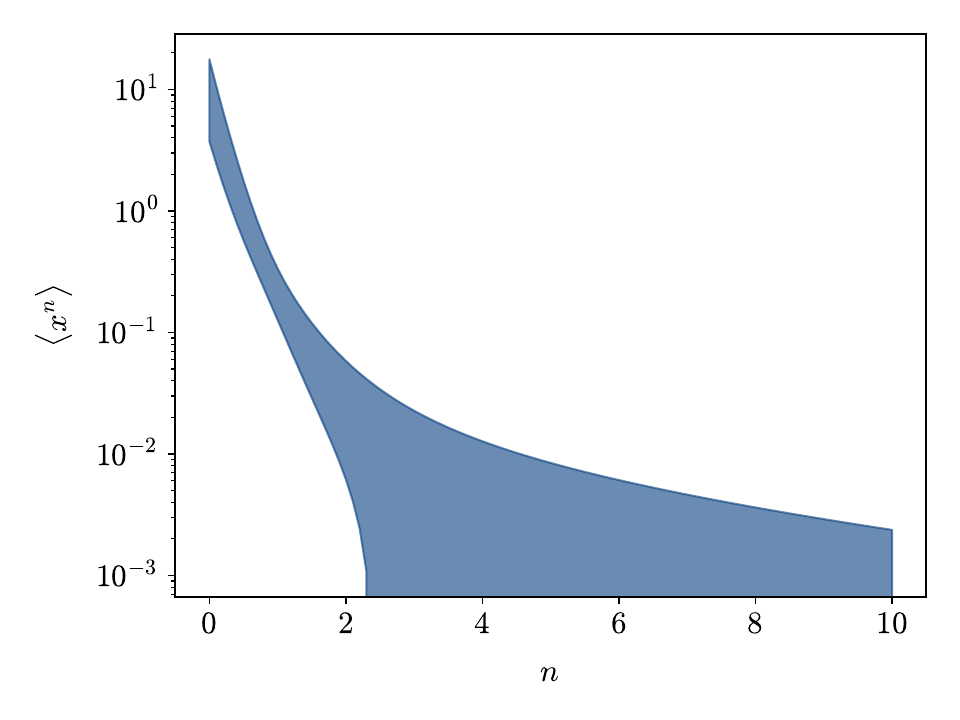}
        \caption{gluon PDF}
    \end{subfigure}

    \caption{$\langle x ^n\rangle$ plotted against $n$ (interpolated between integer $n$) for the \FANTO dataset}
    \label{fig:FANTO_mom_truth}
\end{figure*}

 We begin our empirical studies by looking at nucleon/pion up $u$ (valence), gluon, and strange ($s$) (sea) quark nucleon PDFs.\footnote{We will use the $u$ quark valence definition: $u_V(x)= u(x)-\bar{u}(x)$, and only consider the nucleon $s$ quark PDF.} Their forms are shown in Figs. \ref{fig:NNPDF_truth} and \ref{fig:FANTO_truth} for the \NNPDF and \FANTO datasets respectively.\footnote{Given $xf(x)$'s integrability across all PDF types in addition to easy visualization, our comparisons between truth and reconstruction will be done using $x f(x)$ as opposed to $f(x)$.}, while their computed moments (interpolated between integer $n$) are shown in Figs. \ref{fig:NNPDF_mom_truth} and \ref{fig:FANTO_mom_truth} from the nucleon and pion, respectively. This assortment of five contains distinct combinations of a PDFs' uncertainty as a function of $x$, and its degree of power divergence at $x\to 0$, as organized in Table \ref{tab:PDF_picture_data}. The \NNPDF valence's leading moment has a $\sim 2\%$ fractional error, and this roughly persists to higher order moments. The \NNPDF gluon PDF has a $\sim 5\%$ uncertainty on the first moment, and this increases for each successive moment. The nucleon $s$ quark PDFs on the other hand are even more unconstrained, and we will briefly study the $s$ (strange) quark PDF from the \NNPDF dataset in this paper, whose minimum fractional uncertainty of moments is $\sim 20 \%$. For the \FANTO dataset however, the leading moments for the valence PDFs have about $\sim20\%$ uncertainties and the minimum fractional uncertainty for gluon moments is $\sim 50\%$. As outlined theoretically in Secs. \ref{subsec:math_theory_disc} and \ref{subsec:prior_bias}, these high uncertainties in data make the reconstruction of \FANTO PDFs very challenging (definitely more so than \NNPDF).  We will cover reconstruction strategies with our GP framework for each element of Table \ref{tab:PDF_picture_data}.

\begin{table}[]
\begin{tabular}{lll}
 &
  Constrained $\langle x^n\rangle$ &
  Unconstrained $\langle x^n \rangle$ \\ \cline{2-3} 
\multicolumn{1}{l|}{Mild $x^\alpha$ divergence} &
  \multicolumn{1}{l|}{\NNPDF valence (easy)} &
  \FANTO valence (moderate) \\ \cline{2-3} 
\multicolumn{1}{l|}{Severe $x^\alpha$ divergence} &
  \multicolumn{1}{l|}{\NNPDF gluon (difficult)} &
  \begin{tabular}[c]{@{}l@{}}\NNPDF $s$ quark,\\  \FANTO gluon (difficult)\end{tabular}
\end{tabular}
\caption{Sorting different datasets into difficulty classes}
\label{tab:PDF_picture_data}
\end{table}

 \subsubsection{Reconstructing $n = 0$}
 \label{subsubsec:extrapolation_nto0}
 We note that valence PDFs satisfy the norm constraint $\langle x^0 \rangle = 1$ or $2$ to a very small (but nonzero) error. It is trivial to add this norm constraint into our framework, but for the time being, we will not enforce it:  we will instead observe how well our moment space reconstruction of $n=0$ matches the truth value: if our reconstruction is statistically consistent with the valence sum rules, adding the norm constraint is not necessary. For \NNPDF, as we will see, there is effectively no difference in enforcing the constraint, but with \FANTO, adding it may potentially result in an underestimation of low $x$ uncertainty: a further discussion of this topic will be held in Sec. \ref{subsec:comments_nto0}.

\subsection{Implementing bias correction in practice}
\label{subsec:mean_adjustment_howto}
\begin{figure*}[h!]
    \centering
    \begin{subfigure}[t]{0.40\textwidth}
        \centering
        \includegraphics[width=\textwidth]{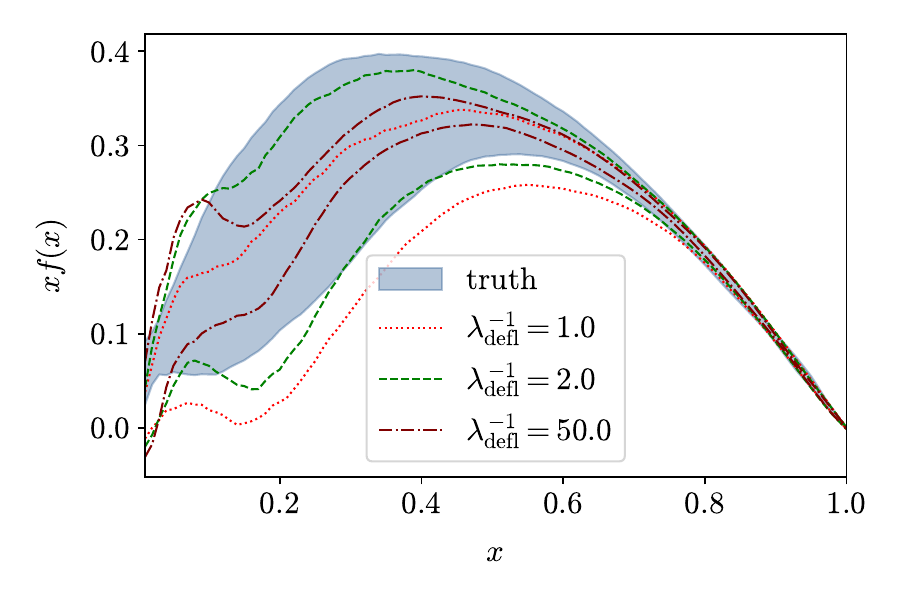}
        
    \caption{}
    \label{fig:FANTO_deflate_xPDF}
    \end{subfigure}
    \begin{subfigure}[t]{0.40\textwidth}
        \centering
        \includegraphics[width=\textwidth]{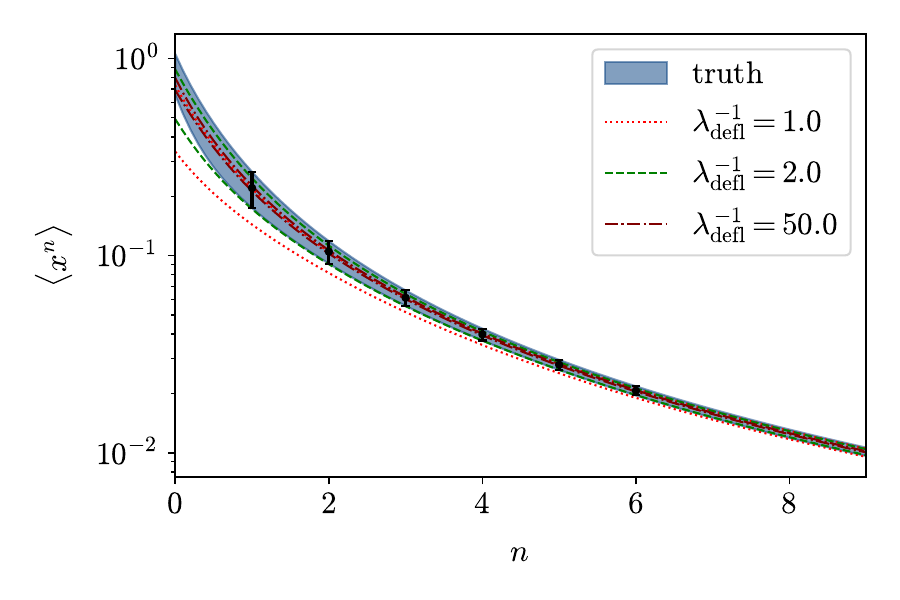}
        
        \caption{}
        \label{fig:FANTO_deflate_momRec}
    \end{subfigure}
    \caption{\FANTO reconstruction at $\nmax=6$ for progressively deflated values of $\Cy$}
    \label{fig:FANTO_deflate_scan}
\end{figure*}

\begin{figure*}[h!]
    \centering
    \begin{subfigure}[t]{0.40\textwidth}
        \centering
        \includegraphics[width=\textwidth]{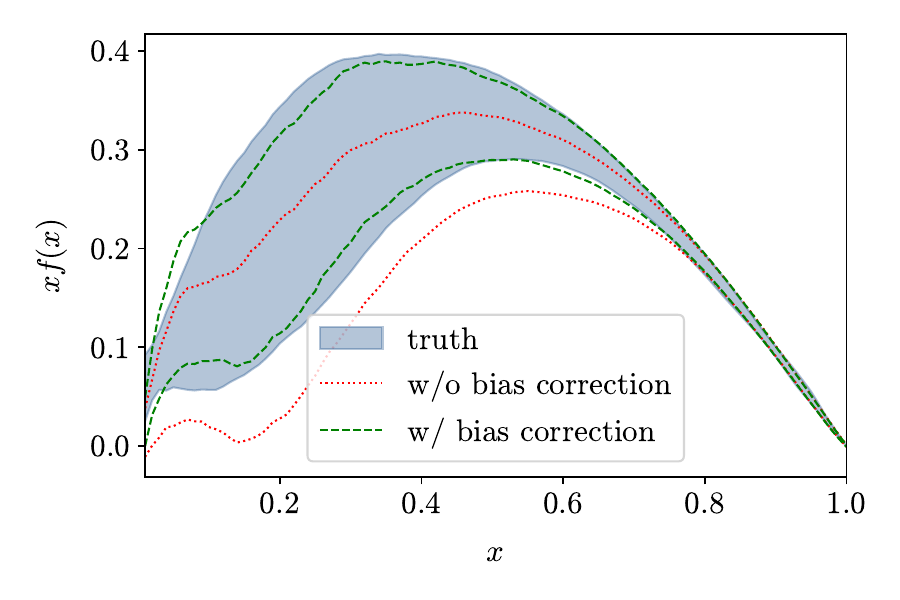}
        
    \caption{}
    \label{fig:FANTO_cDscan_xPDF}
    \end{subfigure}
    \begin{subfigure}[t]{0.40\textwidth}
        \centering
        \includegraphics[width=\textwidth]{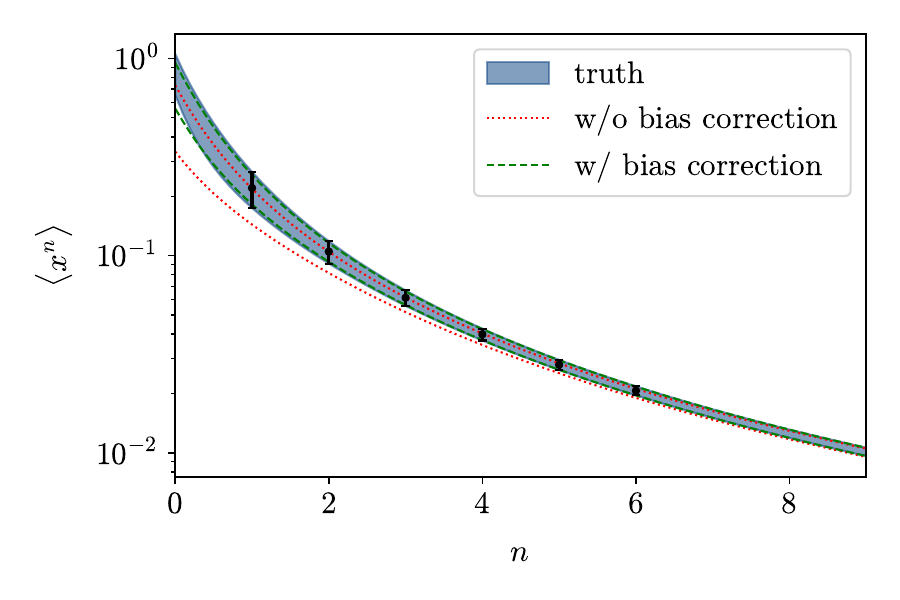}
        
        \caption{}
        \label{fig:FANTO_cDscan_momRec}
    \end{subfigure}
    \caption{\FANTO reconstruction at $\nmax=6$ with and without mean bias adjustment }
    \label{fig:FANTO_cDscan}
\end{figure*}

As mentioned in Sec. \ref{subsec:prior_bias}, upon introduction of any error associated with data into our analysis, our reconstruction is subject to an additional source of error other than $\K$ and $\FK$. In particular, our posterior mean will be biased from our true mean by an amount given by \ref{eq:post_bias}. We propose \textit{bias correction} to overcome this issue, with a prescription detailed in \ref{subsubsec:bias_corr} that takes a controlled deflation of $\Cy \to \lim_{\lambda\to\infty} \frac{\Cy}{\lambda}$, effectively eliminating the second, data-sensitive term in Eq. \ref{eq:master}. As a result, our reconstructed mean posterior converges to its unbiased value. In our validation tests, we can visually inspect whether this extrapolation strategy is needed. As a dramatic example, let us first consider the \FANTO valence dataset, with rather large uncertainties for $x\ll 1$, and relatively large errors for low moments. Here, across a large range of covariance kernels, we will see some sort of bias. We will use a novel kernel\footnote{Within this self-contained section, there is no need to know what the details of this kernel are, just know that it works. It will be expounded upon later} and set $\nmax = 6$. In Fig \ref{fig:FANTO_deflate_scan}, we see an egregious error for $\lambda =1$. As we deflate the covariance with the sequence $\lambda =\{1,2,50\}$, we see a convergence to the mean posterior. In theory, we can take an extrapolation, to narrow down the true value of the unbiased mean posterior. If we have a ground truth model we can perform an extrapolation in data space as well taking the limit $\nmax \to \infty$. In addition to this, we may take several covariance kernels and conduct extrapolations with them to obtain a set of mean posteriors $\{\m_\mathrm{UB}\}_\K$ from which a model averaged unbiased mean posterior $\tilde \m_\mathrm{UB}$ may be created. We will leave this model averaging step for a subsequent study: as we see from Fig. \ref{fig:FANTO_cDscan_momRec}, for $\lambda=50$, our reconstructed mean already has zero effective bias in data space (which we will be able to see at all times), so we can let this be our $\m_\mathrm{UB}$. In our studies we usually find that $\lambda = 50$ or $\lambda=100$ nearly completely eliminates bias.

Now suppose we conduct an analysis and obtain a reconstruction $\PDFv$ with mean $\m$. We may then adjust our reconstruction by simply adjusting the posterior mean by the difference $\Delta \m \coloneqq (\m_\mathrm{UB} - \m)$. We thereby have a mean adjusted reconstruction that eliminates the mean bias induced by experimental error. In Figure \ref{fig:FANTO_cDscan} we observe this bias correction just as described. \textit{When our reconstruction plots are mean adjusted, it will be explicitly mentioned.} Of course, there are a handful instances where mean adjustment may not be immediately useful. The most obvious case is when the data point indicates multi-modality of the model distribution. (An example would be if noisy data does not clearly follow a monotonically decreasing trend.) This would imply that there are many possible values of $\m_\mathrm{UB}$. Whether the numerous forms of $\m_\mathrm{UB}$ are in the same neighborhood or not determines whether a bias correction extrapolation can be performed or cannot be performed, respectively. The only way to determine this is through empirical means.

Note that in theory, it is possible to mix and match values of $\m_\mathrm{UB}$ for different kernels, and different values $\nmax$. For straightforwardness, we will only calculate $\m_\mathrm{UB}$ for a given $\nmax$ from a number of data-points $n =\nmax$ in all of our settings.

\section{Reconstructing valence quark PDFs}
\label{sec:valence_recon}

We will now detail different strategies that can be used to reconstruct valence quark PDFs given moments, beginning with a discussion on the role of a kernel's $x$ autocorrelation length in reconstruction, then subsequently exploring strategies for reconstructing both constrained and unconstrained valence PDFs.

\subsection{Kernels, $x$ correlation scales, and trade-offs}

Covariance kernels play a central role in determining our PDF reconstruction, and reiterating what was mentioned earlier, a key feature of each kernel is its ability to capture the $x$ correlation scale(s) of $f(x)$. With the OOTB kernels we have defined in \ref{subsec:kernels_ootb} \footnote{all the kernels mentioned except the Gibbs (arbitrary auto-correlation) variant}, this is reflected in the parameter $\ell$.

Large $x$ correlation scale (LCS) kernels are sensitive to lengths scales on the order $\Delta x\sim1$, and small $x$ correlation scale (SCS) kernels are sensitive to lengths scales $\Delta x\ll 1$. There are tradeoffs in using one type of kernel over the other: SCS kernels tend to capture small scale structure in $f(x)$, but often take too much data to offer any appreciable resolution, and as a result the reconstructed variance will tend to be overestimated, sometime with no convergence to the true model variance.  LCS kernels will often converge to within $\pm1\sigma$ of the model PDF with remarkably few data-points, but at the cost of being unable to resolve small scale structure. We will classify medium $x$ correlation scale (MCS) kernels as having $x$ correlation somewhere in between the two.

A possible ``best of both worlds" scenario is to make Gibbs-style kernels that are inherently sensitive to numerous $x$ correlation scales. Before examining these kernels, we will examine how different OOTB kernels with different $x$ correlation scales affect the quality/convergence of our reconstruction, and we will use \NNPDF and \FANTO valence PDF data to demonstrate this. We find that \textit{OOTB kernels with either long or short scaled $x$ correlation lengths are powerful in their own ways, but are systematically limited (adding more data/increasing data quality will not improve the quality of the GPR fit) }.

\subsection{Kernels with short $x$ correlation scales}
To view the effects of a SCS kernels, let us consider the Mat\`ern kernel at $\nu=\frac{1}{2}$. From illustrations in Fig 4.2 of \cite{Rasmussen2006Gaussian} we notice how the behavior of the kernel is very jagged, like that resembling a stock option or a discrete random walk, i.e. the $x$ correlation length of this kernel is quite short. To demonstrate its ability in fitting data, we fit the \NNPDF valence dataset with a GPR with this kernel. We use a uniform mean prior, pick $\sigma\in[0,20]$ and $\ell\in[0,50]$, and enforce the $x^\alpha$ prior in our kernel.\footnote{For the \MATERN{1/2} in particular, it was rather tough to efficiently tune our sampler without the $x^\alpha$ kernel prior.} The $\sigma,\; \ell$ bounds are fairly general and will be in place for all valence kernels unless explicitly mentioned.

\begin{figure*}[h!]
    \centering
    \begin{subfigure}[t]{0.40\textwidth}
        \centering
        \includegraphics[width=\textwidth]{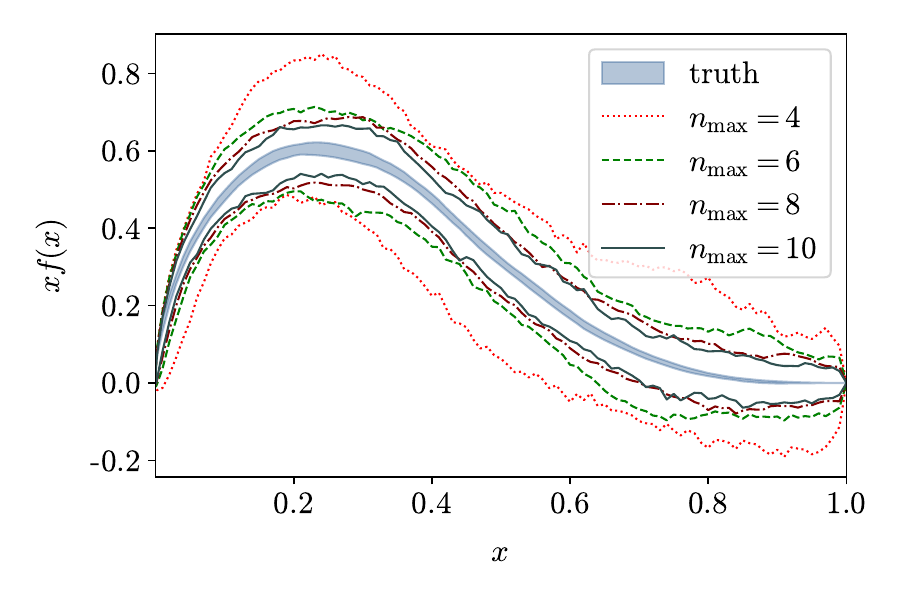}
        \caption{}
        \label{fig:NNPDF_MATERN12_xPDF}
    \end{subfigure}
    \begin{subfigure}[t]{0.40\textwidth}
        \centering
        \includegraphics[width=\textwidth]{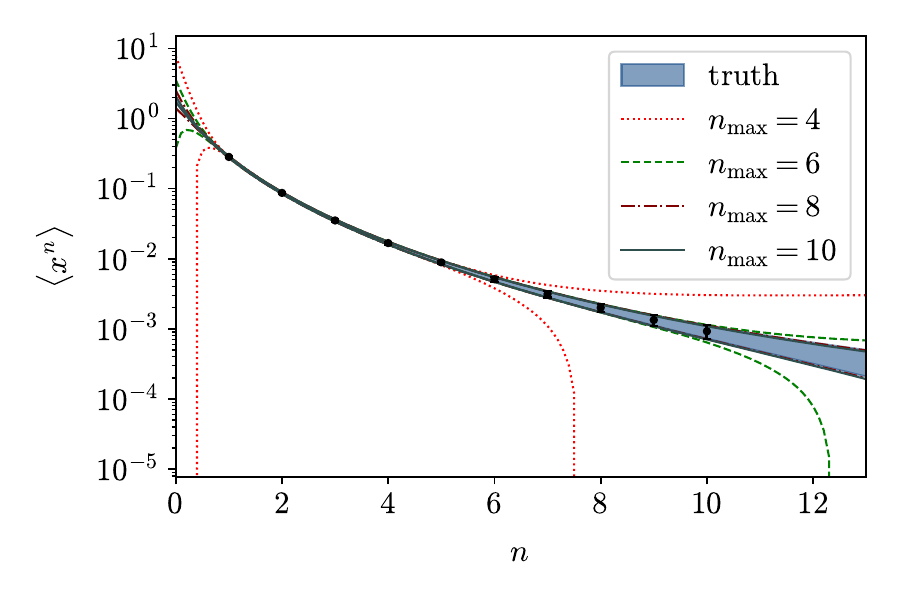}
        
        \caption{}
        \label{fig:NNPDF_MATERN12_momRec}
    \end{subfigure}
    \caption{\NNPDF dataset reconstruction with \MATERN{1/2}.}
    \label{fig:NNPDF_MATERN12}
\end{figure*}

\begin{figure*}[h!]
    \centering
    \begin{subfigure}[t]{0.40\textwidth}
        \centering
        \includegraphics[width=\textwidth]{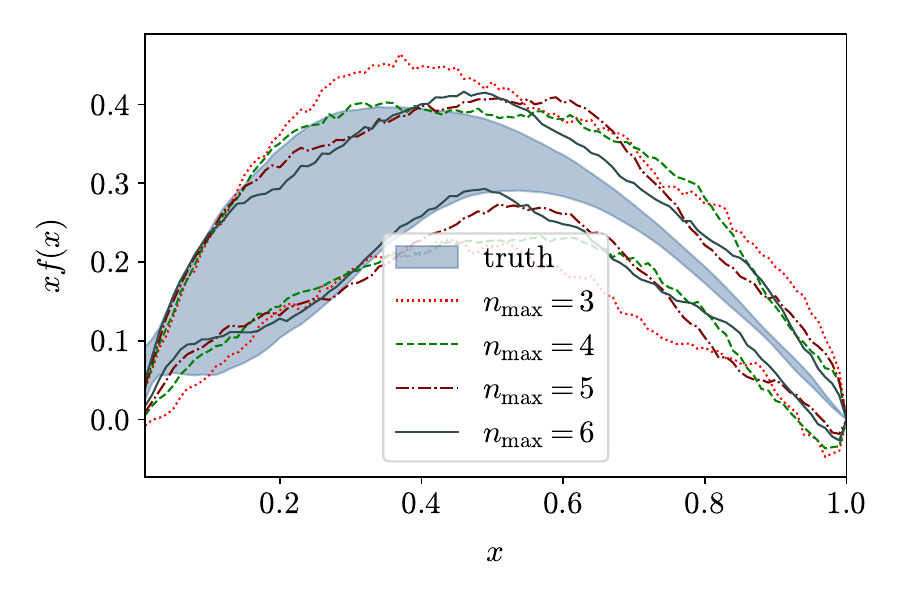}
        
    \caption{}
    \label{fig:FANTO_MATERN12_xPDF}
    \end{subfigure}
    \begin{subfigure}[t]{0.40\textwidth}
        \centering
        \includegraphics[width=\textwidth]{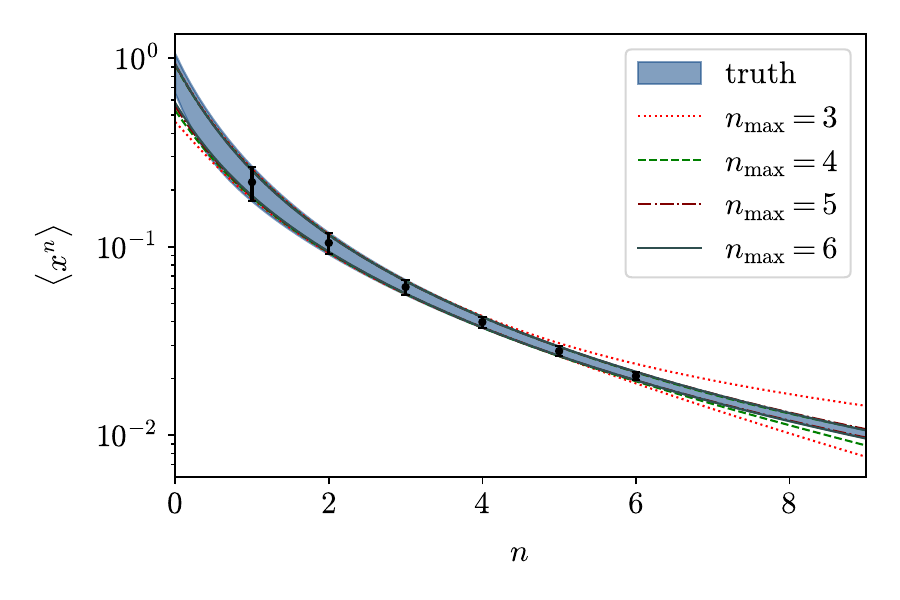}
        
        \caption{}
        \label{fig:FANTO_MATERN12_momRec}
    \end{subfigure}
    \caption{\FANTO dataset reconstruction with \MATERN{1/2}. \biascorrnote}
    \label{fig:FANTO_MATERN12}
\end{figure*}

In Fig \ref{fig:NNPDF_MATERN12}, we scan $\nmax \in {4,6,8,10}$ and plot our reconstructions for this kernel. While we see some convergence around the model truth, we see that the variance in reconstruction does not seem tend towards the true variance, even for a large number of moments ($\nmax=10$). If we look at \ref{fig:NNPDF_MATERN12_momRec} we see that the reconstructed moment variance is nearly identical to the truth variance. In other words, the truncated moment problem posed by this dataset is not unique, and our GPR found another solution with higher oscillatory modes in $x$ space that fit the data well: the higher oscillatory forms lead to an inflated reconstruction variance in $f(x)$, and features like the "tapering" of the PDF uncertainty at $x\to 1$ will not be learned with more data unless a different kernel with a long intrinsic $x$ correlation length is used. This problem is also apparent in applying \MATERN{1/2} to the \FANTO dataset: shown in \ref{fig:FANTO_MATERN12}, we see that the region of high $x$ is dominated by the high variance behavior of the kernel, and shows no signs of converging for a reasonable amount of moments.

\subsection{Kernels with long $x$ correlation scales}

\begin{figure*}[h!]
    \centering
    \begin{subfigure}[t]{0.40\textwidth}
        \centering
        \includegraphics[width=\textwidth]{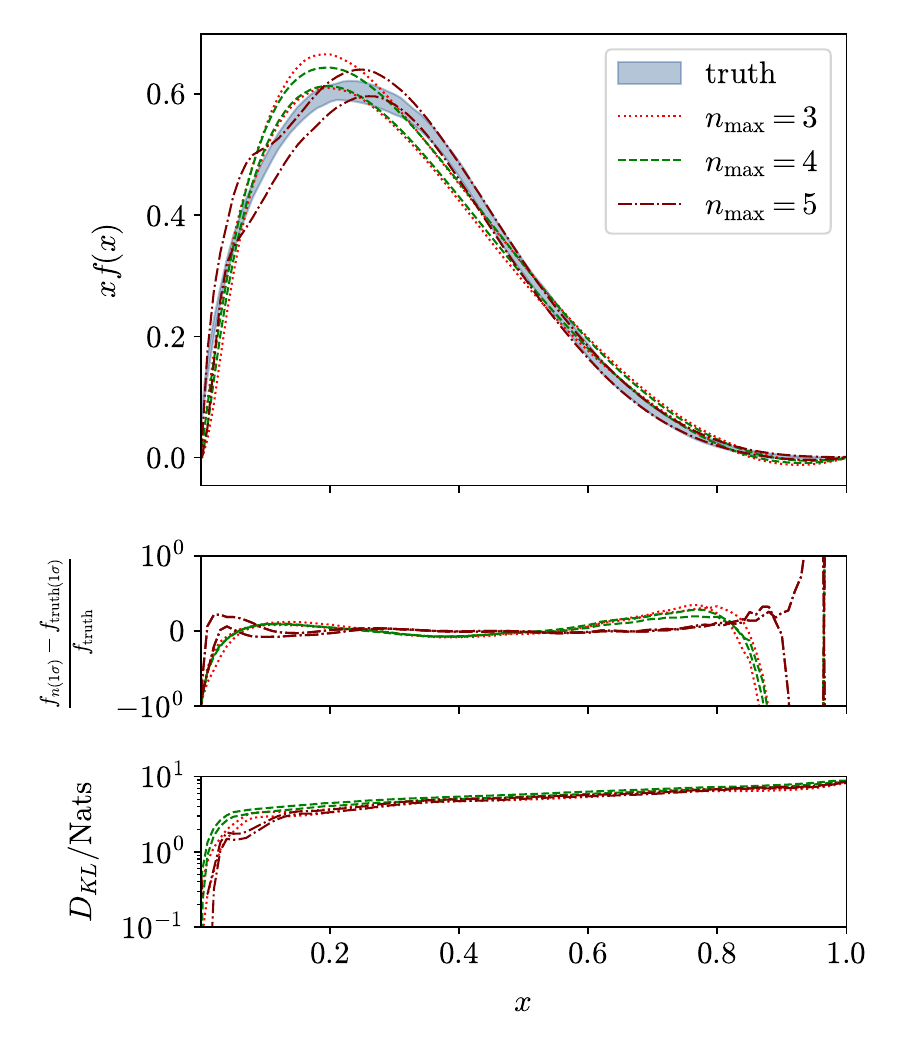}
        
    \caption{}
    \label{fig:xPDF_LSEoSE_JAM}
    \end{subfigure}
    \begin{subfigure}[t]{0.40\textwidth}
        \centering
        \includegraphics[width=\textwidth]{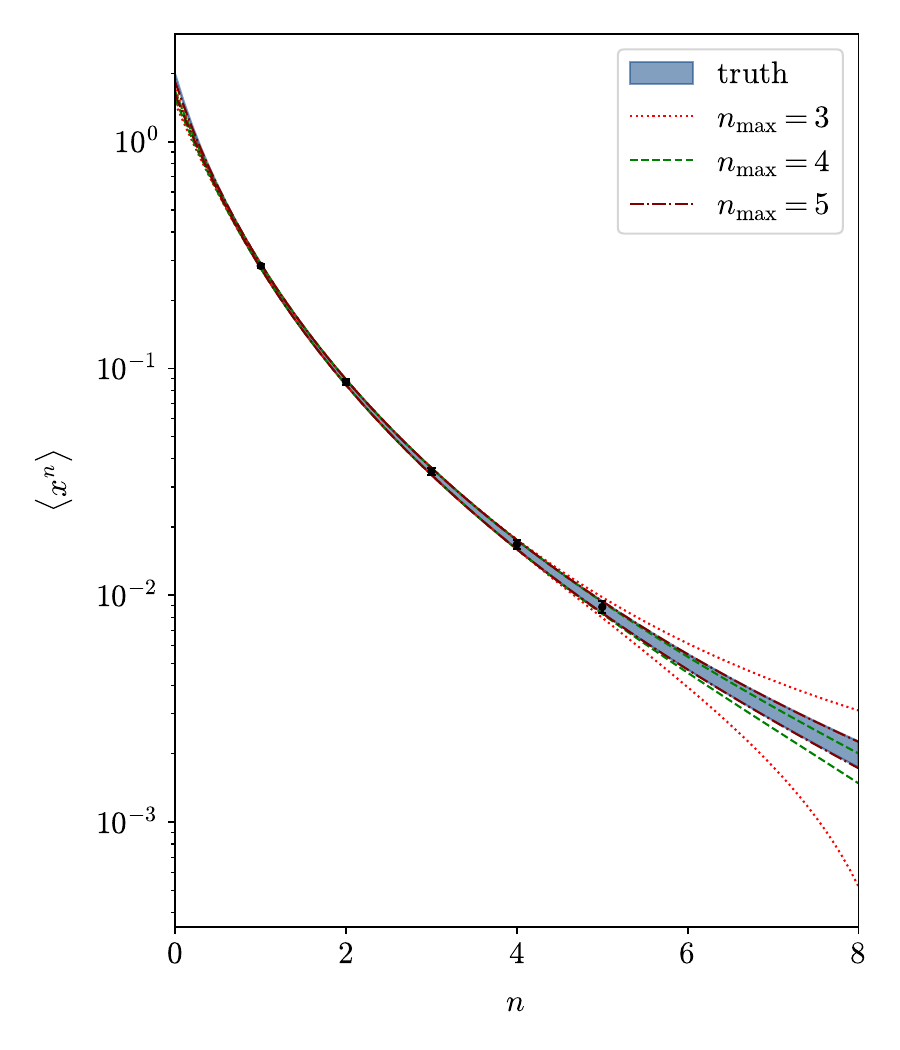}
        
        \caption{}
        \label{fig:momRec_LSEoSE_NNPDF}
    \end{subfigure}
    \caption{LSE performance on \NNPDF dataset. \biascorrnote}
    \label{fig:NNPDF_LSE_momScan}
\end{figure*}

We now turn from SCS kernels to LCS kernels. Specifically, we will focus on the LSE kernel as described in Sec \ref{subsec:kernels_ootb}. As mentioned earlier, LCS kernels make assumptions of low variance as small $x$ scales, and thus converge far quicker than SCS kernels, at the cost of not picking up fine details that may be present in a truth model. For the \NNPDF set in particular, LCS kernels may have a hard time learning the ``tapering" at high $x$. As an example, let us look at the LSE kernels ability to reconstruct data with $\nmax\in\{3,4,5\}$, as shown in Fig. \ref{fig:NNPDF_LSE_momScan}, lifting the $x^\alpha$ previously places for the \MATERN{1/2}. We can see how convergence is achieved much faster than the Mat\`ern short $x$ correlation kernels in the previous section. 

An interesting feature we see is a kink/deviation that develops around $x\sim0.1$. This can be understood if we observe closely the behavior at $x\sim1$. The taper in the truth model is not matched by the rather straight reconstruction simply due to the LSE kernel's  inability to probe those distances. However, with the data constraint $\FK \PDFv = \y$ needing to be matched, the kernel compensates for this inability to resolve fine behavior at $x\sim1$ by adding in lower frequency behavior to fulfill the data constraint. This problem will not be resolved with more/higher quality data and is a systematic error associated with the choice of a long correlation length kernel. 

\begin{figure*}[h!]
    \centering
    \begin{subfigure}[t]{0.40\textwidth}
        \centering
        \includegraphics[width=\textwidth]{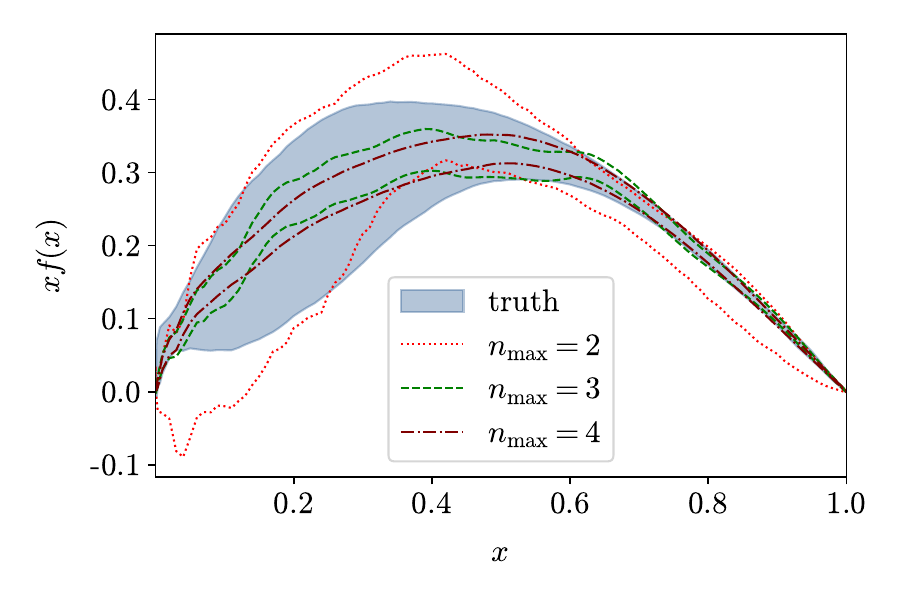}
    \end{subfigure}
    \begin{subfigure}[t]{0.40\textwidth}
        \centering
        \includegraphics[width=\textwidth]{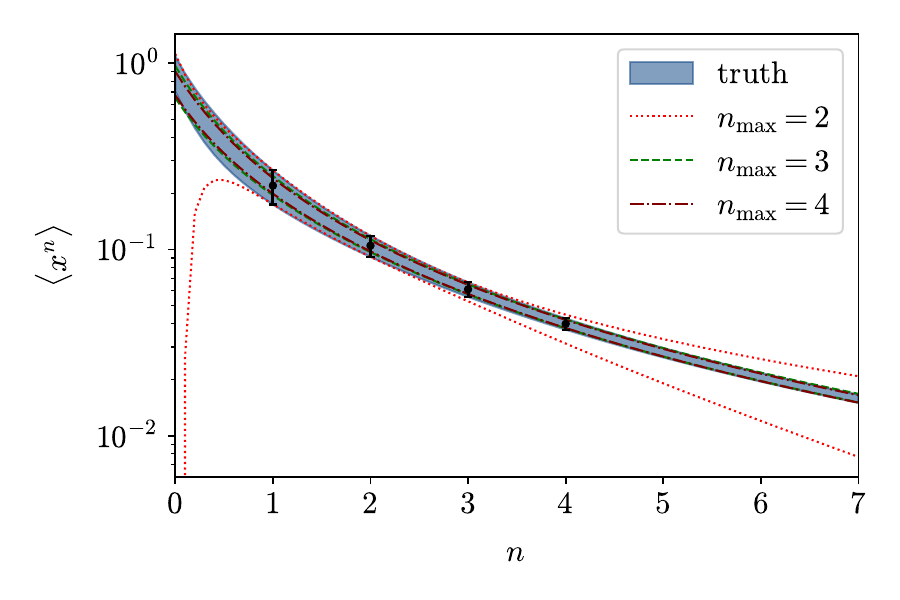}
    \end{subfigure}
    \caption{LSE kernel applied on \FANTO data. Notice the egregious underestimation of uncertainties at low $x$ \biascorrnote }
    \label{fig:FANTO_LAS}
\end{figure*}

If we use the LSE (this problem persists for SE for the same reason) kernel on the \FANTO dataset, we arrive at a poor reconstruction, as is shown in Fig. \ref{fig:FANTO_LAS}: the kernel ends up overfitting to the more constrained high $x$ region (which is informed by the increasingly higher moments), and completely misses the large variance at low $x$. 

\subsection{Uniform vs Parametric mean priors}

\begin{figure*}[h!]
    \centering
    \begin{subfigure}[b]{0.30\textwidth}
        \centering
        \includegraphics[width=\textwidth]{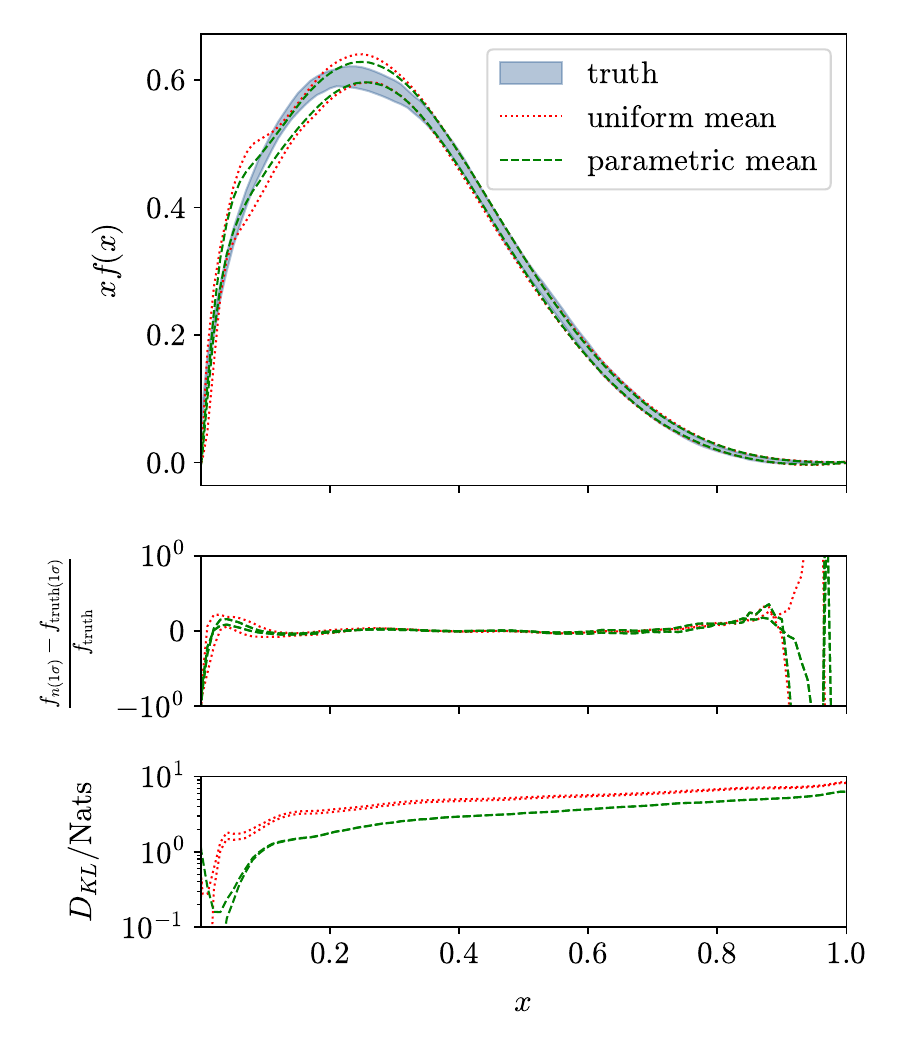}
    \end{subfigure}
    \begin{subfigure}[b]{0.30\textwidth}
        \centering
        \includegraphics[width=\textwidth]{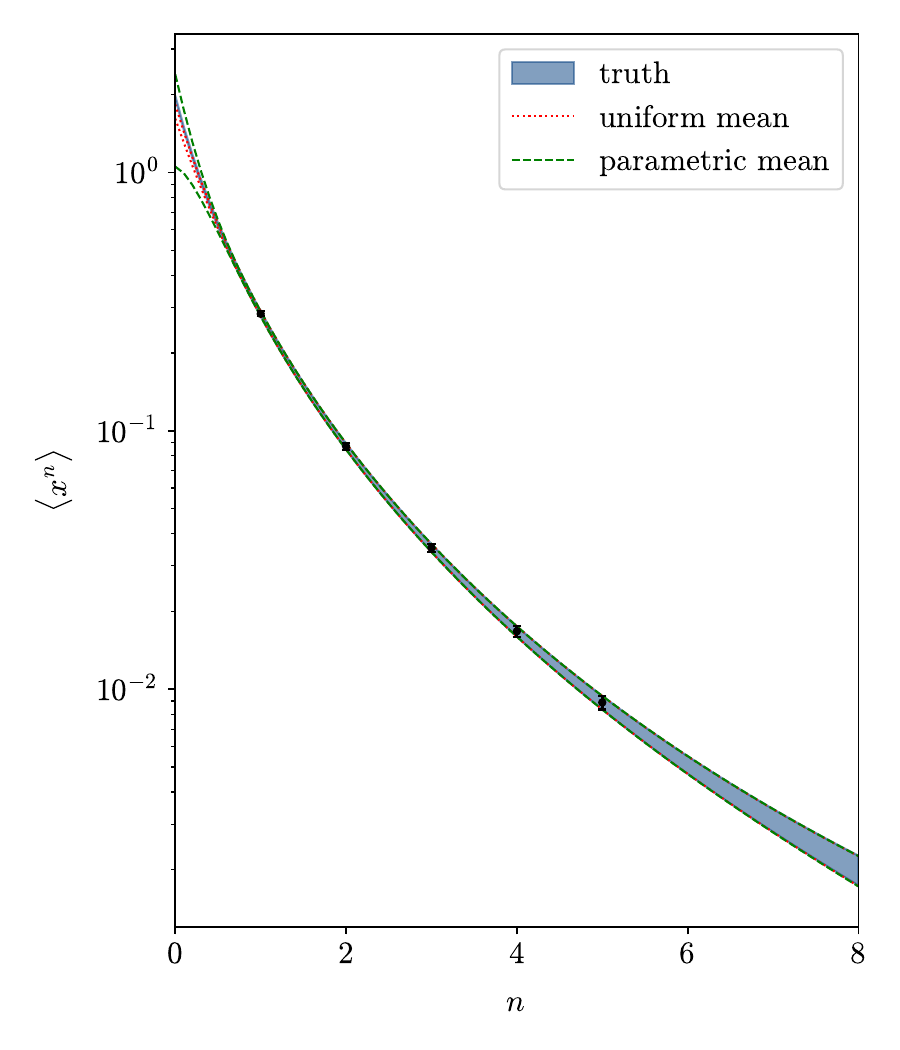}
    \end{subfigure}
    
    \begin{subfigure}[b]{0.25\textwidth}
        \centering
        \includegraphics[width=\textwidth]{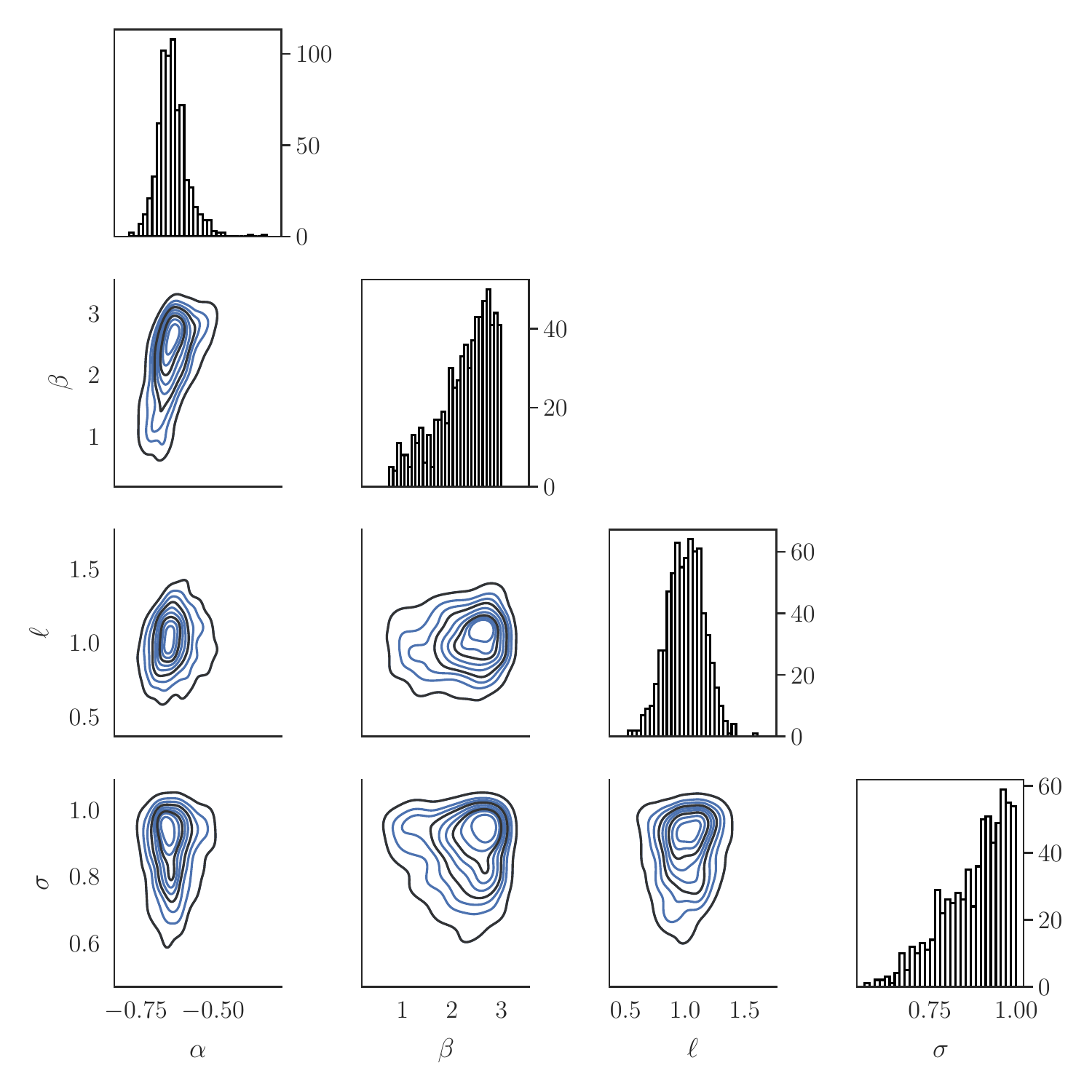}
        \caption{hyperparameter plot for parametric mean prior case}
    \end{subfigure}
    
    \caption{different mean priors effects on LSE performance on \NNPDF dataset at $\nmax=5$}
    \label{fig:NNPDF_mupriorScan}
\end{figure*}

The natural question of whether uniform or parametric mean priors would result in a better fit has an almost immediate answer: the parametric prior may converge upon sampled hyper-parameters better than in the uniform case, but the inherently more constrained form of the mean prior suggests that more of the convergence to the posterior takes place within the sampling step as opposed to the Gaussian Process step, which will result in a lower information gain than for the uniform mean prior case. The degree to which it is lower than in the uniform case can only be computed empirically. As an example, Fig. \ref{fig:NNPDF_mupriorScan} compares different mean priors' performances on the \NNPDF valence dataset : we see a significantly higher information gain in the uniform mean as opposed to the parametric mean case. In the extreme event of the sampled distribution peaking at $\sigma=0$ for the parametric mean case, all information gain will occur during the sampling phase, effectively turning the regression to a non Gaussian parametric fit.

\subsection{Kernels with medium $x$ correlation scales}
\begin{figure*}[h!]
    \centering

    \begin{subfigure}[t]{0.40\textwidth}
        \centering
        \includegraphics[width=\textwidth]{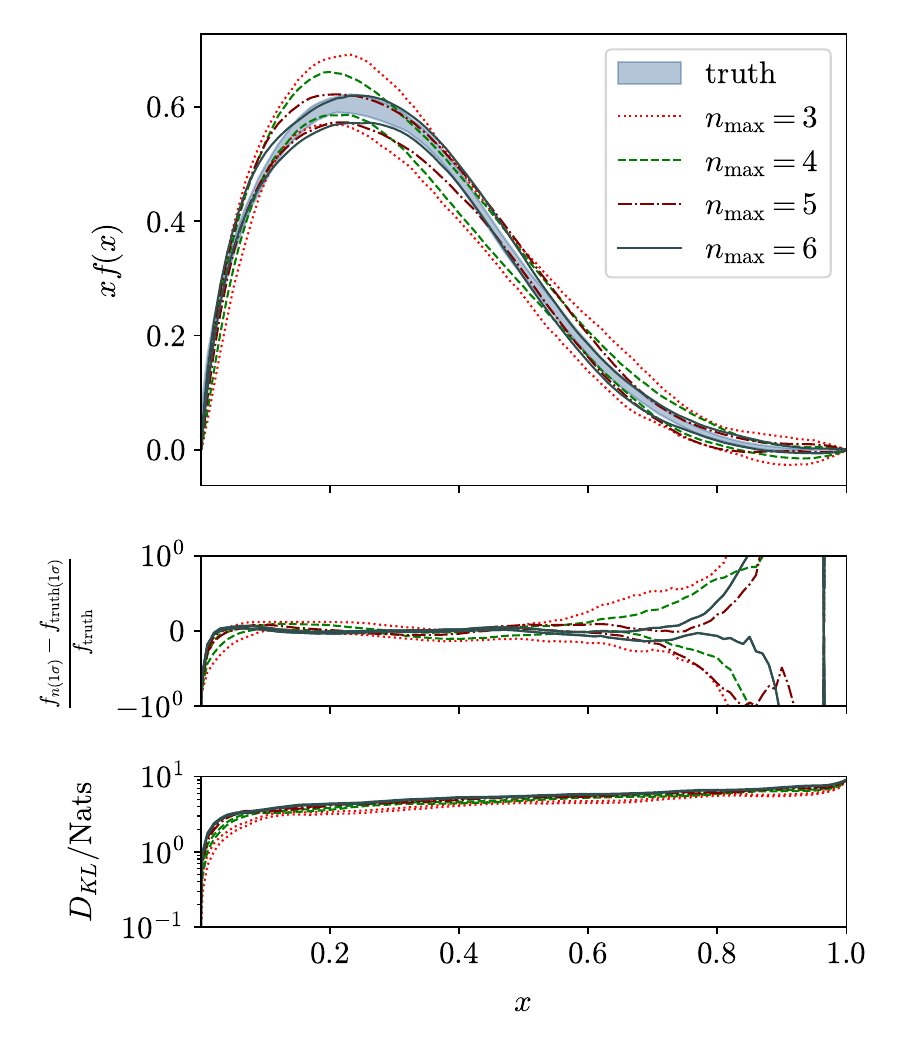}
        \caption{}
        \label{fig:xPDF_NNPDF_Matern32_cD}
    \end{subfigure}
    \begin{subfigure}[t]{0.40\textwidth}
        \centering
        \includegraphics[width=\textwidth]{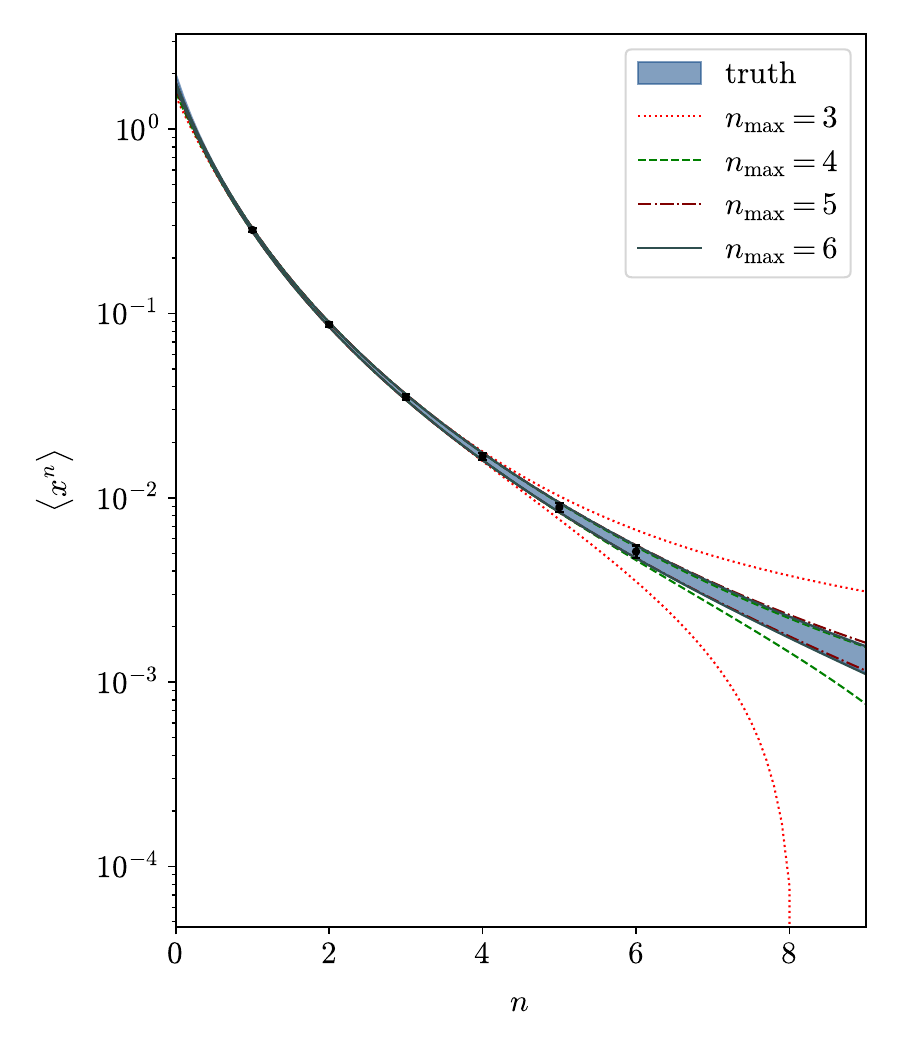}
        \caption{}
        \label{fig:MomRec_NNPDF_Matern32_cD}
    \end{subfigure}
    
    \caption{\MATERN{3/2} kernel used on \NNPDF dataset. \biascorrnote}
    \label{fig:NNPDF_Matern32_cD}
\end{figure*}

\begin{figure*}[h!]
    \centering

    \begin{subfigure}[t]{0.40\textwidth}
        \centering
        \includegraphics[width=\textwidth]{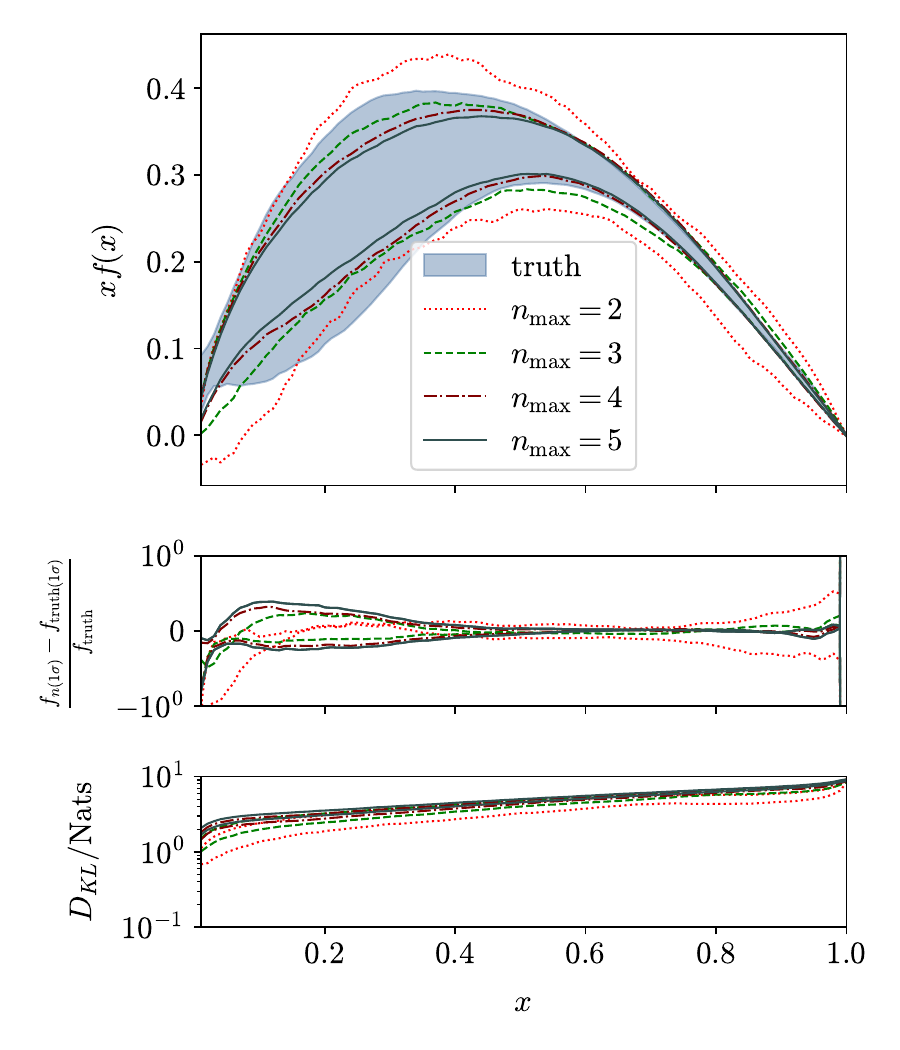}
        \caption{}
        \label{fig:xPDF_FANTO_Matern32}
    \end{subfigure}
    \begin{subfigure}[t]{0.40\textwidth}
        \centering
        \includegraphics[width=\textwidth]{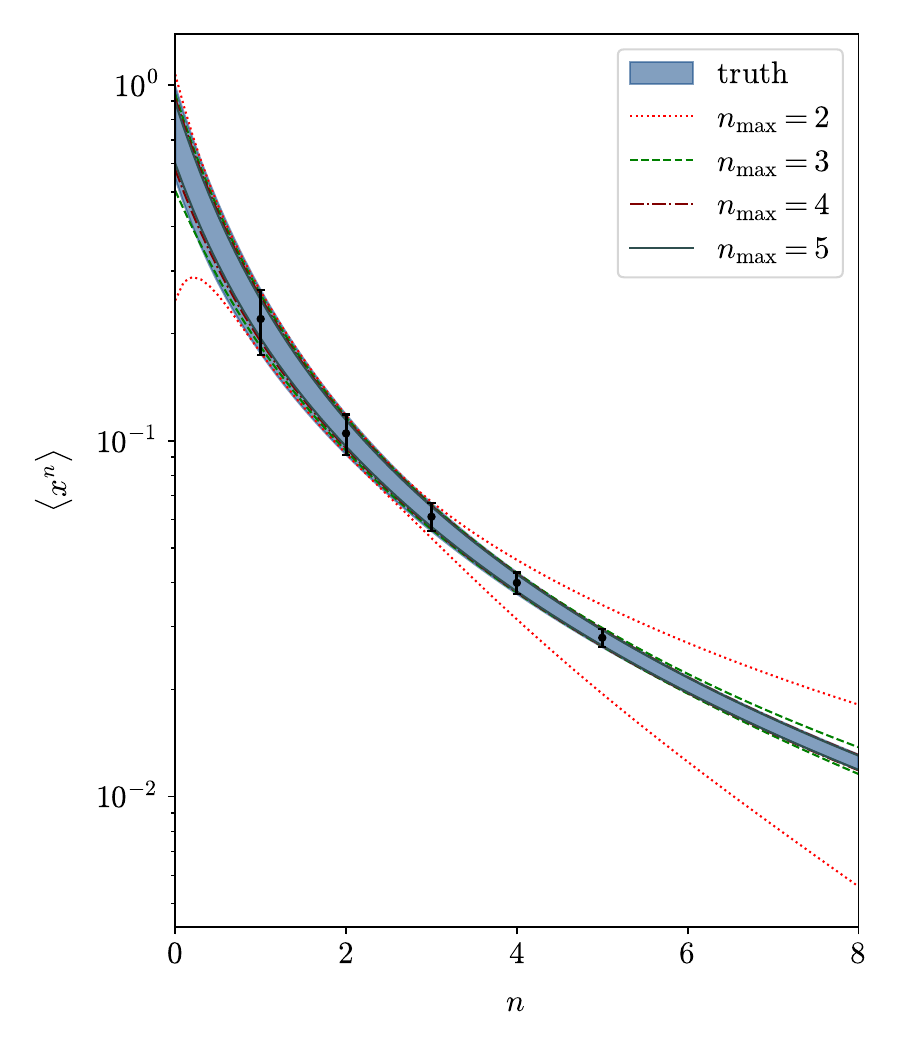}
        \caption{}
        \label{fig:MomRec_FANTO_Matern32}
    \end{subfigure}
    
    \caption{\MATERN{3/2} kernel used on \FANTO dataset. \biascorrnote}
    \label{fig:FANTO_Matern32}
\end{figure*}

Having discussed both SCS and LCS kernels, we will now turn to MCS kernels, in search of an intermediary $x$ correlation length that can mitigate the problems present in both SCS and LCS kernels. Having previously looked at two extremes of the Mat\`ern kernel ($\nu = \frac{1}{2}$ and $\nu=\infty$), we now will look at an intermediate $\nu$ value, i.e. $\nu=\frac{3}{2}$. On the \NNPDF dataset shown in Fig \ref{fig:NNPDF_Matern32_cD} (mean adjusted) we see good convergence, and altering the $\nu$ smoothing parameter from $\nu=\frac{1}{2}\to \frac{3}{2}$ mitigates the problem in Fig. \ref{fig:NNPDF_MATERN12} of the variance being permanently inflated due to the GP converging to a higher frequency solution of the moment problem. Additionally, notable gains over the (L)SE kernels are also seen, as the ``tapering" at high $x$ is fully learned for sufficiently high $\nmax$. We also see that satisfactory convergence to the truth model occurs at values of $\nmax$ that lie in between those used with LCS or SCS kernels. When we look at results with the \FANTO dataset in Fig. \ref{fig:FANTO_Matern32}, however, we run into the same problems encountered previously. The \MATERN{3/2} kernel does moderately well in gauging the low $x$ uncertainty when trained on only the first 2 moments. As we add more moments, our GPR fits increasingly to higher moments, and consequently, to higher $x$, and in the process underestimates low $x$ variance. This inability to capture variance in uncertainty and multiscale $x$ correlation in the \FANTO
 dataset has evaded successful modeling by any of the OOTB kernels we have chosen as of yet.

\subsection{The impact of the power divergence $x^\alpha$ kernel prior}
\label{subsec:xa_impact}
\begin{figure}[h!]

    \centering
    \begin{subfigure}[t]{0.3\textwidth}
        \centering
        \includegraphics[width=\linewidth]{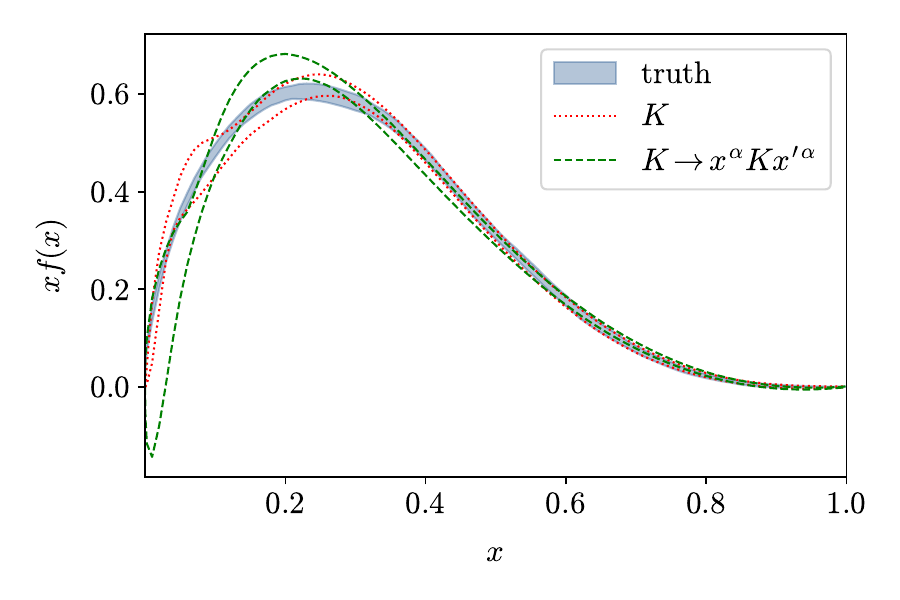}
    \end{subfigure}\hfill
    \begin{subfigure}[t]{0.3\textwidth}
        \centering
        \includegraphics[width=\linewidth]{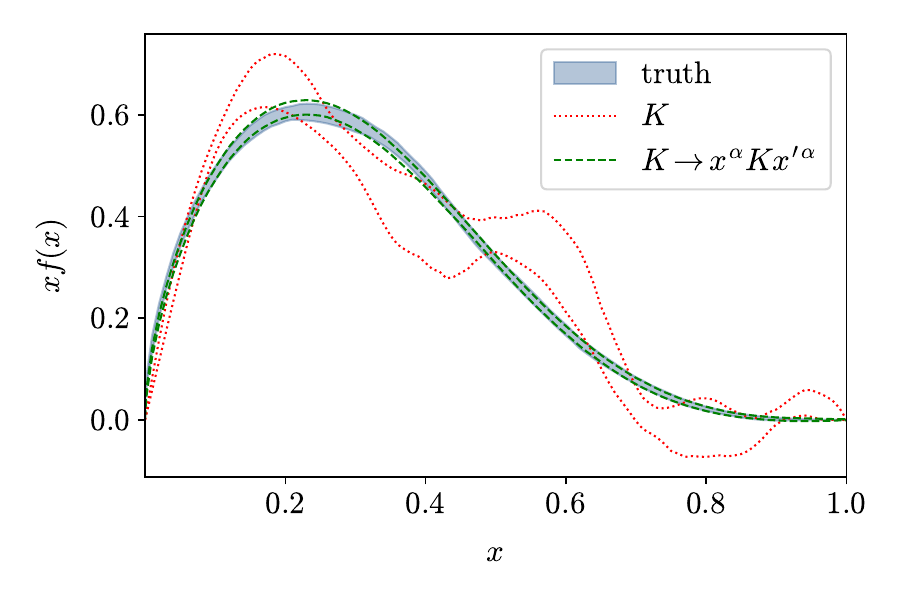}
    \end{subfigure}\hfill
    \begin{subfigure}[t]{0.3\textwidth}
        \centering
        \includegraphics[width=\linewidth]{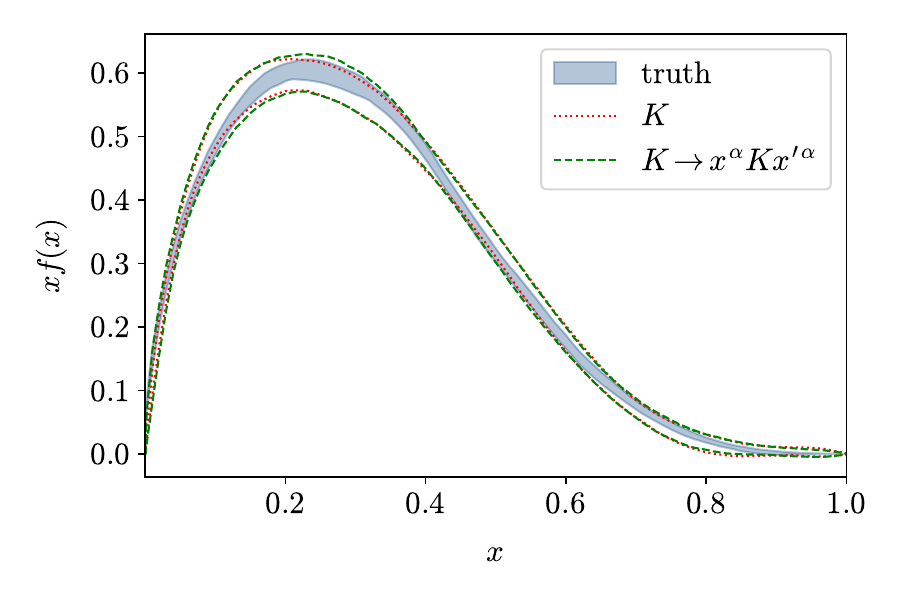}
    \end{subfigure}
    
    \centering
    \begin{subfigure}[t]{0.3\textwidth}
        \centering
        \includegraphics[width=\linewidth]{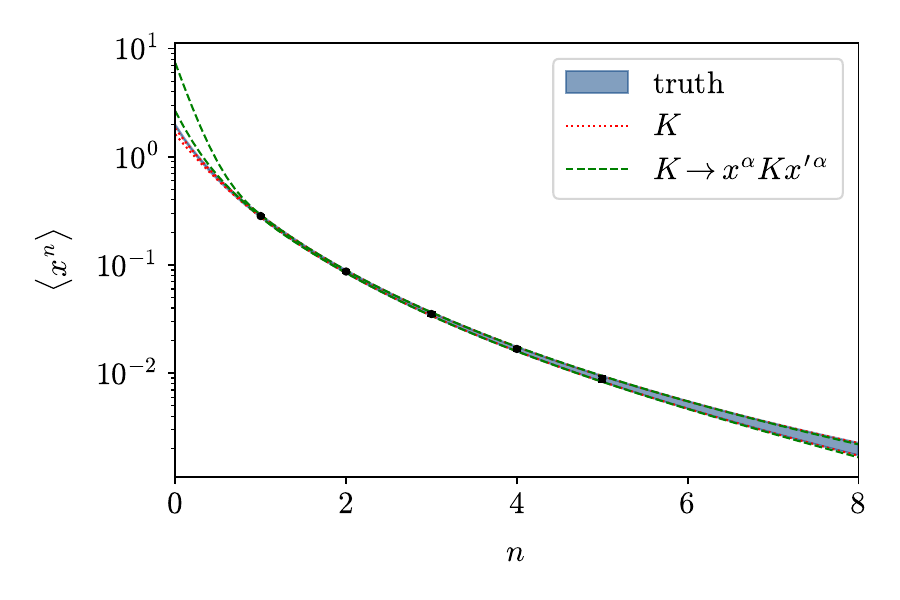}
    \end{subfigure}\hfill
    \begin{subfigure}[t]{0.3\textwidth}
        \centering
        \includegraphics[width=\linewidth]{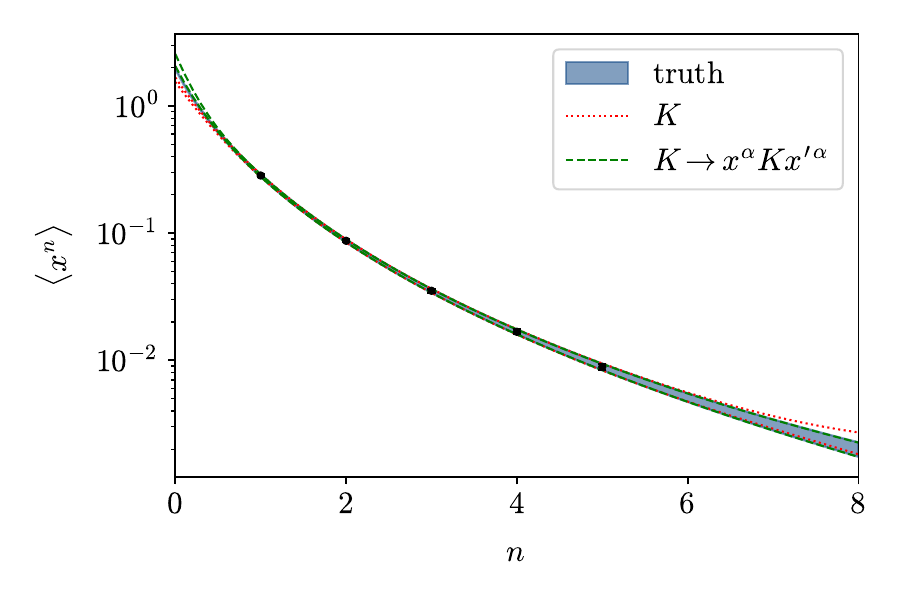}
    \end{subfigure}\hfill
    \begin{subfigure}[t]{0.3\textwidth}
        \centering
        \includegraphics[width=\linewidth]{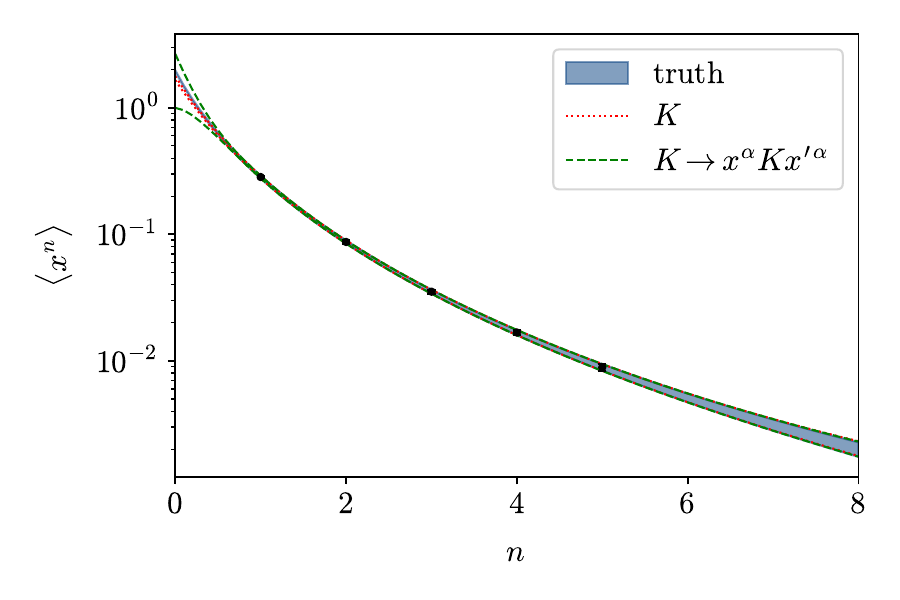}
    \end{subfigure}\hfill

    \centering
    \begin{subfigure}[t]{0.27\textwidth}
        \centering
        \includegraphics[width=\linewidth]{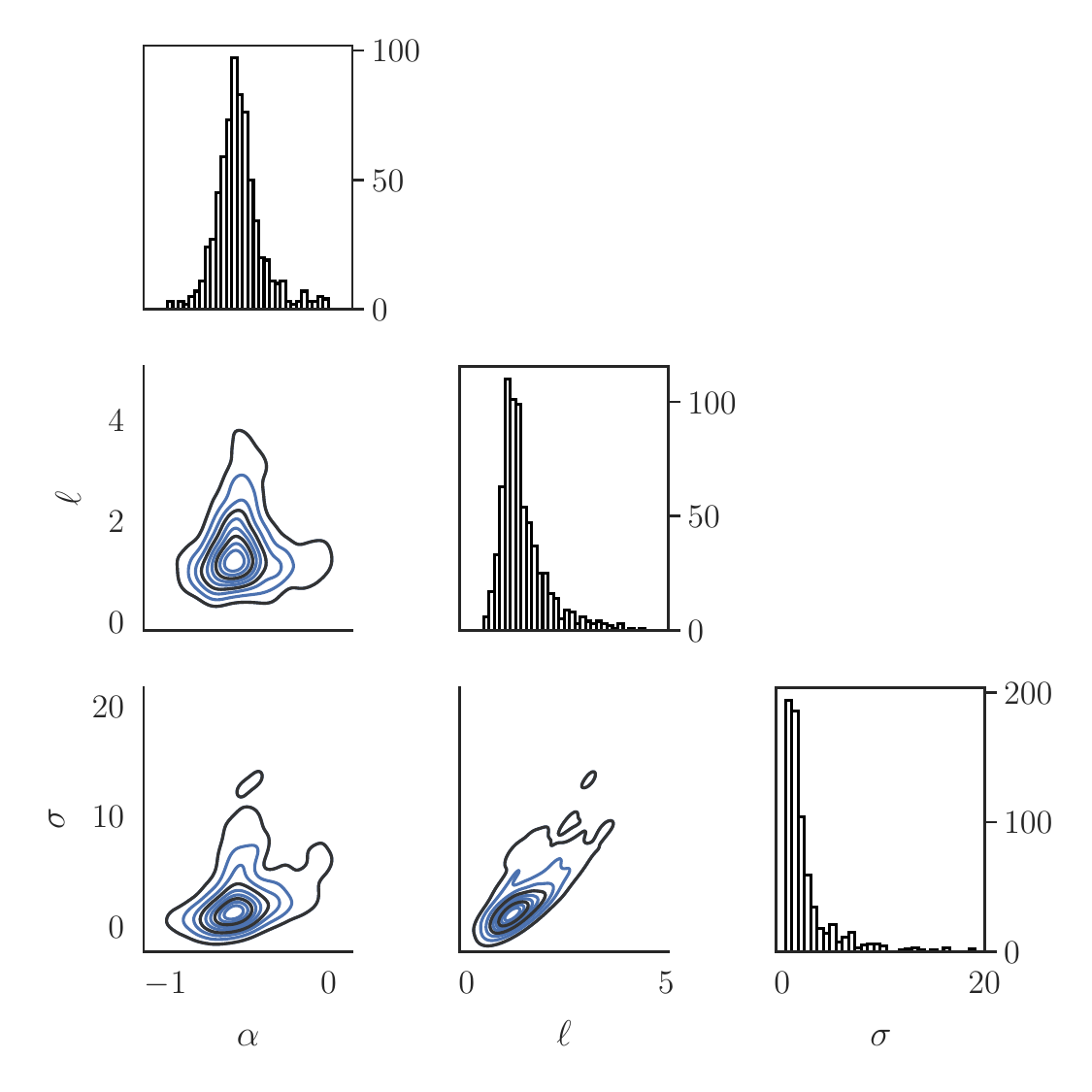}
        \caption{LSE}
    \end{subfigure}\hfill
    \begin{subfigure}[t]{0.27\textwidth}
        \centering
        \includegraphics[width=\linewidth]{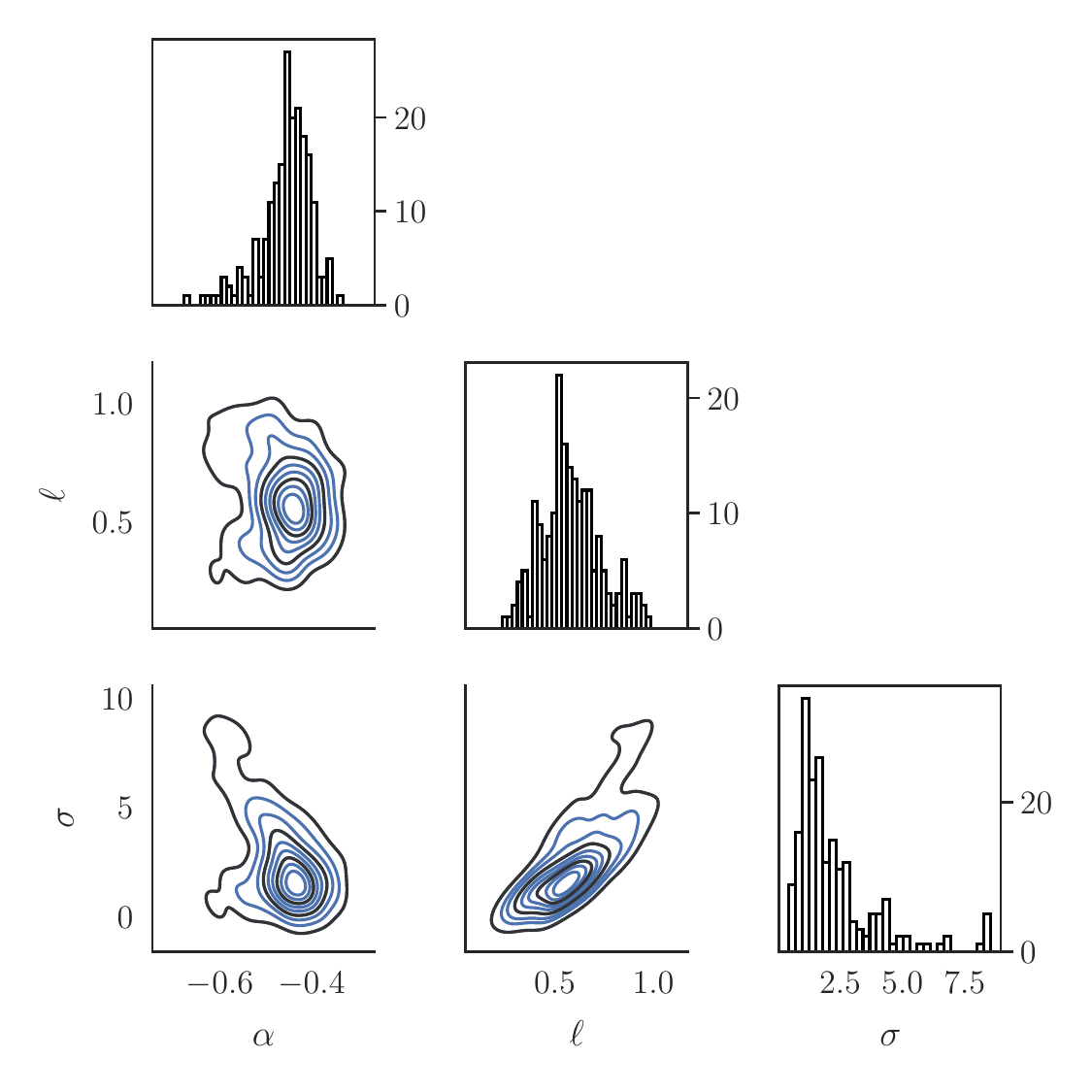}
        \caption{SE}
    \end{subfigure}\hfill
    \begin{subfigure}[t]{0.27\textwidth}
        \centering
        \includegraphics[width=\linewidth]{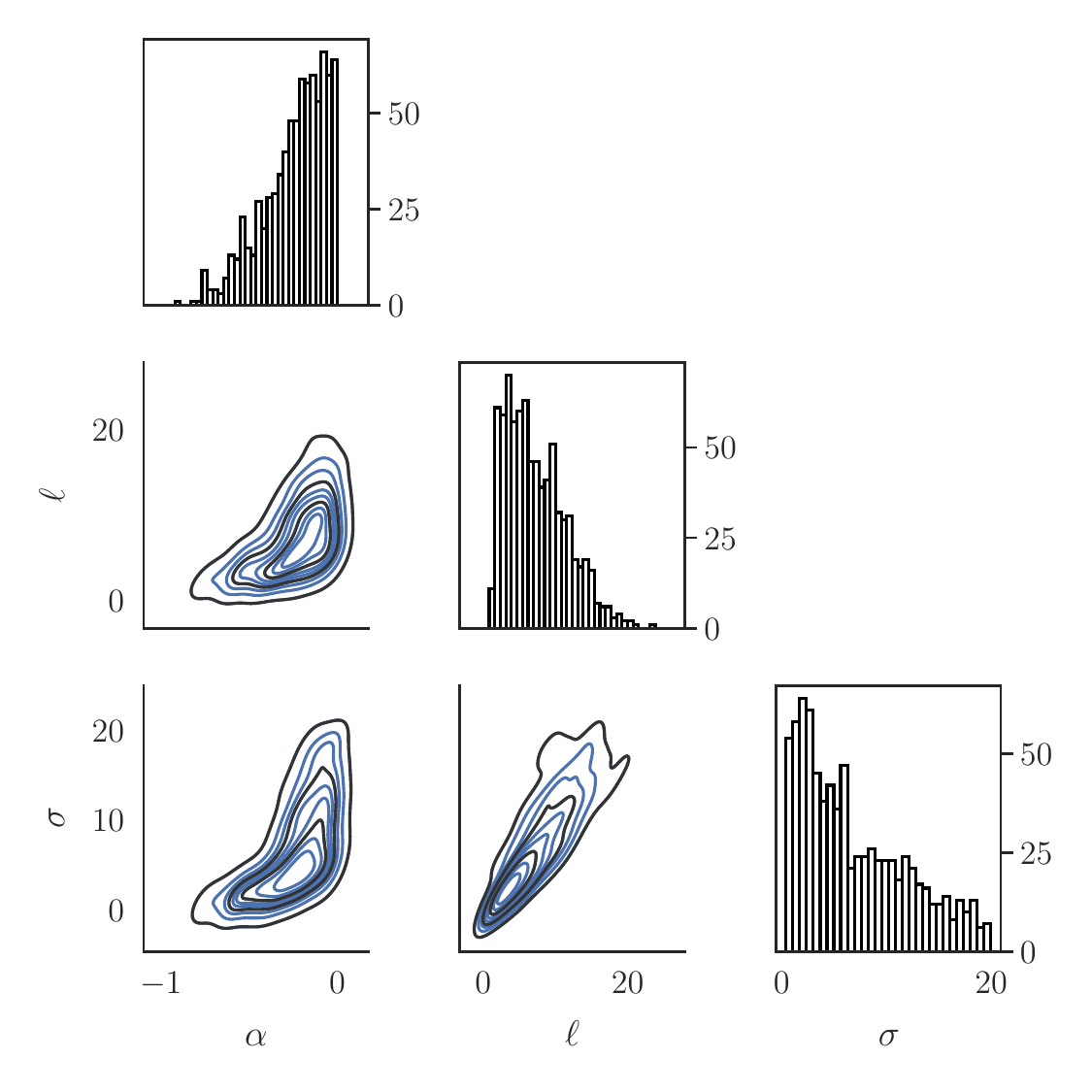}
        \caption{\MATERN{3/2}}
    \end{subfigure}\hfill

    \caption{Comparison of different kernels on \NNPDF data at $\nmax = 5$, with and without $x^\alpha$ kernel prior. \biascorrnote}
    \label{fig:NNPDF_alphaScan}
\end{figure}

For MCS and LCS kernels we have tried on the \NNPDF valence dataset up until now, we have left out the $x^\alpha$ kernel prior, and have observed behavior ranging from acceptable to excellent for a large range of kernel choice. If our sampled value for $\alpha$ is sharply peaked at a given value, adding this prior will constrain our fit, if not, it will add a degree of uncertainty to our posterior reconstruction, particularly at $x\ll1$. When we add this prior to some of our OOTB kernels, as shown in Fig. \ref{fig:NNPDF_alphaScan}, we do not get large deviations in our posterior reconstruction. Following from Sec. \ref{subsec:math_theory_disc}, it is clear that the power divergence in valence datasets is simply not steep enough for the divergence prior to make too much of a difference. We will see later how for other types of PDF data, the power divergence becomes a necessary prior rather than an optional one.

\subsection{Controlling $x$ correlation lengths with Gibbs-style kernels}

As seen thus far with the \FANTO valence dataset, the lack of control of $x$ correlation leads to systematic issues that do not resolve themselves with amelioration in data space. One way to address this issue is to define kernels that are piecewise in $x$-space like studied in Ref. \cite{medrano_gaussian_2025}; if we have a known model that behaves differently in different region of $x$, this may be useful to implement. We will instead turn to model-agnostic Gibbs-type kernels introduced in Sec. \ref{para:Gibbs}, which offer a way to automatically ``learn" the correlation length $\ell(x)$ by appropriately scaling the basis functions that make up whatever kernel we are using, thereby maintaining our kernels positive (semi)definiteness. As can be seen in Fig 4.6 in Ref. \cite{Rasmussen2006Gaussian}, a tiny ``blip'' downwards in our $\ell$ function at point $x_0$ will correspond to a tiny ``blip" in the correlation in our prior function at that location. For any more complicated a scenario, it can be hard to intuit the exact form of $\ell(x)$ that should be associated with a truth model, especially given the choice of base kernels we can use. For the purposes of this entire study, our ``base kernel"\footnote{the kernel off of which we make the Gibbs extension} will be the LSE kernel; it contains the longest scale $x$ correlation of the kernels we have discussed. Given that the LSE kernel reconstructs the long scale correlation behavior of our PDFs well, our hope is that $\ell(x)$ will now be easier to tune than if other kernels are used. 

In Sec. \ref{subsec:glue_constrained}, for gluon PDFs, we will explore how purposefully crafted $\ell(x)$ functions can circumvent the problems we encounter with OOTB kernels given the un-integrable power divergences at low $x$. The choice of $\ell(x)$ is surprisingly robust in dealing with issues in regression \textit{provided that we know the underlying model  of our reconstruction}. This adds a model bias, but a path can be laid out where multiple models are evaluated with data and tested/weighted under the metrics given in \ref{subsec:validation}. If we have constrained/integrable data like in the valence PDF case, what would be more ideal would be
to have a model agnostic method that naturally adapts to a wide range of data. In the following section we present a rather simple model that does just that, and demonstrate its effectiveness in modeling the underlying truth of the \FANTO valence dataset that has thus far evaded proper reconstruction.

\subsection{Automatic Gibbs adjustment}

In Ref. \cite{candido_bayesian_2024}, the Gibbs kernel with $\ell(x)= x + \delta$\footnote{$\delta$ being a small regularizing parameter} was explored, with the goal of enforcing relative low $x$ correlation at $x\ll 1$ and high $x$ correlation $x \sim 1$. We seek to generalize this choice of $\ell(x)$ to ``learn" our correlation function using as little prior information as possible. The method(s) we present below is rudimentary, and while there are a number of improvements that can be added to it in time, results are already robust, and can scale between different models.

\paragraph{Degree one polynomial} Our first step in automating our Gibbs adjustment is to let the end points be free points, thereby letting our $x$ correlation function be an arbitrary\footnote{within bounds set by $p(\Theta)$} polynomial of degree 1. This means that our correlation function is $\ell(x) = ax+b(1-x)$, where $a$ and $b$ are parameters to be sampled. 

\begin{figure*}[h!]
    \centering

    \begin{subfigure}[t]{0.40\textwidth}
        \centering
        \includegraphics[width=\textwidth]{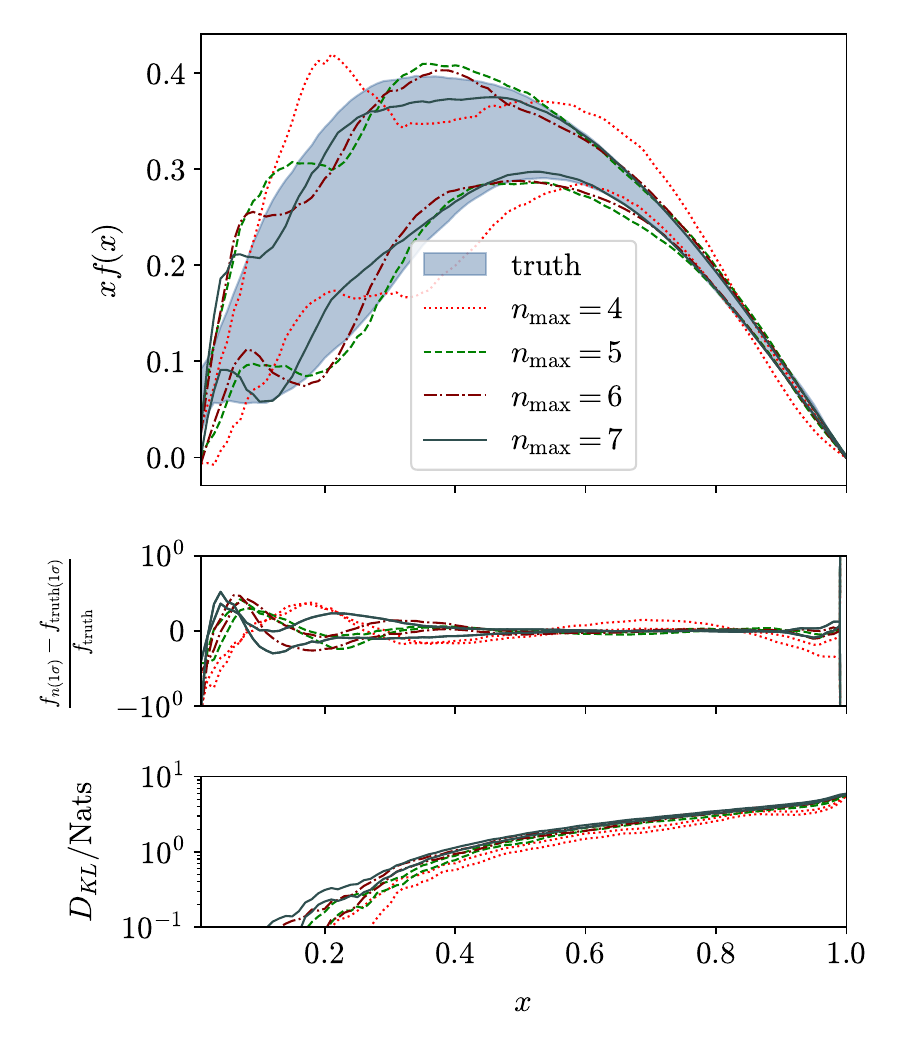}
        \caption{}
        \label{fig:xPDF_FANTO_GIBBS_d1_cD}
    \end{subfigure}
    \begin{subfigure}[t]{0.40\textwidth}
        \centering
        \includegraphics[width=\textwidth]{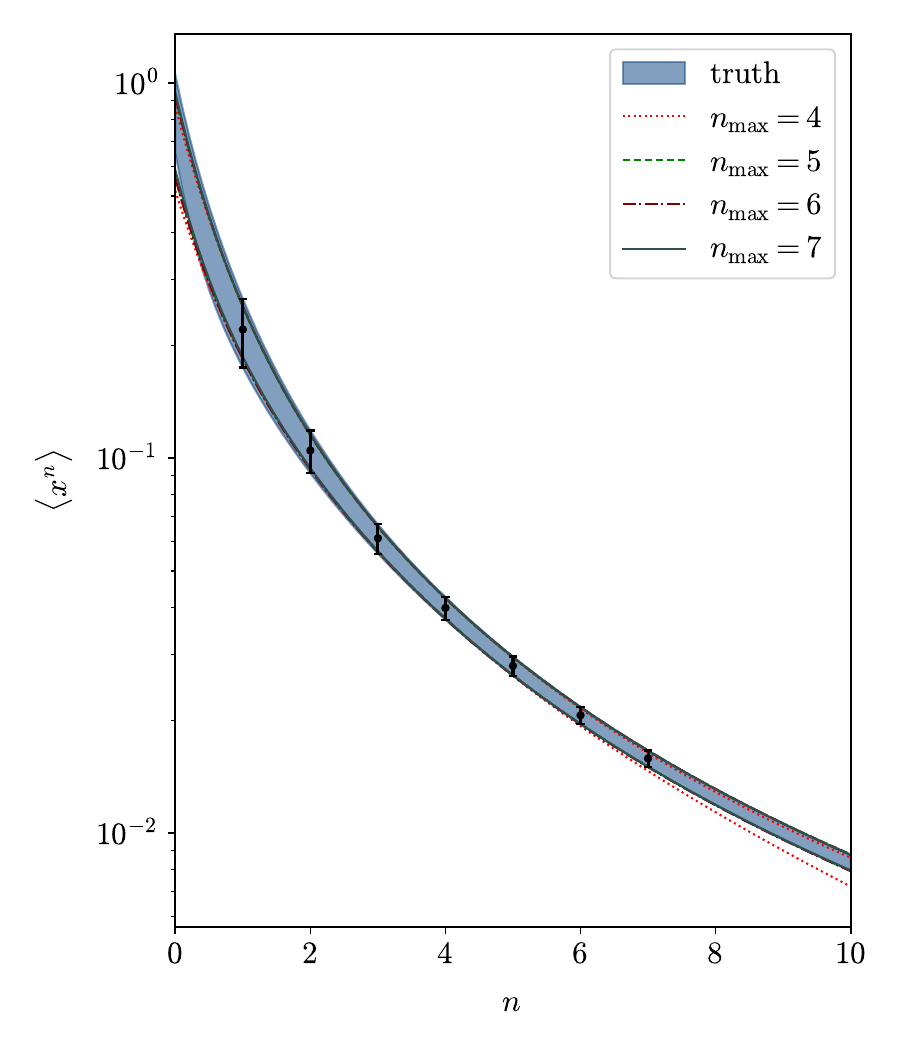}
        \caption{}
        \label{fig:MomRec_FANTO_GIBBS_d1_cD}
    \end{subfigure}
    
    \caption{\GIBBS{1} kernel used on \FANTO dataset. \biascorrnote}
    \label{fig:FANTO_GIBBS_d1_cD}
\end{figure*}

The \FANTO dataset has large uncertainties for low $n$ and small uncertainties for high $n$ making OOTB kernels particularly hard to control for a variable $n$ number of moment data points. In line with the considerations detailed in Ref. \cite{candido_bayesian_2024}, we would naively expect more uncertainty as $x\to 0$, so we let $a\in[0.01,0.5]$ and $b\in[0.01,1]$ when constructing our degree one polynomial. Applying this, we can see in Fig. \ref{fig:FANTO_GIBBS_d1_cD} how beginning at $\nmax = 4$ we already have a slightly deflated variance. As we increase the amount of data points, we see the reconstruction in both data and $x$ space both deflate further. While this progression offers significant improvements over OOTB kernels, it does indicate that we need a kernel with extra degrees of generality that can accurately sense when not to under estimate the variance of low moment data when more precise high moment data is added.

\paragraph{Degree two polynomial}
\label{para:FANTO_GIBBS}
We now graduate from a polynomial of degree one to one of degree two in an effort to afford more flexibility to our $x$ correlation function. In addition to keeping our endpoints free, we introduce a mid-point $m$ on the $x$ axis, letting $\ell(m)$ be an extra variable $c$. With these three points, a parabola is completely specified.  

\begin{figure*}[h!]
    \centering

    \begin{subfigure}[t]{0.40\textwidth}
        \centering
        \includegraphics[width=\textwidth]{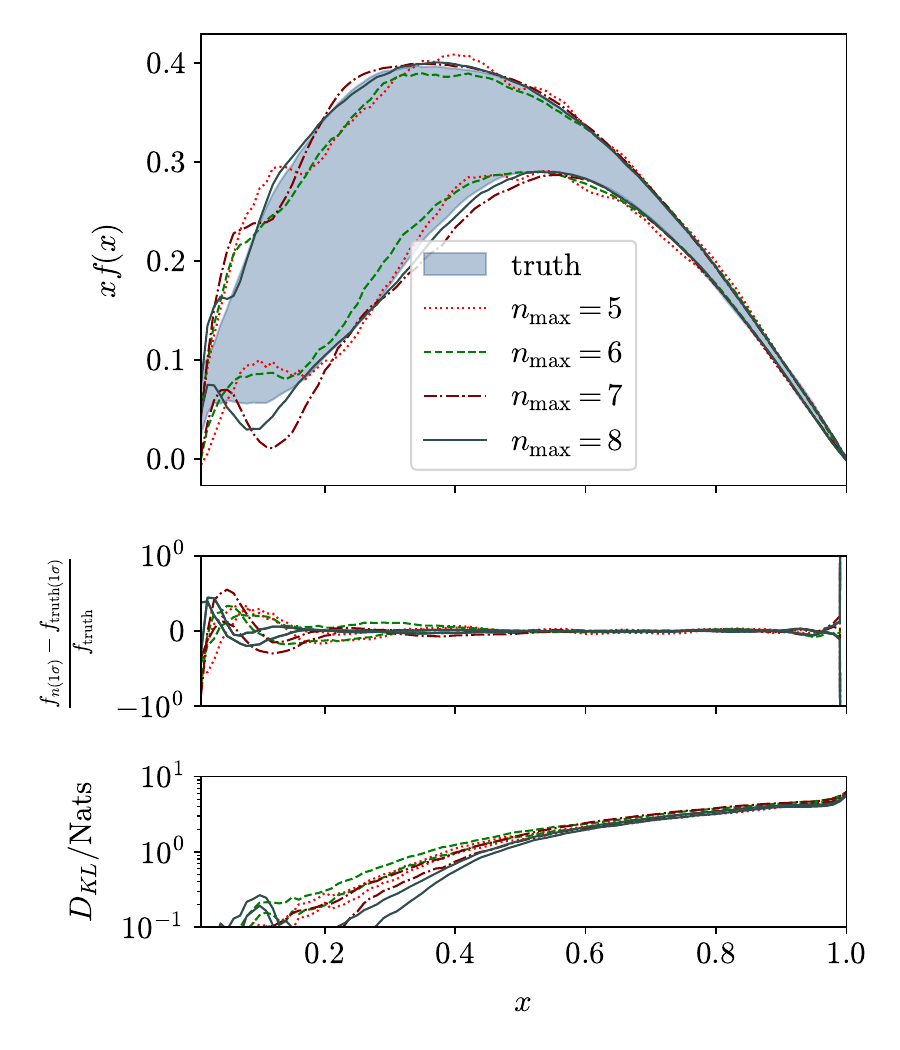}
        \caption{}
        \label{fig:xPDF_FANTO_GIBBS_d2_cD}
    \end{subfigure}
    \begin{subfigure}[t]{0.40\textwidth}
        \centering
        \includegraphics[width=\textwidth]{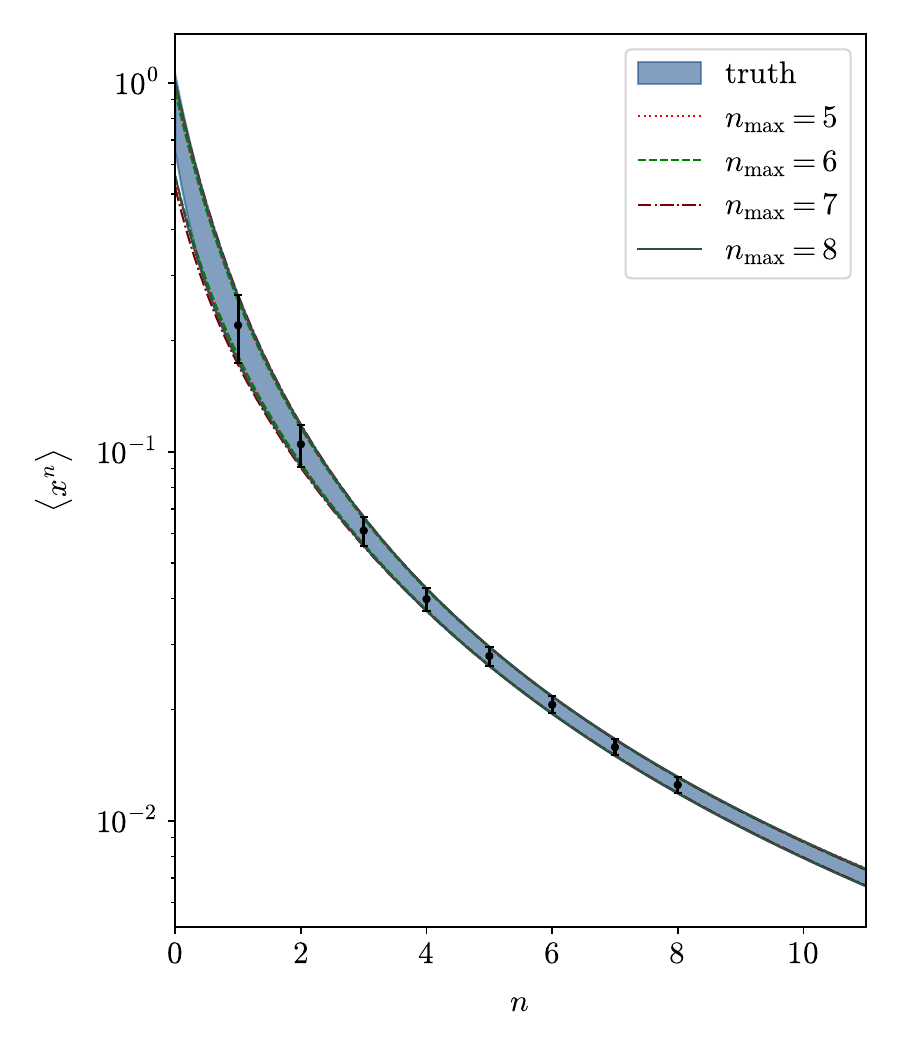}
        \caption{}
        \label{fig:MomRec_FANTO_GIBBS_d2_cD}
    \end{subfigure}
    
    \caption{\GIBBS{2} kernel used on \FANTO dataset. \biascorrnote}
    \label{fig:FANTO_GIBBS_d2_cD}
\end{figure*}

When we apply this kernel to the \FANTO dataset, we need to tune $p(\Theta)$, i.e. set the bounds of the polynomials defining $\ell(x)$. Just by looking at our data, we do not know explicitly that uncertainty at low $x$ is much larger than at high $x$, but we can infer this from the low moments having much higher uncertainties than high moments. From this, we define $p(\Theta)$ first and specify the ranges of $a$,$m$, and $c$: $a\in[0.01,0.5]$, $m\in[0.01,0.1]$, and $c\in[0.01,0.8]$. This selection is ad-hoc, but a) is informed by the data we given and b) our selection further enforces physics-rooted priors, i.e. our uncertainty for low-$x$ is far larger than it is for high-$x$. We can tune bounds appropriately to achieve more satisfactory results if need be. Results for this kernel are shown in Fig. \ref{fig:FANTO_GIBBS_d2_cD} (with bias correction). Here we see a successful reconstruction (i.e. one that does not get highly biased towards progressively higher $\nmax$). The oscillations  at low-to-mid $x$ are not exactly the same between truth and reconstruction, but the reconstruction in data space shows negligible error in mean/variance. To continue off the discussion in Sec. \ref{subsec:xa_impact} of whether the $x^\alpha$ kernel prior makes any difference, plots are shown in Sec. \ref{sec:overflow_FANTO} to confirm that there is no appreciable impact. 

\begin{figure*}[h!]
    \centering

    \begin{subfigure}[t]{0.40\textwidth}
        \centering
        \includegraphics[width=\textwidth]{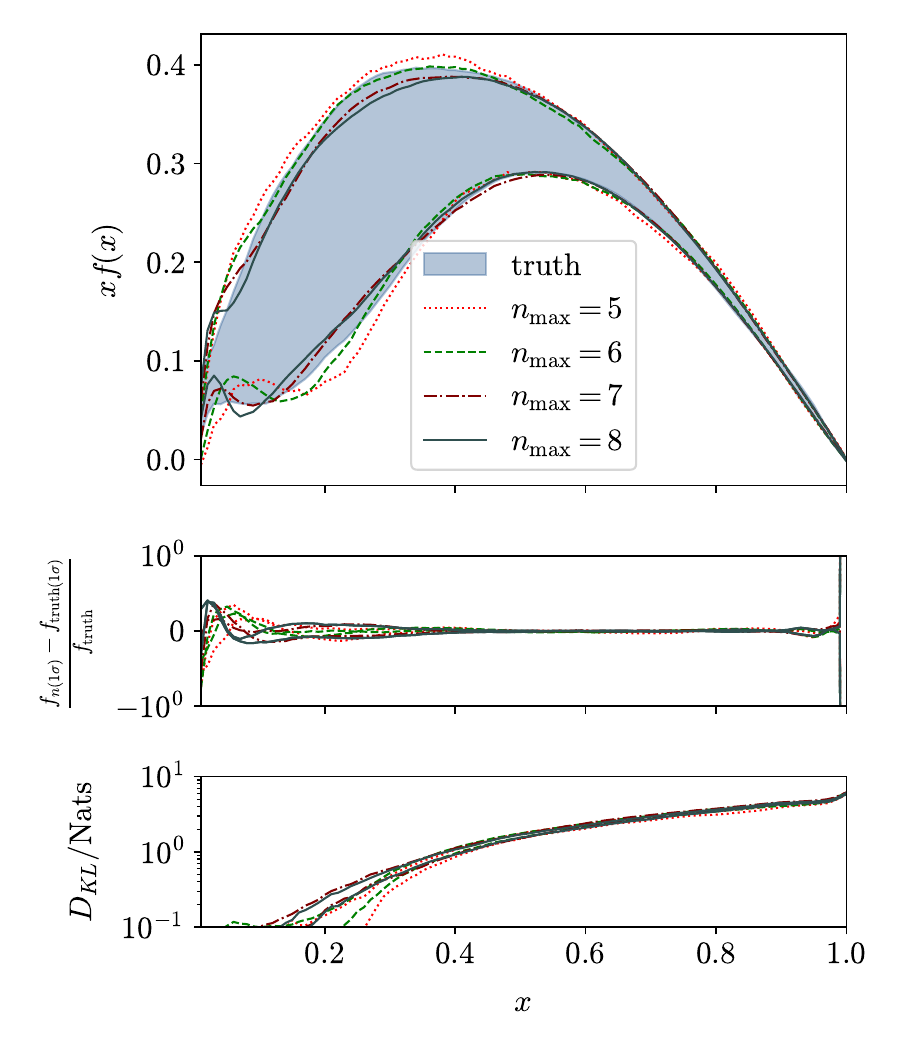}
        \caption{}
        \label{fig:xPDF_FANTO_GIBBS_d2_wm_cD}
    \end{subfigure}
    \begin{subfigure}[t]{0.40\textwidth}
        \centering
        \includegraphics[width=\textwidth]{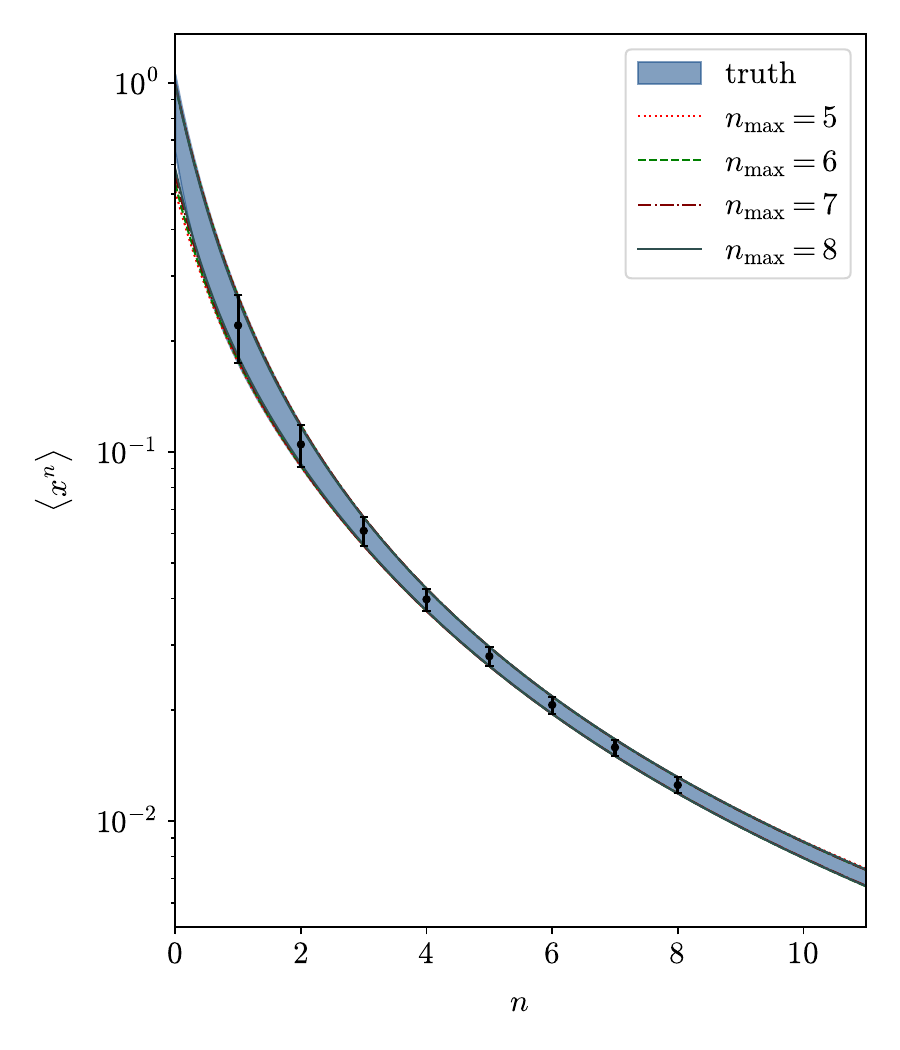}
        \caption{}
        \label{fig:MomRec_FANTO_GIBBS_d2_wm_cD}
    \end{subfigure}
    
    \caption{\GIBBS{2} kernel with modified (suboptimal) $p(\Theta)$ used on \FANTO dataset. \biascorrnote}
    \label{fig:FANTO_GIBBS_d2_wm_cD}
\end{figure*}

\paragraph{Robustness in tuning $p(\Theta)$} Our reconstruction in $x$-space is robust to changes in $p(\Theta)$\footnote{remember that in our study we are only considering $p(\Theta)$ following uniform distributions}. As an example, let us redefine our bounds for variables to be $a,b \in[0.01,1]$, $c\in [0.01,1.5]$, $m\in [0.01,0.5]$. Rather than limiting $a$'s bounds, we stretch the bounds on $c$, thereby still enforcing behavior at $x\sim0$ to be very uncorrelated at higher $x$. Results are shown in Fig. \ref{fig:FANTO_GIBBS_d2_wm_cD}. We note very good convergence for $\nmax \in [5,6]$, and, as we increase $\nmax$, we see signs of narrowing variance for low $x$. This would indeed be less preferable than the immediately preceding \GIBBS{2} kernel, but more preferable than all OOTB kernels and the \GIBBS{1} kernel. The large errors in data space imply multi-modality of the underlying $f(x)$, and if we had no truth model to perform closure tests, both \GIBBS{2} kernels discussed would have very similar performances.

\paragraph{Higher degree polynomials}

We could in principle attempt to set even higher degrees of polynomials to be our $x$ correlation functions. One choice for a third degree polynomial could be modifying our degree two setup by additionally letting the derivative at point $(m,c)$ be an free parameter, and then letting $\ell(x)$ consist of two cubic splines  in the regions $x<m$  and $x\geq m$. Yet another choice would be to add another point in the normal domain $(m',c')$. This approach of adding points which are free to move in the domain of $x$ is not obviously better than the approach of keeping the $x$ value of these midpoints fixed and only sampling their $\ell$ values. A future direction of study would be to gain a more analytical understanding of the role certain choices of $\ell(x)$ play in reconstruction and thereby exploiting these choices on datasets.

\subsection{Robustness of grid choice for valence PDFs}

\begin{figure*}[h!]
    \centering

    \begin{subfigure}[t]{0.40\textwidth}
        \centering
        \includegraphics[width=\textwidth]{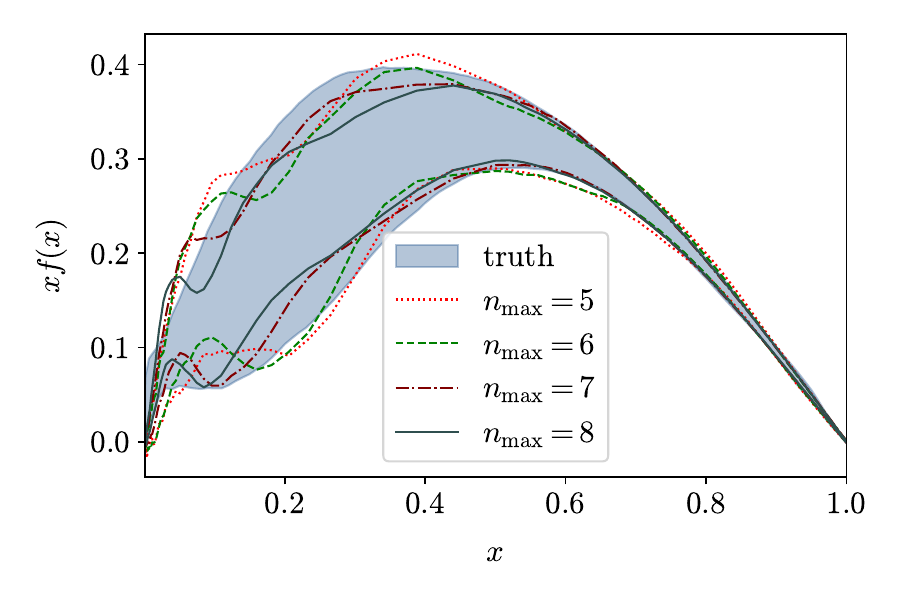}
    \end{subfigure}
    \begin{subfigure}[t]{0.40\textwidth}
        \centering
        \includegraphics[width=\textwidth]{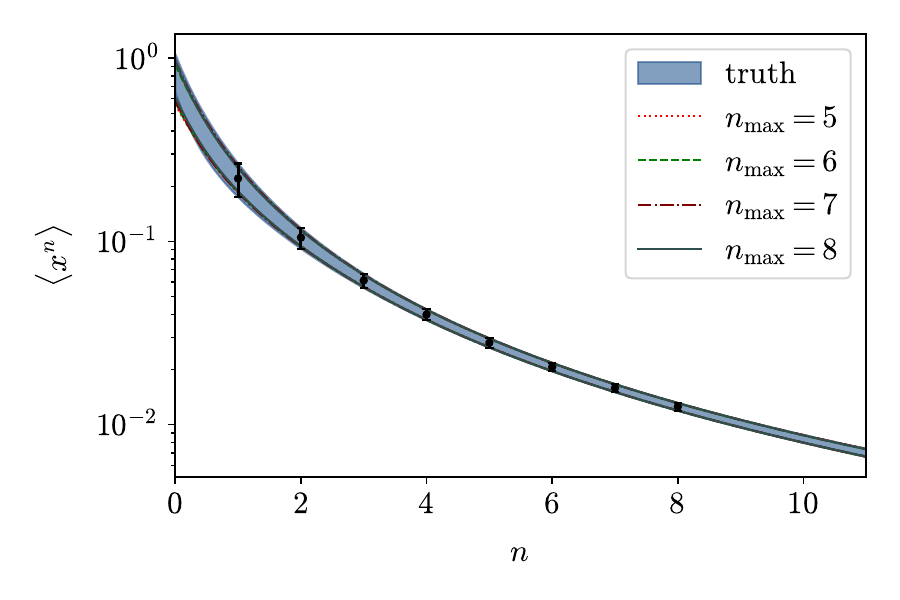}
    \end{subfigure}
    
    \caption{Grid scheme 2 used on \FANTO dataset with \GIBBS{1} kernel. \biascorrnote}
    \label{fig:FANTO_GIBBS_d1_LOG}
\end{figure*}

\begin{figure*}[h!]
    \centering

    \begin{subfigure}[t]{0.40\textwidth}
        \centering
        \includegraphics[width=\textwidth]{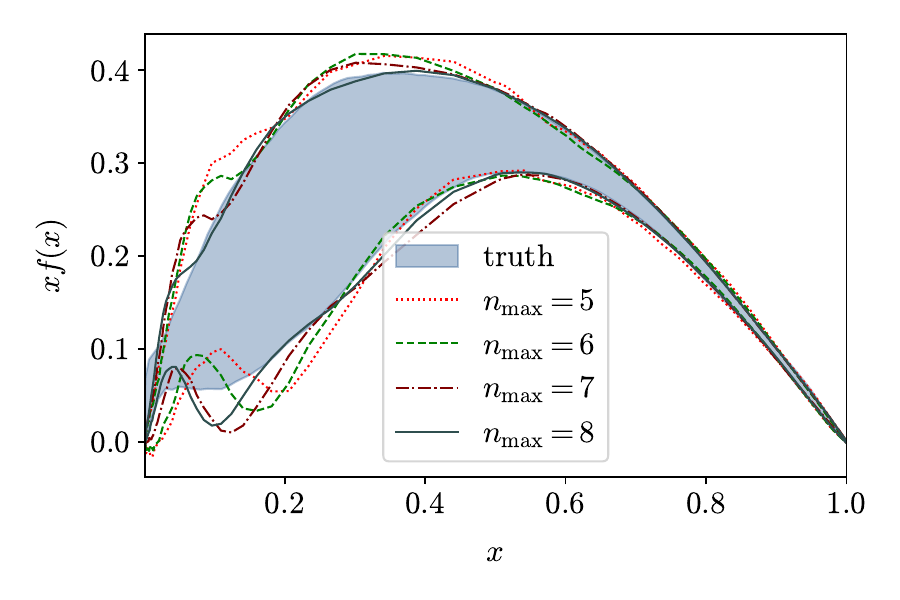}
    \end{subfigure}
    \begin{subfigure}[t]{0.40\textwidth}
        \centering
        \includegraphics[width=\textwidth]{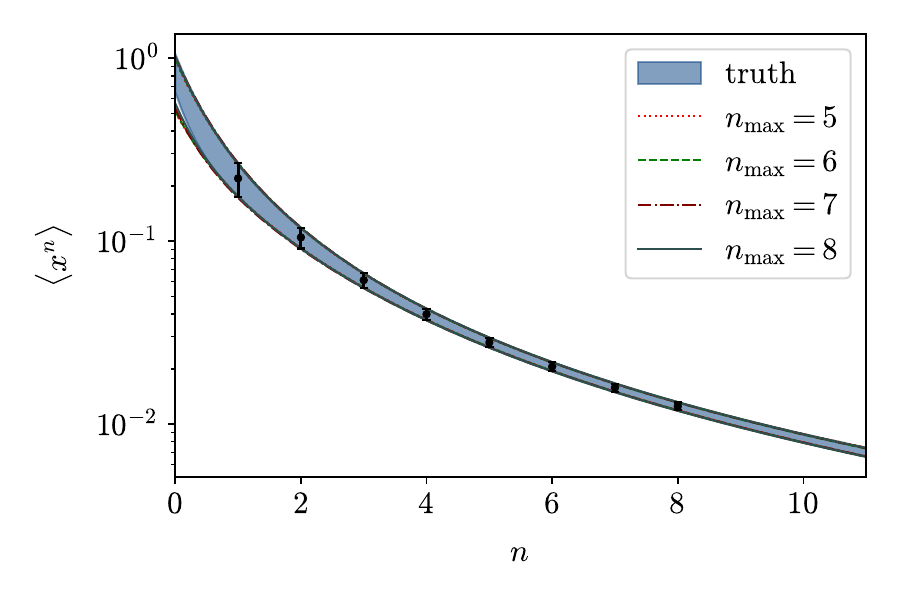}
    \end{subfigure}
    
    \caption{Grid scheme 2 used on \FANTO dataset with \GIBBS{2} kernel. \biascorrnote}
    \label{fig:FANTO_GIBBS_d2_LOG}
\end{figure*}

We have thus far only discussed results with a hybrid yet mostly linear grid (grid scheme 1 as detailed in Sec. \ref{subsec:grid_details}). The results for the two Gibbs style kernels with grid scheme 2 (a mostly logarithmic grid scheme) are shown in Figs. \ref{fig:FANTO_GIBBS_d1_LOG} and \ref{fig:FANTO_GIBBS_d2_LOG} and show generally the same features as in grid scheme 1; our valence reconstructions are evidently robust to grid scheme choice. The manageability of the divergence at $x\ll1$ we see for the valence case will no longer hold in the gluon/sea quark case, which we will discuss next.

\subsection{Comments on $n=0$ reconstruction and enforcing norm constraints}
\label{subsec:comments_nto0}

\begin{figure*}[h!]
    \centering

    \begin{subfigure}[t]{0.40\textwidth}
        \centering
        \includegraphics[width=\textwidth]{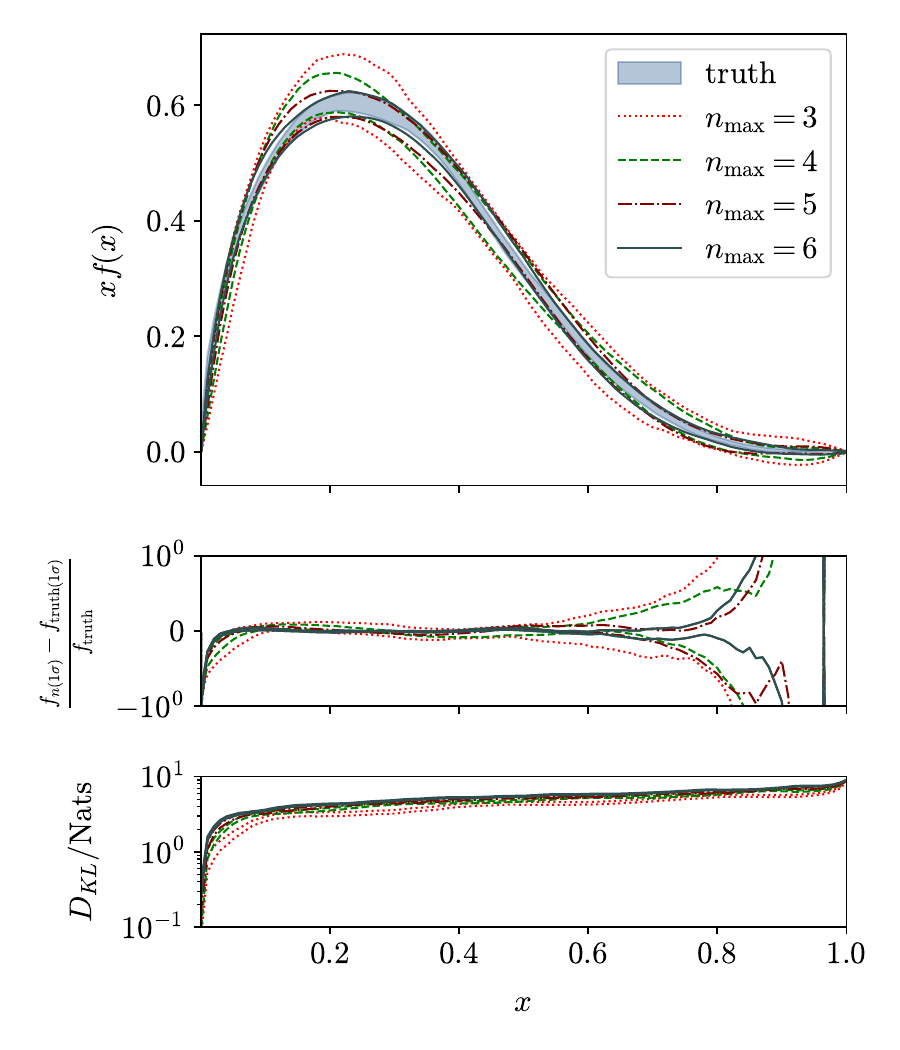}
        \caption{}

    \end{subfigure}
    \begin{subfigure}[t]{0.40\textwidth}
        \centering
        \includegraphics[width=\textwidth]{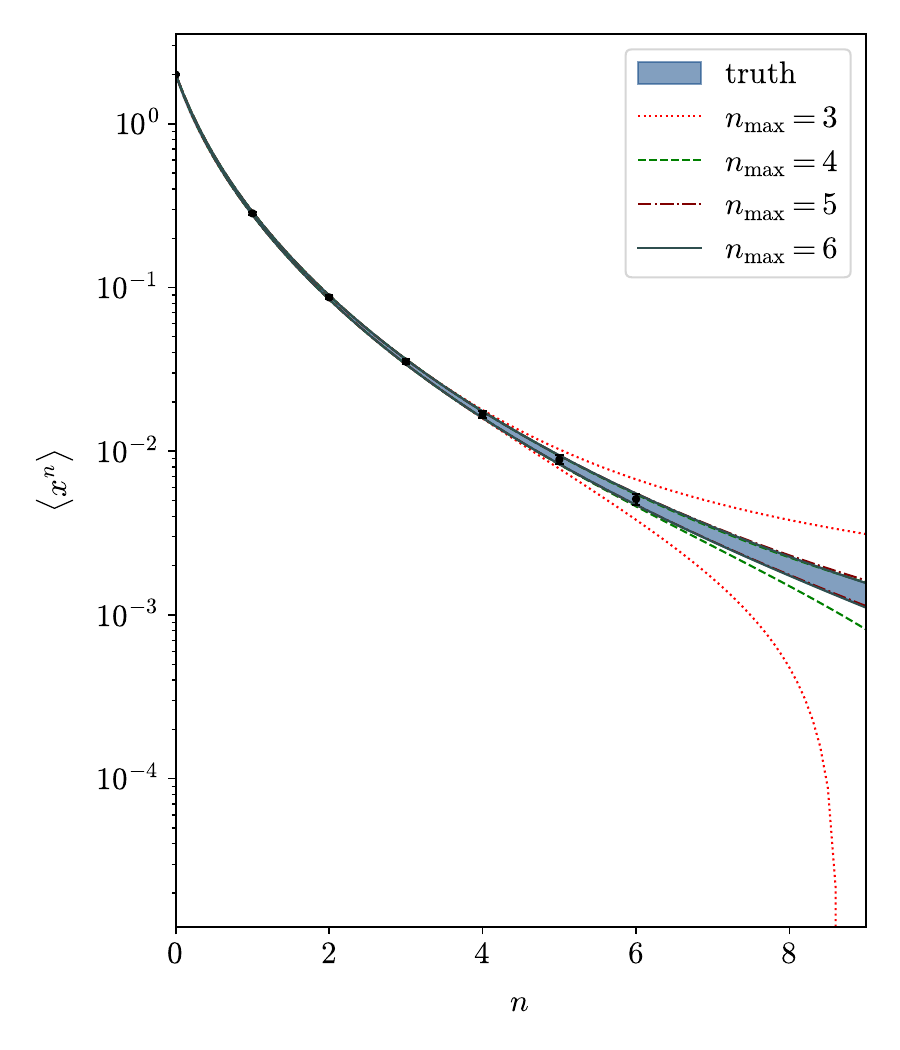}
        \caption{}
    \end{subfigure}
    
    \caption{\MATERN{3/2} kernel used on \NNPDF dataset with norm constraint enforced. \biascorrnote}
    \label{fig:NNPDF_MATERN32_EN}
\end{figure*}

In Sec. \ref{subsubsec:extrapolation_nto0} we noted how we would observe how well the $n=0$ reconstruction would be. We see that across both valence PDF sets, a broad range of kernels achieve an extrapolation that is consistent with the ground truth, the only exception to this being the LSE kernel with the $x^\alpha$ divergence kernel prior at $\nmax=5$. Additionally, extrapolation with the \MATERN{1/2} kernel was particularly noisy, as expected. For all other kernels with a relatively longer $x$ correlation scale, the lack of norm information makes little to no difference in reconstruction in $x$ space. To further drive home this point, consider the plots shown in Fig. \ref{fig:NNPDF_MATERN32_EN}: the norm enforcement evidently makes no discernible difference in PDF reconstruction in $x$ space; in moment space, we see how it predictably improves the $n\to 0$ extrapolation. It is possible for the added constraint to make sampling of hyperparameter more difficult (although we have not seen this problem firsthand), and in this case, the constraint can have its uncertainty widened, or dropped entirely. With a dataset like \FANTO valence, such a constraint could 
also be readily added, however, the addition of a norm constraint would add an extra scale of uncertainty that would likely require a slight modification to the bounds in our Gibbs style kernel. Our results are robust as is, with sufficiently satisfactory $n\to 0$ extrapolation, so we will omit this norm constraint for the \FANTO valence dataset. Drawing from our discussion in Sec. \ref{subsec:math_theory_disc}, we can develop an analytical intuition for this lack of necessity for a norm constraint: the integrability of $f(x)$ ensured that the blow ups at $x\sim 0$ are manageable, and relatively few moments would therefore be necessary to achieve a suitable reconstruction. Thus, even without imposing the valence sum rule, our reconstructions in $x$ and moment space were successful.

\section{Reconstructing gluon/sea quark PDF datasets}
\label{sec:glue_sea_recon}
 Now we turn to gluon and sea PDFs. Here, there is no more norm constraint: unlike the valence PDF case, there is no clear extrapolation of $\langle x^n\rangle$ to $n\to 0$. The sea quarks and gluons both take part in the PDF momentum sum rule and so this constraint can be used, but the sea/gluon moment computation does not rely on previously calculating the rest of the flavor decomposition, and so we will only consider directly fitting their moments to a PDF form. This means that without prior expectation of $f(x)$'s behavior, there will likely be a multitude of very different functions in $x$ space that all satisfy the reconstruction in moment space. This poses a significant problem, as we know that the majority of the gluon/sea quark physics of interest occurs in the $x\ll 1$ regime. We will outline strategies to deal with these power divergent PDFs that can result in a good reconstruction, albeit not at a level competitive with valence PDF reconstructions.\footnote{As a visualization note, we will truncate that $y$ axis range of many of the sea/gluon plots for easier viewing.}

\subsection{Fitting constrained gluon PDF datasets}
\label{subsec:glue_constrained}
 Stemming from our discussion in Sec. \ref{subsec:math_theory_disc}, we expect that the greater the degree of power divergence in $f(x)$, the greater the amount of moments would be needed, and this issue would persist regardless of the data precision. For the case of gluons and sea quarks, due to no a priori assumptions of $f(x)$'s integrability, this will be a persistent issue. To faithfully reconstruct $f(x)$, we must introduce some extra specifications and added priors that were not strictly necessary in the valence PDF case, both to prevent un-physical solutions and to filter for solutions that are representative of gluonic/sea quark physics. 

 \paragraph{Hyper-parameters}

 The choice of $p(\Theta)$ bounds has a significant impact of reconstruction, but for now we will keep $\sigma\in[0,20]$ and $\ell\in[0,50]$.
 
\paragraph{Logarithmic spacing}

As we can see from the gluon/sea PDFs in Sec. \ref{subsec:datasets}, the majority of the features in $f(x)$ are contained in the $x\ll 1$ region. Unlike the valence case, the choice of grid spacing that changes the flexibility of our grid at low $x$ will have a significant impact on our reconstruction. This calls for taking into account how different grid spacings affect our reconstruction. Refer to Sec. \ref{subsec:grid_details} for more details.

\begin{figure*}[h!]
    \centering

    \begin{subfigure}[t]{0.40\textwidth}
        \centering
        \includegraphics[width=\textwidth]{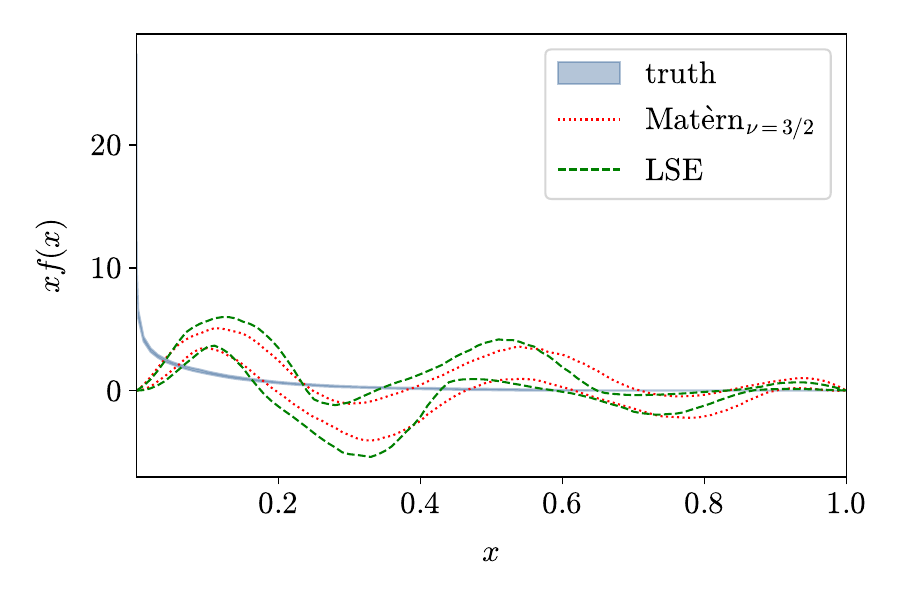}
        \label{fig:NNPDF_gluon_CovScan:xPDF}
    \end{subfigure}
    \begin{subfigure}[t]{0.40\textwidth}
        \centering
        \includegraphics[width=\textwidth]{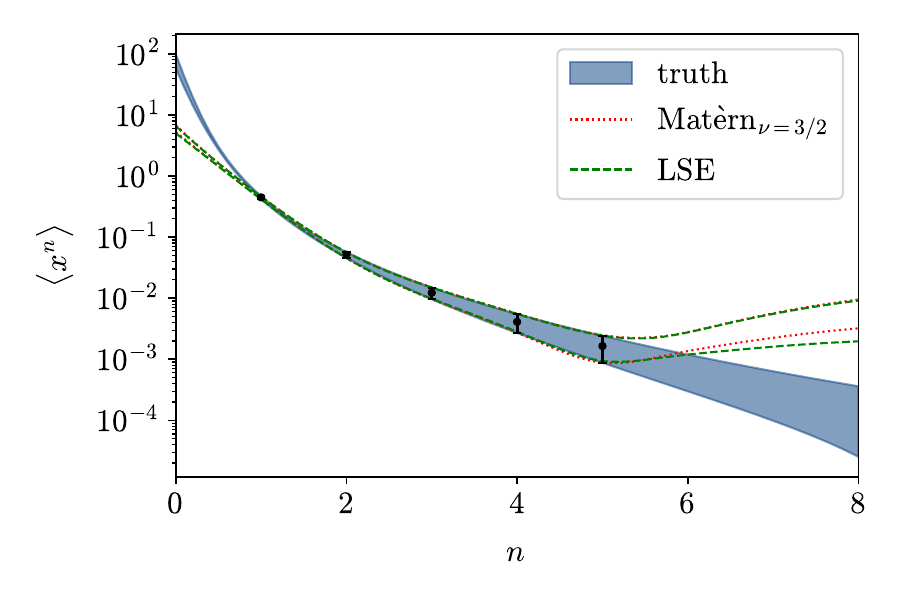}
        \label{fig:NNPDF_gluon_CovScan:MomRec}
    \end{subfigure}
    
    \caption{An example with the \NNPDF gluon PDF of extrapolation inconsistent with the Hausdorff moment condition that leads to a poor $x$ space reconstruction}
    \label{fig:NNPDF_gluon_CovScan}
\end{figure*}

\begin{figure*}[h!]
    \centering

    \begin{subfigure}[t]{0.40\textwidth}
        \centering
        \includegraphics[width=\textwidth]{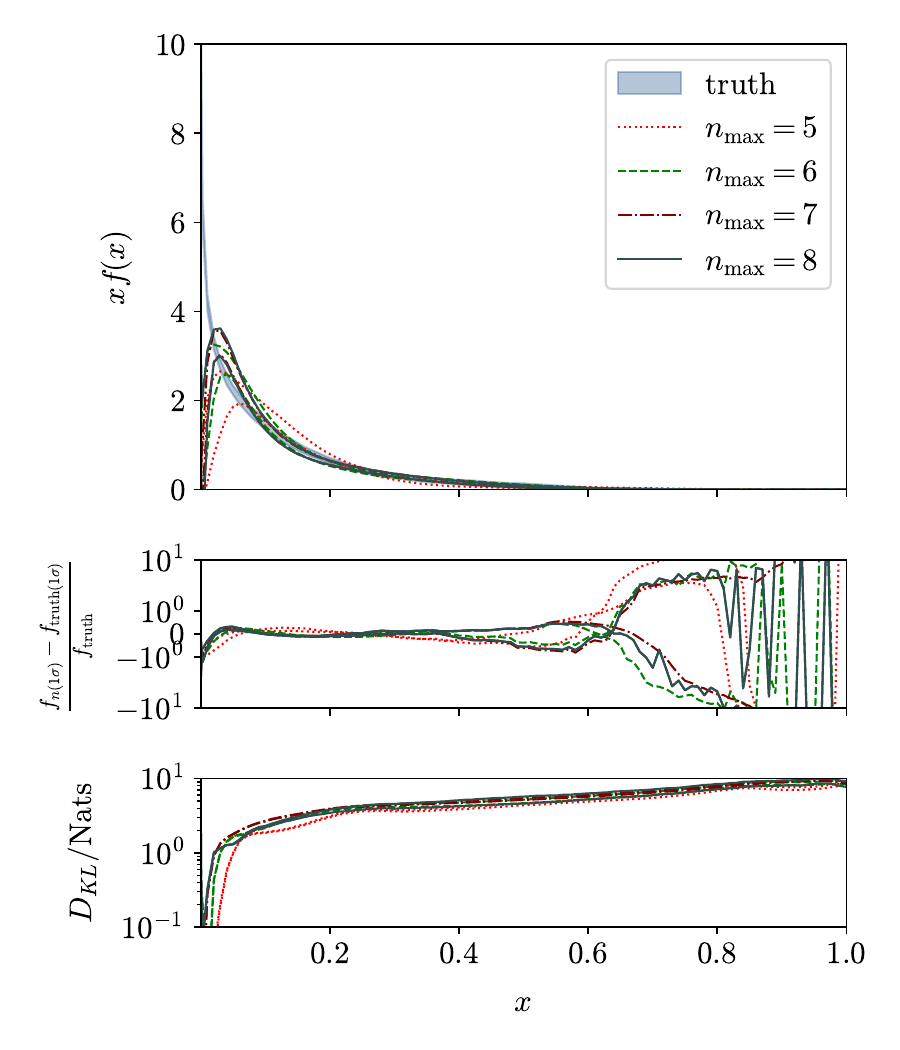}
    \end{subfigure}
    \begin{subfigure}[t]{0.40\textwidth}
        \centering
        \includegraphics[width=\textwidth]{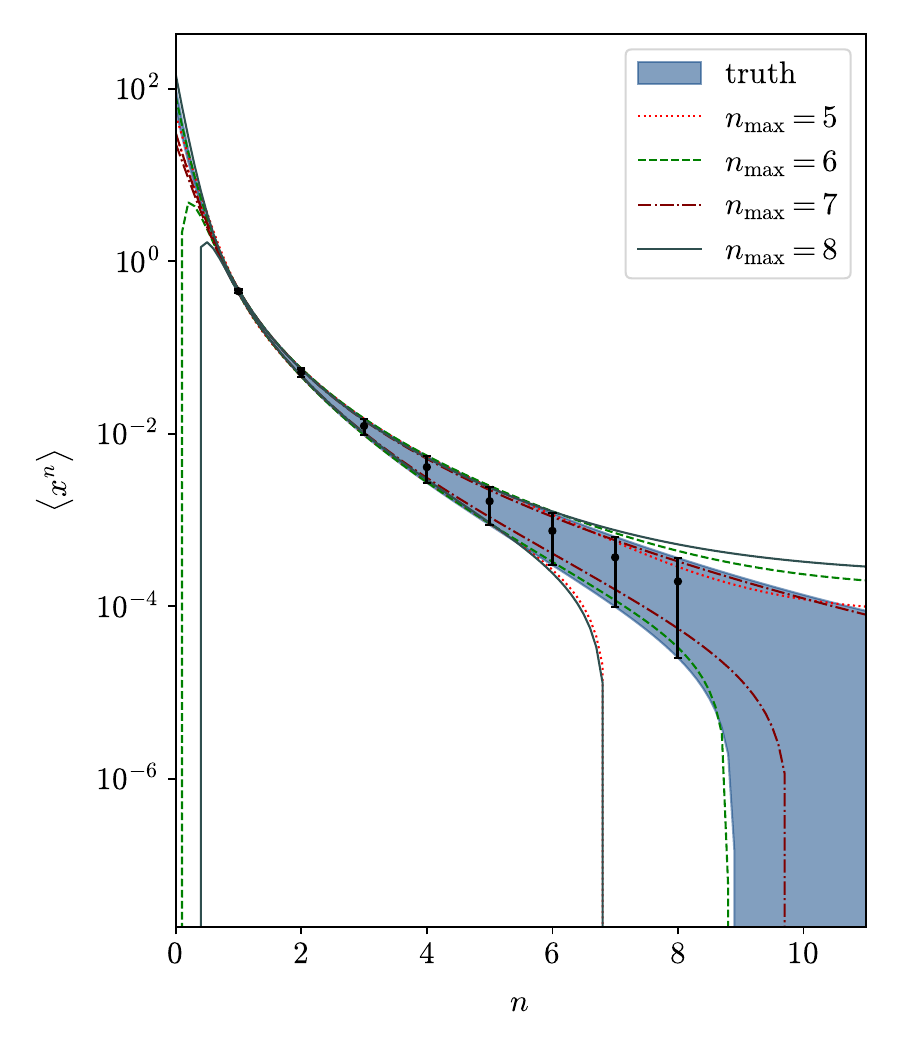}
    \end{subfigure}
    
    \caption{LSE kernel performance with $x^\alpha$ kernel prior on the \NNPDF gluon PDF. \biascorrnote}
    \label{fig:NNPDF_gluon_LSE_momScan}
\end{figure*}

\begin{figure*}[h!]
    \centering

    \begin{subfigure}[t]{0.40\textwidth}
        \centering
        \includegraphics[width=\textwidth]{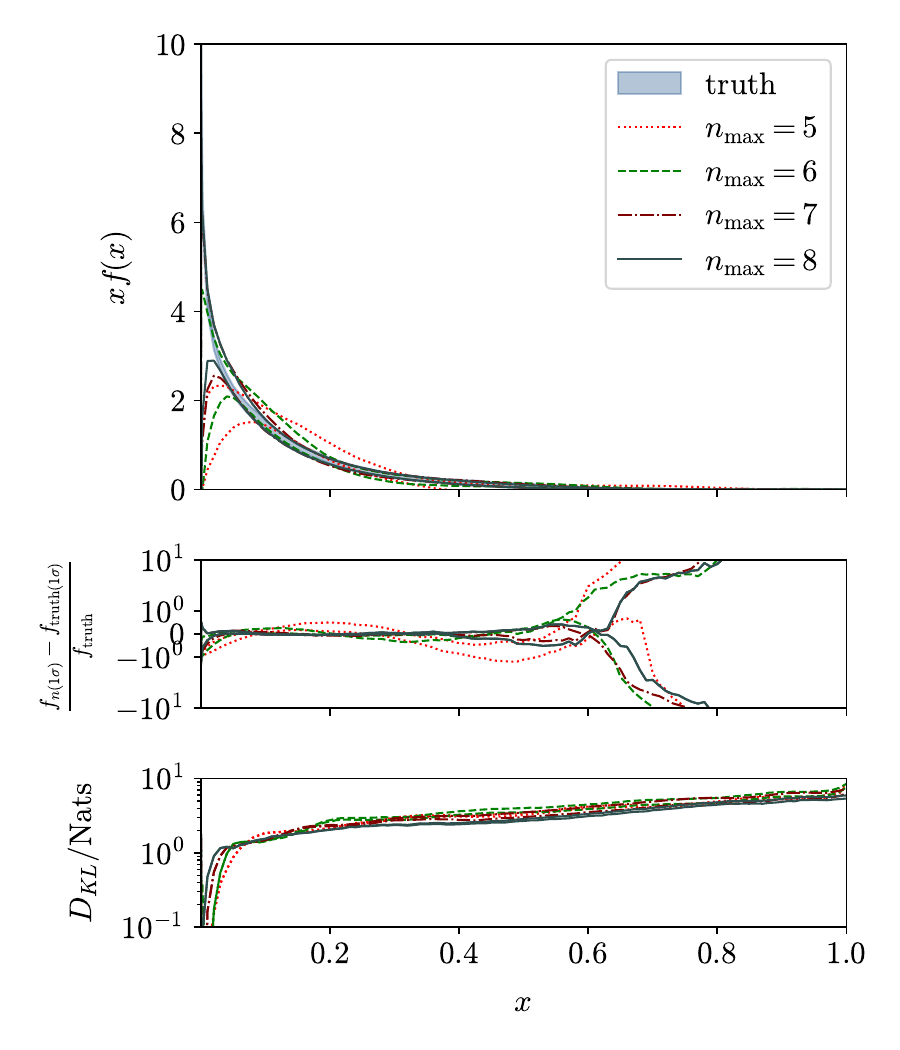}
    \end{subfigure}
    \begin{subfigure}[t]{0.40\textwidth}
        \centering
        \includegraphics[width=\textwidth]{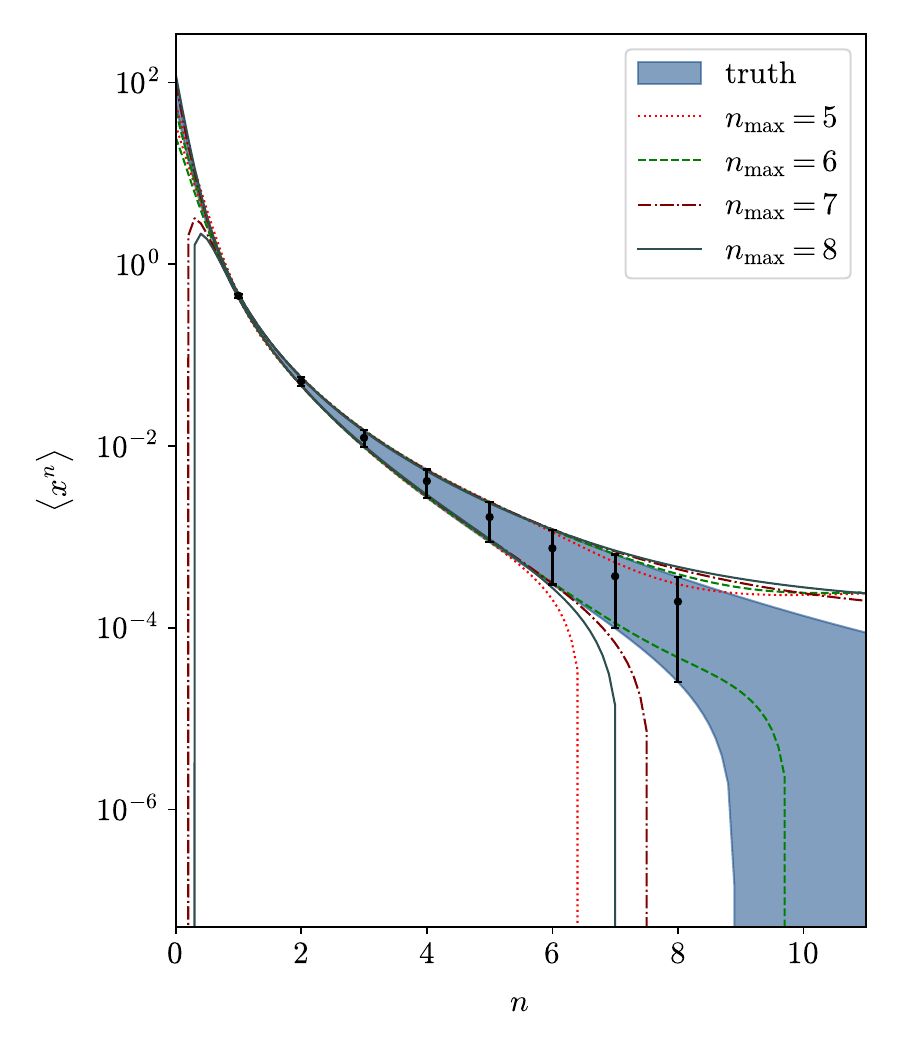}
    \end{subfigure}
    
    \caption{\MATERN{3/2} kernel performance with $x^\alpha$ kernel prior on the \NNPDF gluon PDF. \biascorrnote}
    \label{fig:NNPDF_gluon_MATERN32_momScan}
\end{figure*}

 \paragraph{Power divergence $x^\alpha$ prior} The first important prior is the power divergence enforcement $x^\alpha$ in the kernel. To see how important it is, we plot the effect, at $\nmax=5$, of using some OOTB kernels to fit moment data in Fig. \ref{fig:NNPDF_gluon_CovScan} with grid scheme 1 (the issue persists for grid scheme 2). We see that they all converge to heavily oscillatory solutions in Fig. \ref{fig:NNPDF_gluon_CovScan} that are caused by ``incorrect" extrapolations in moment space that do not obey the Hausdorff moment condition, as can be seen in Fig. \ref{fig:NNPDF_gluon_CovScan}.  Sticking with grid scheme 1 and now enforcing the $x^\alpha$ kernel prior effectively resolves this issue for the LSE and \MATERN{3/2} kernels, as can be seen in Figs. \ref{fig:NNPDF_gluon_LSE_momScan} and \ref{fig:NNPDF_gluon_MATERN32_momScan} respectively.

With the $x^\alpha$ kernel prior we see that for $\nmax>6$ for both the LSE and \MATERN{3/2} on grid scheme 1, we get good reconstruction in data space. In $x$ space, for the LSE kernel, there seems to be convergence to within $1\sigma$ of truth everywhere except for $x\ll 1$, where it consistently seems to ignore the power divergence. Unfortunately in our case, the degree of multi-modality means that the convergence to the correct power divergence behavior will never be achieved, thus showing that our LSE kernel is limited in performance. With the \MATERN{3/2} kernel on the other hand, our reconstruction converges in the $x\ll 1$ region to the mean truth, albeit with much higher uncertainty. Given moment data with no norm condition, this is desirable, as it highlights the ``indeterminable" nature of the $x\ll 1$ region. It is hard to tell whether the \MATERN{3/2}'s good performance is a persistent feature of the kernel suited for constrained gluon-like PDFs or just a coincidence. Regardless, let us devise Gibbs-style kernels that will give us a direct handle into matching gluon moment data to gluon-like $f(x)$ behavior. Given the degree of multi-modality we already face with these types of datasets, using the model agnostic Gibbs spline kernels would be hopeless unless we concertedly tune the prior ranges of hyper-parameters governing those kernels. Instead, we device the following Gibbs form, which we term a \textit{cosh dip} kernel:

\begin{equation}
    \ell(x) =\frac{\ell_0}{\frac{a}{\cosh(bx)}+1}
    \label{eq:cosh_dip}
\end{equation}

\begin{figure}
    \centering
    \begin{subfigure}[t]{0.40\textwidth}
        \centering
        \includegraphics[width=\textwidth]{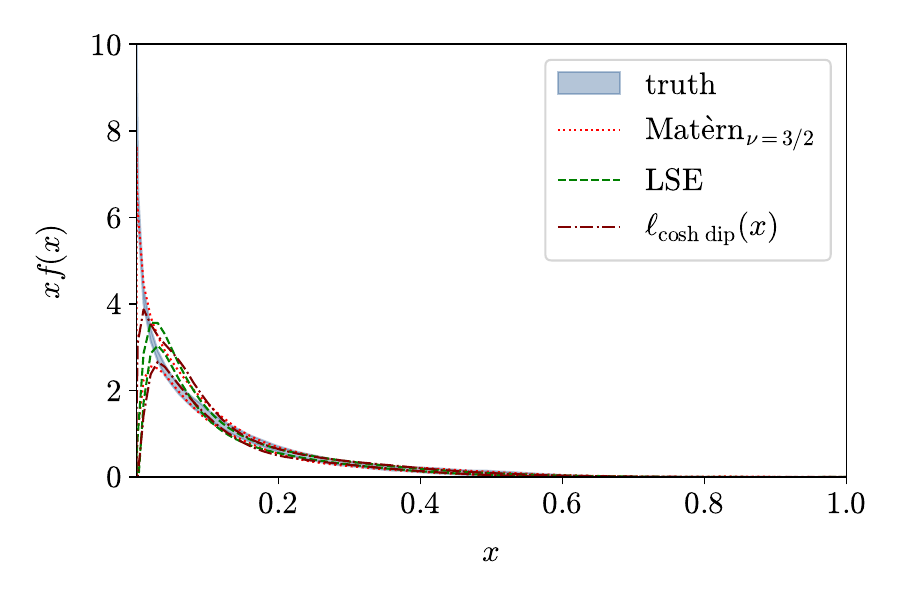}
    \end{subfigure}
    \begin{subfigure}[t]{0.40\textwidth}
        \centering
        \includegraphics[width=\textwidth]{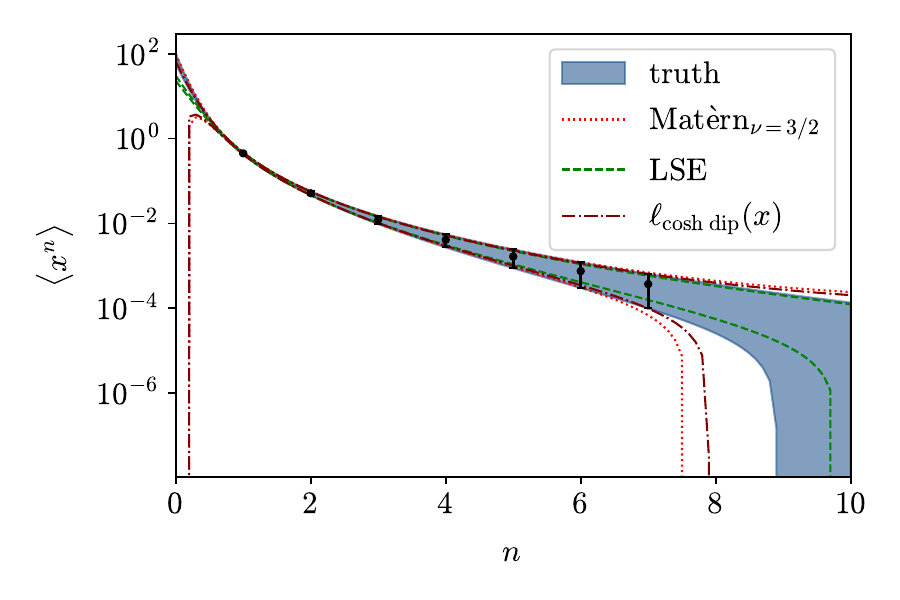}
    \end{subfigure}

    \caption{Comparison of different kernels on the \NNPDF gluon PDF under grid scheme 1 }
    \label{fig:GluonCompareGrid1}
\end{figure}

Here, Eq \ref{eq:cosh_dip} enforces a constant $x$ correlation of amount $\ell_0$ everywhere except at $x\ll 1$, where a ``dip" occurs, whose depth and width are controlled by $a$ and $b$ respectively. The dip in $x$ correlation signifies the fundamentally different behavior of $f(x)$ in the high/low $x$ regions. The rather un-constrained range for the hyper parameters we select are: $a,b\in[0.1,10]$. To get the maximum amount of utility out of this kernel, we will need to rely on the flexibility of (the logarithmic) grid scheme 2. To see this, observe Fig \ref{fig:GluonCompareGrid1}, where we see unremarkable performance from the cosh dip kernel. 

\subsubsection{Switching to a logarithmic grid}
\begin{figure*}[h!]
    \centering

    \begin{subfigure}[t]{0.20\textwidth}
        \centering
        \includegraphics[width=\textwidth]{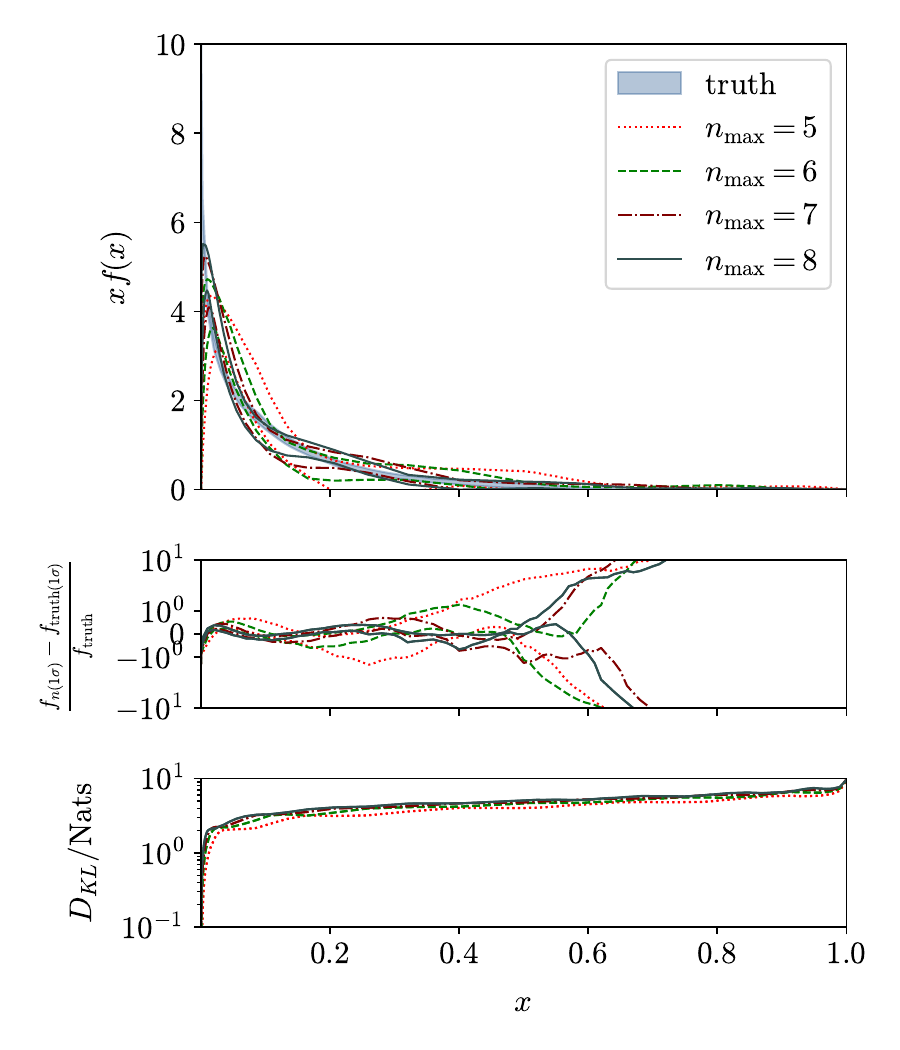}
    \end{subfigure}
    \begin{subfigure}[t]{0.20\textwidth}
        \centering
        \includegraphics[width=\textwidth]{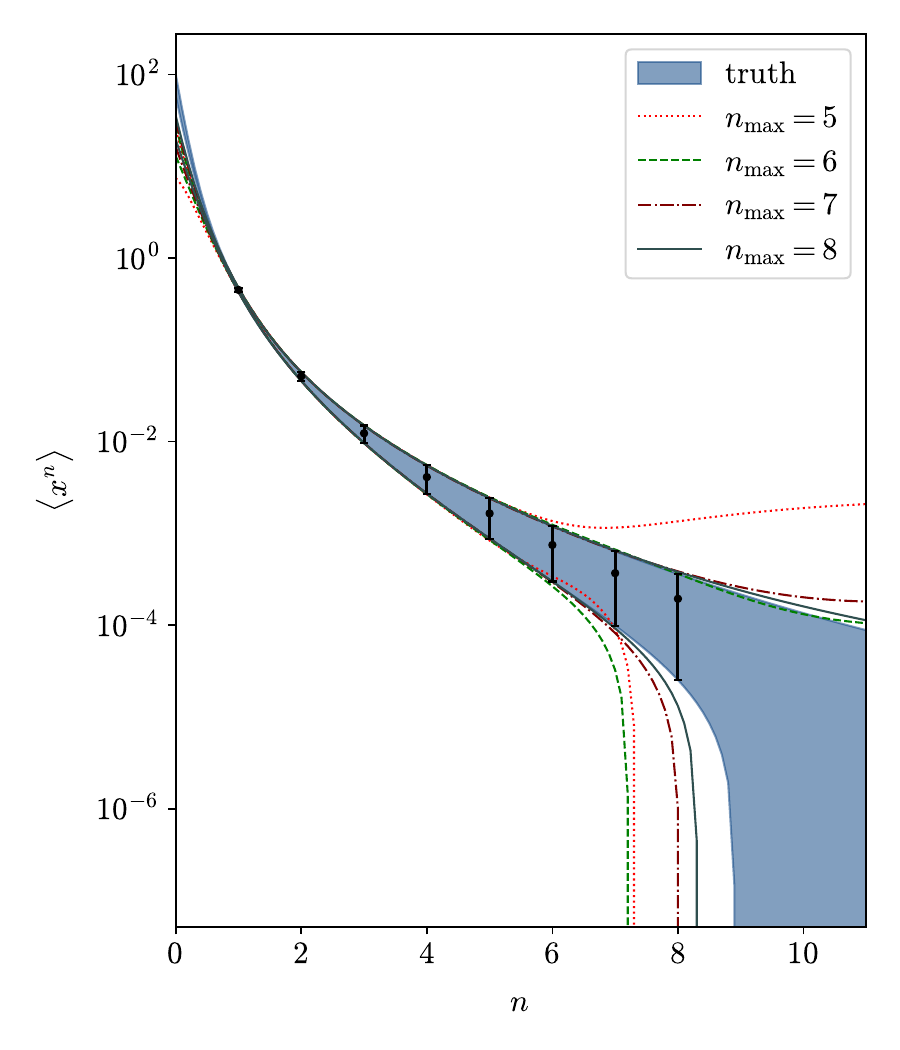}
    \end{subfigure}
    \begin{subfigure}[t]{0.20\textwidth}
        \centering
        \includegraphics[width=\textwidth]{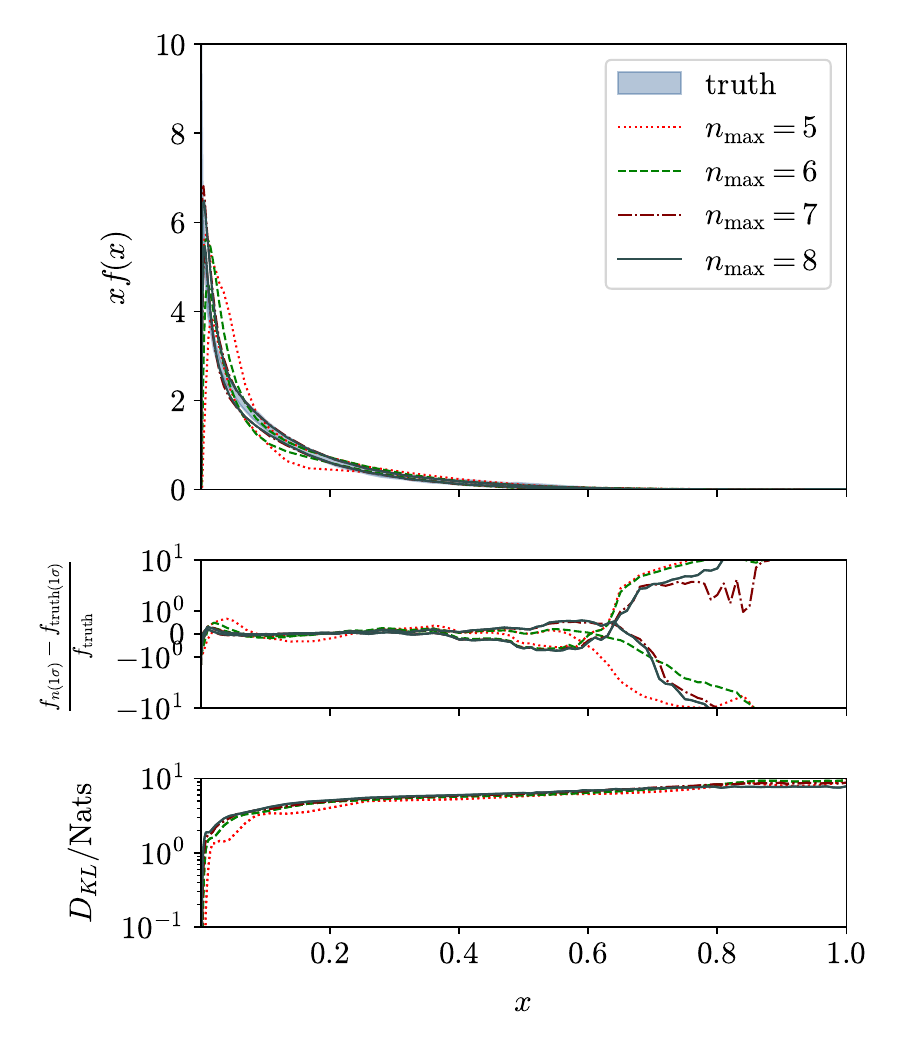}
    \end{subfigure}
    \begin{subfigure}[t]{0.20\textwidth}
        \centering
        \includegraphics[width=\textwidth]{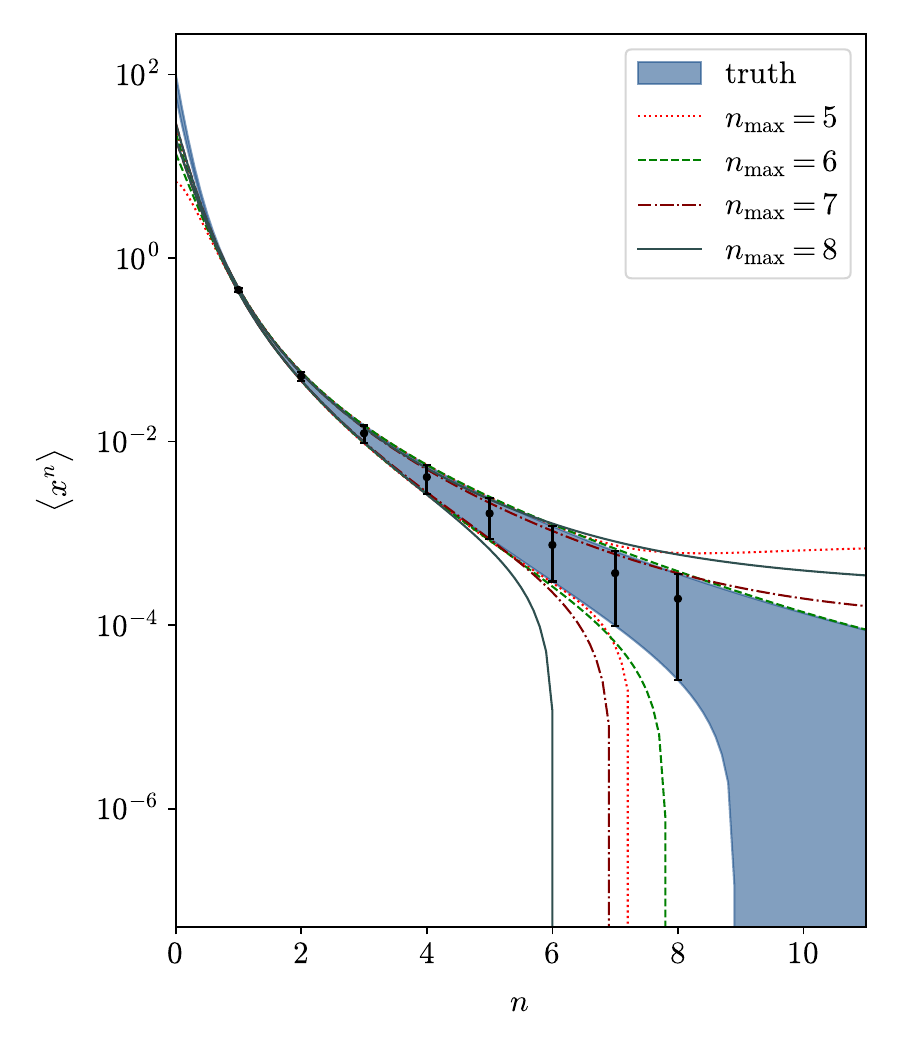}
    \end{subfigure}
    \caption{\MATERN{3/2} reconstruction of \NNPDF gluon PDF with $\sigma$ restriction. \biascorrnote}
    \label{fig:NNPDF_gluon_MATERN32_ohs_mmScan}
\end{figure*}

\begin{figure*}[h!]
    \centering

    \begin{subfigure}[t]{0.30\textwidth}
        \centering
        \includegraphics[width=\textwidth]{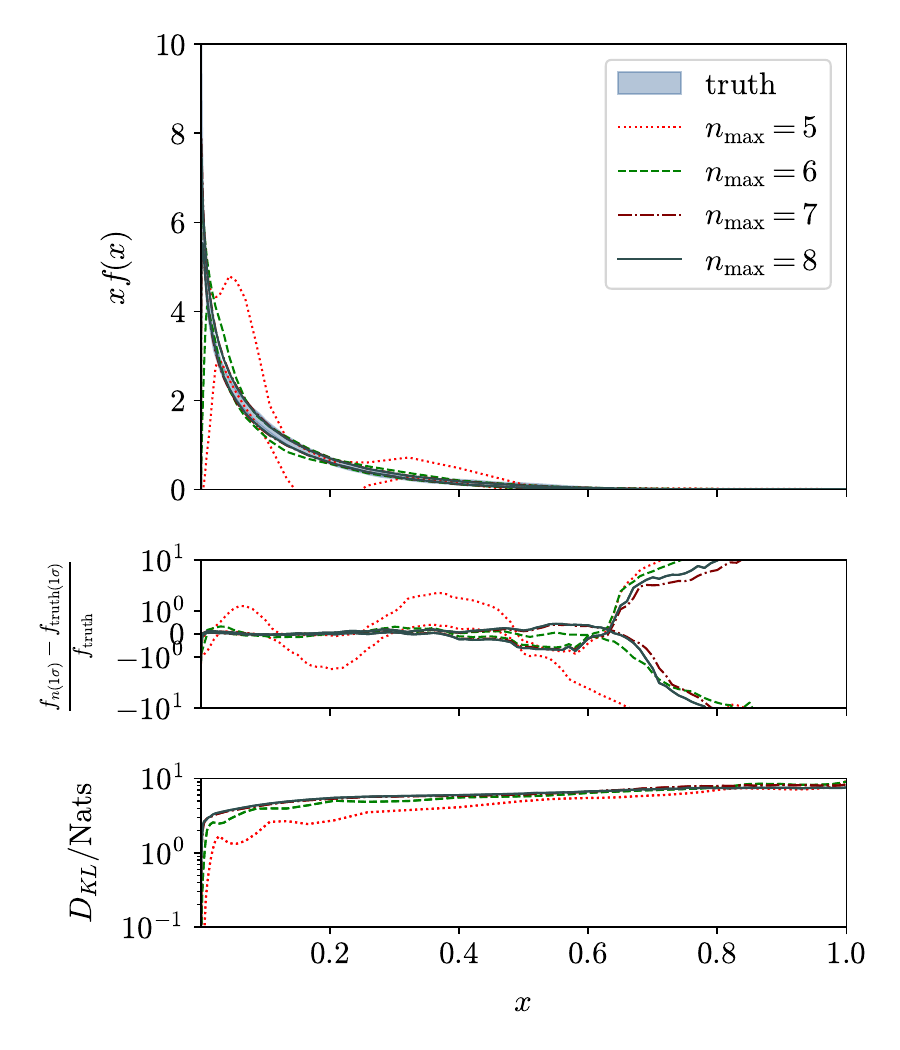}
    \end{subfigure}
    \begin{subfigure}[t]{0.30\textwidth}
        \centering
        \includegraphics[width=\textwidth]{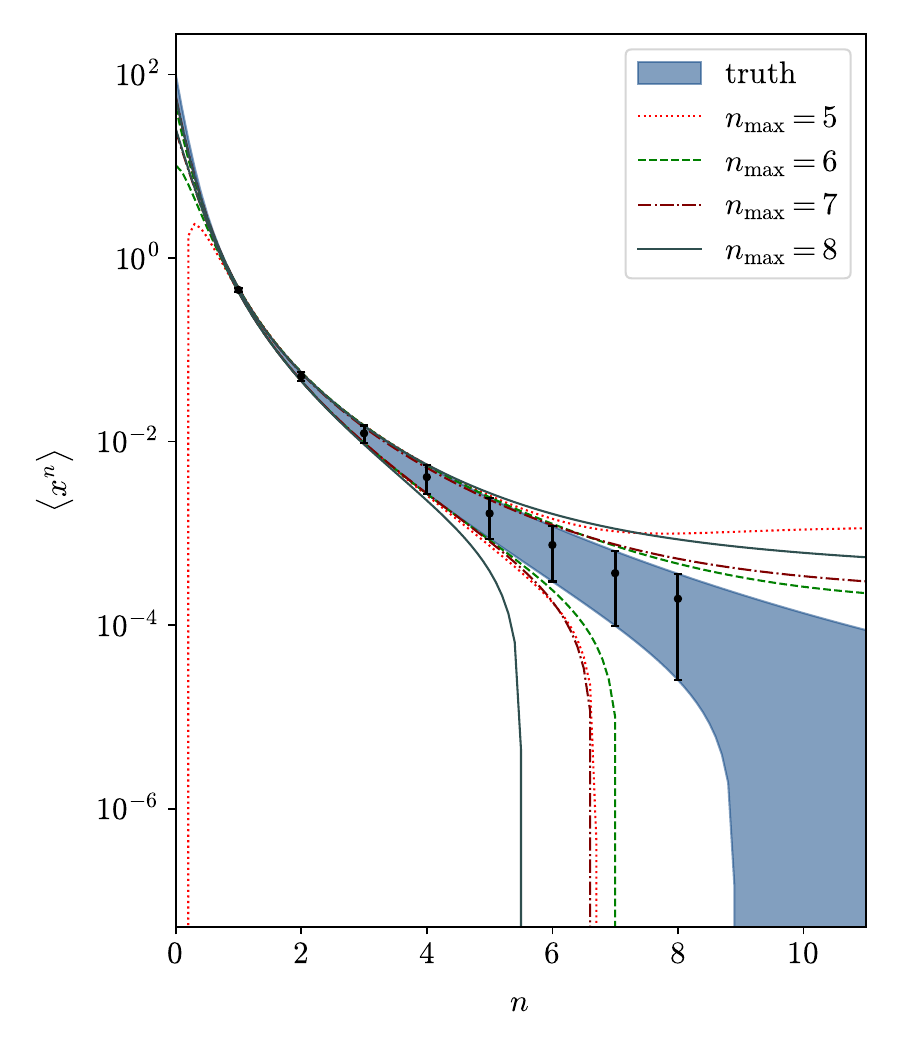}
    \end{subfigure}
    \begin{subfigure}[t]{0.20\textwidth}
        \centering
        \includegraphics[width=\textwidth]{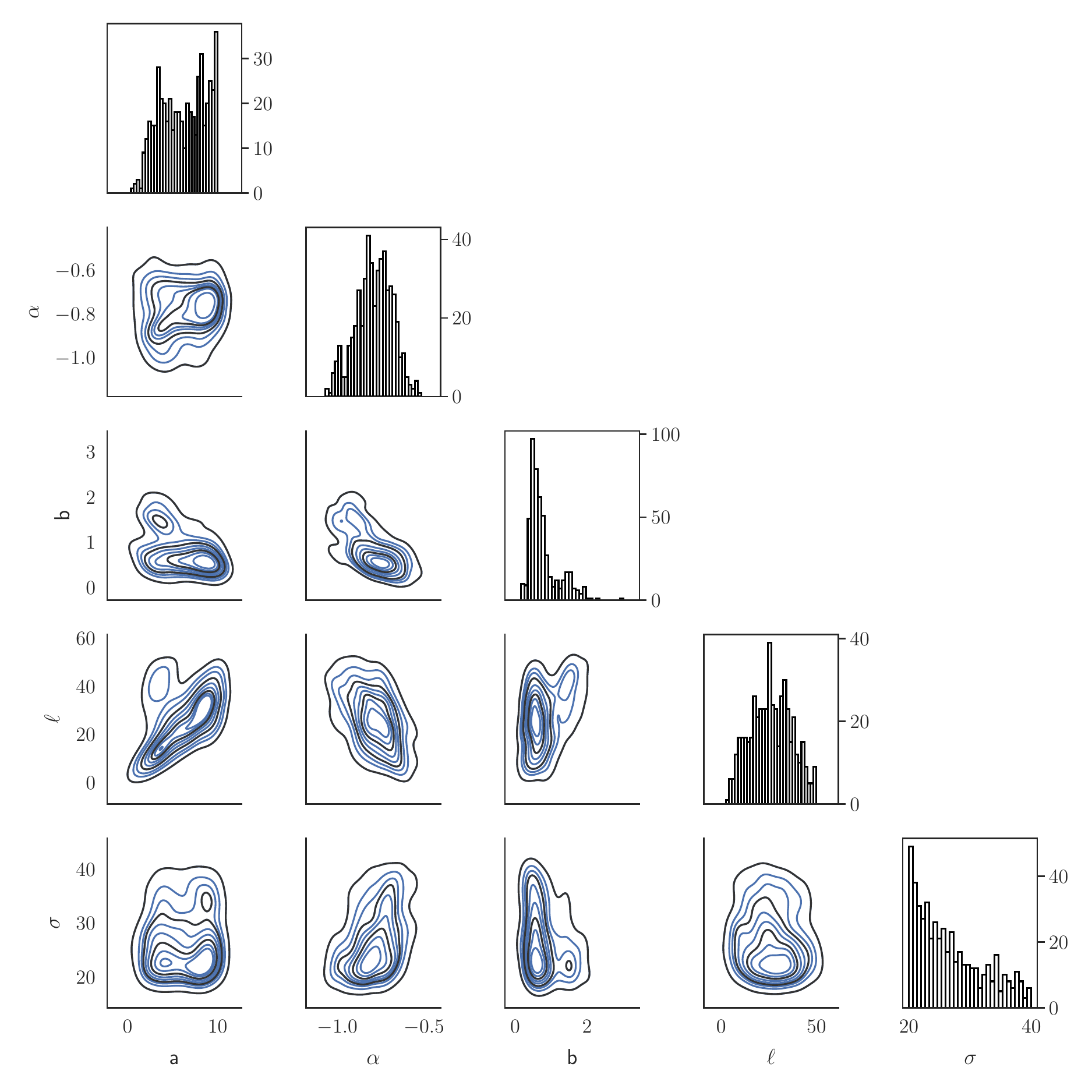}
    \end{subfigure}
    \caption{cosh dip reconstruction of \NNPDF gluon PDF with $\sigma$ restriction. \biascorrnote}
    \label{fig:NNPDF_gluon_GIBBS_fancy_ohs_mmScan}
\end{figure*}

To observe reconstructed gains with the cosh dip kernel, we need to first switch to grid scheme 2 to provide more flexibility for movement in the low $x$ region. With grid scheme 2, we see a dramatic difference in reconstruction for all three kernels, as shown in Figures \ref{fig:NNPDF_gluon_MATERN32_log},\ref{fig:NNPDF_gluon_GIBBS_fancy_log},\ref{fig:NNPDF_gluon_LSE_log} in the appendix. They all reconstruct the $x\ll 1$ region within inflated uncertainty and agree with the truth in $x$ space. What seems to be a reasonably successful reconstruction in $x$ space is mirrored by an extremely unconstrained reconstruction in data space, with only the first moment reconstructed with an error of order $\sim 100\%$; the first moment plays by far the largest role in reconstructing constrained gluonic PDFs. For our purposes, we can try to achieve a better reconstruction by noticing how the $\sigma$ parameter is sampled for the above kernels. We see that despite providing a wide range that $\sigma$ can be sampled from, our sampler converges to very low values of $\sigma$. This prompts us to limit the range by which $\sigma$ is sampled to exclude low values of $\sigma$, which we will henceforth call \textit{$\sigma$ restriction}. Let us restrict our $\sigma$ to now be within $\sigma\in[20,40]$. Results are shown in Figs. \ref{fig:NNPDF_gluon_MATERN32_ohs_mmScan},\ref{fig:NNPDF_gluon_GIBBS_fancy_ohs_mmScan}; $\sigma$ restriction is quite effective with all kernels, and thus we see that in conjunction with a logarithmic grid spacing, the power divergence kernel prior, our framework shows promise in reconstructing constrained gluon PDFs. The cosh dip kernel yields great information gain and looking at the triangle plot associated with it, we see the sampled distribution of $b$ has a clear peak, potentially indicating this kernel's effectiveness in reconstruction on other gluonic datasets.

\subsection{Strategies to fit the ``moderately" unconstrained gluon/sea quark PDFs}
The multi-modality in $x$ space encountered in the constrained gluon moments seen in the previous subsection will only increase with the less constrained moments that we see in the \NNPDF $s$ quark PDF. The uncertainty levels in this dataset are termed ``moderately" unconstrained as they lie somewhere between the \NNPDF gluon PDF and \FANTO gluon dataset.

\begin{figure}
    \centering
    \begin{subfigure}[t]{0.40\textwidth}
        \centering
        \includegraphics[width=\textwidth]{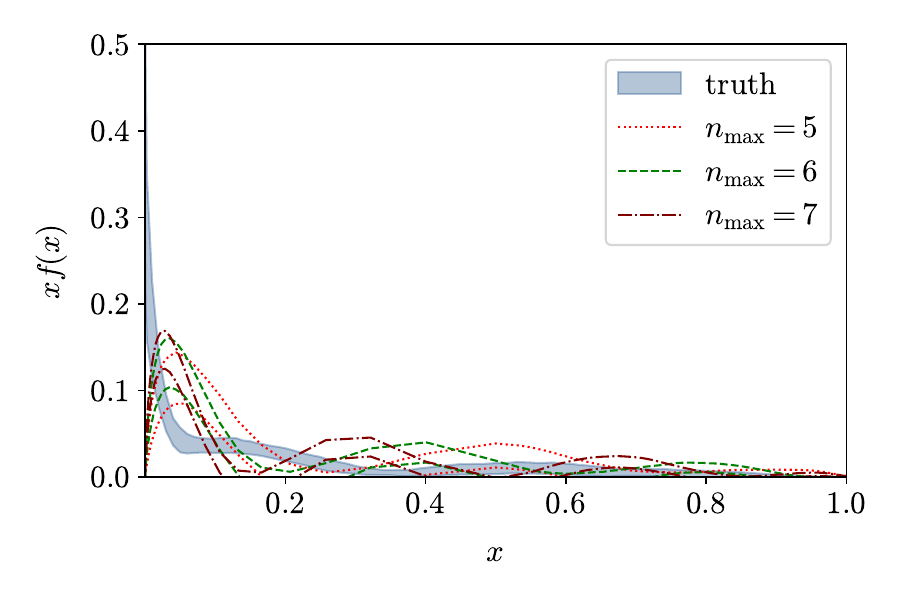}
    \end{subfigure}
    \begin{subfigure}[t]{0.40\textwidth}
        \centering
        \includegraphics[width=\textwidth]{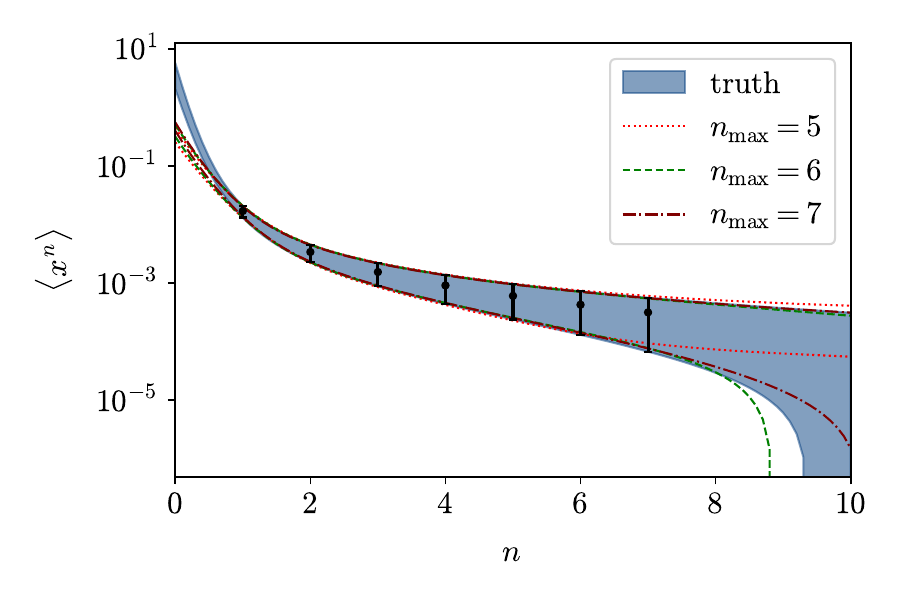}
    \end{subfigure}
    \caption{\MATERN{3/2} reconstruction of \NNPDF $s$ quark PDF}
    \label{fig:NNPDF_sea_MATERN}
\end{figure}

\begin{figure}
    \centering
    \begin{subfigure}[t]{0.40\textwidth}
        \centering
        \includegraphics[width=\textwidth]{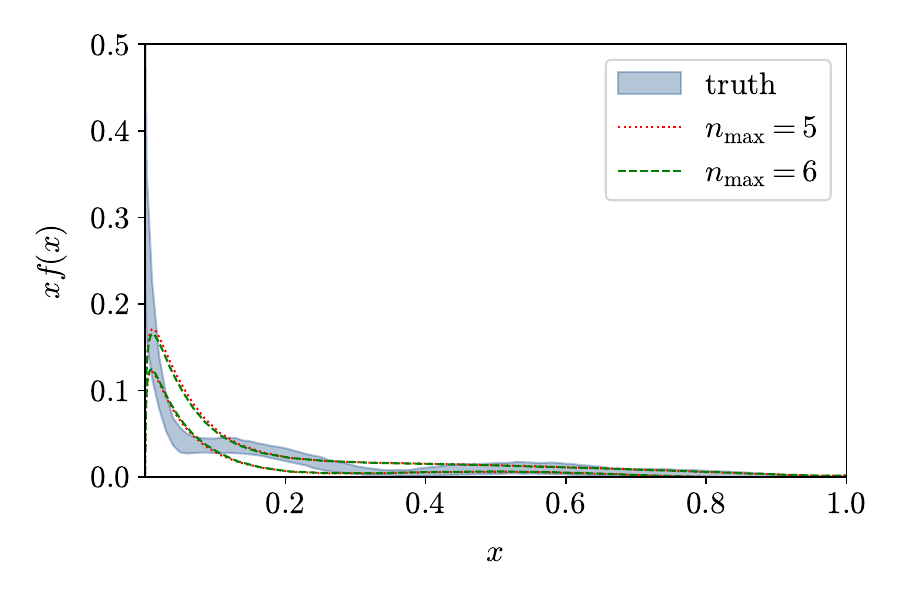}
    \end{subfigure}
    \begin{subfigure}[t]{0.40\textwidth}
        \centering
        \includegraphics[width=\textwidth]{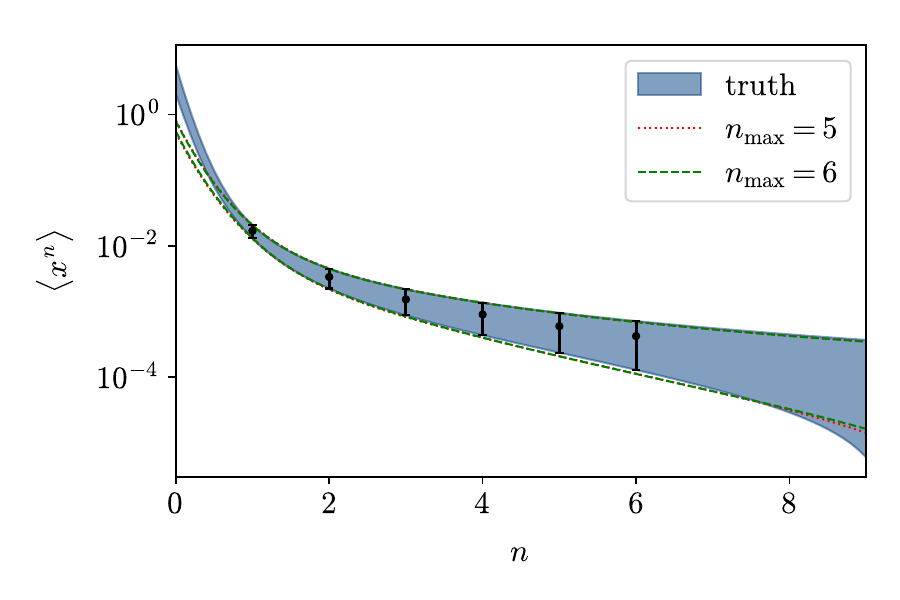}
    \end{subfigure}
    \caption{LSE reconstruction of \NNPDF $s$ quark PDF}
    \label{fig:NNPDF_sea_LSE}
\end{figure}

\begin{figure}
    \centering
    \begin{subfigure}[t]{0.30\textwidth}
        \centering
        \includegraphics[width=\textwidth]{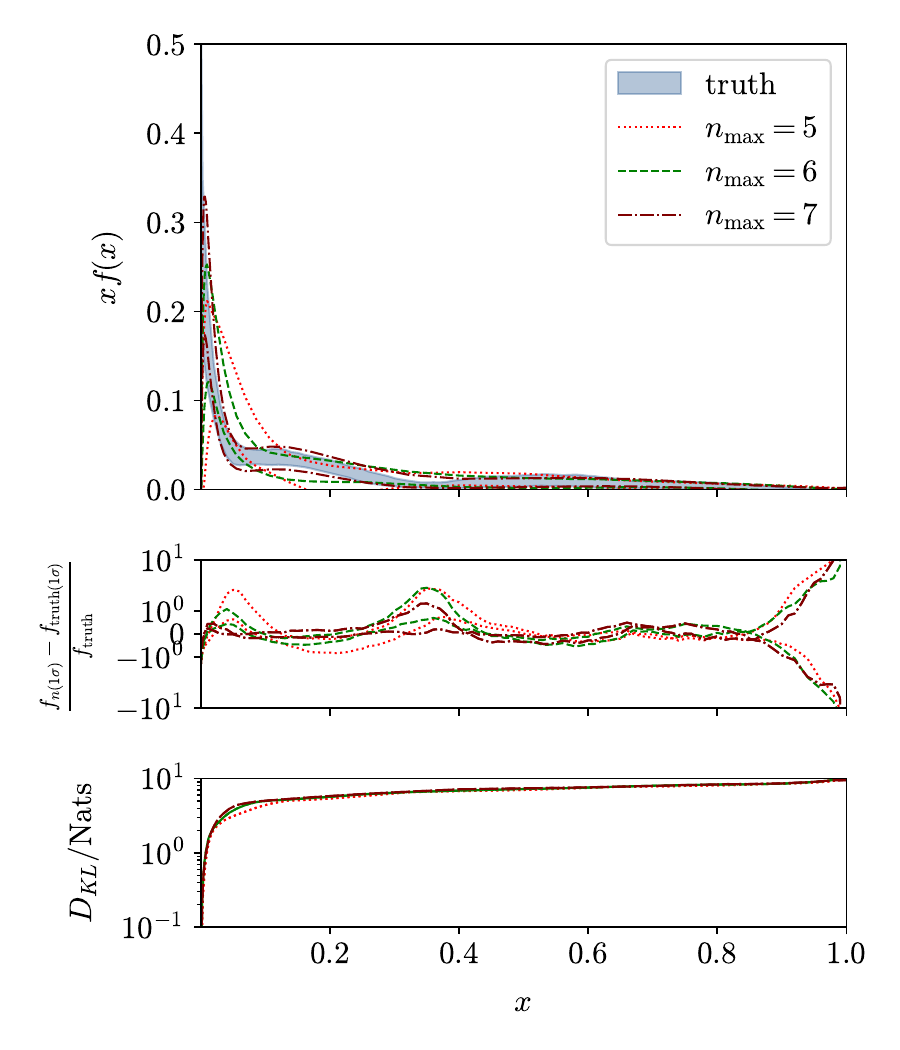}
    \end{subfigure}
    \begin{subfigure}[t]{0.30\textwidth}
        \centering
        \includegraphics[width=\textwidth]{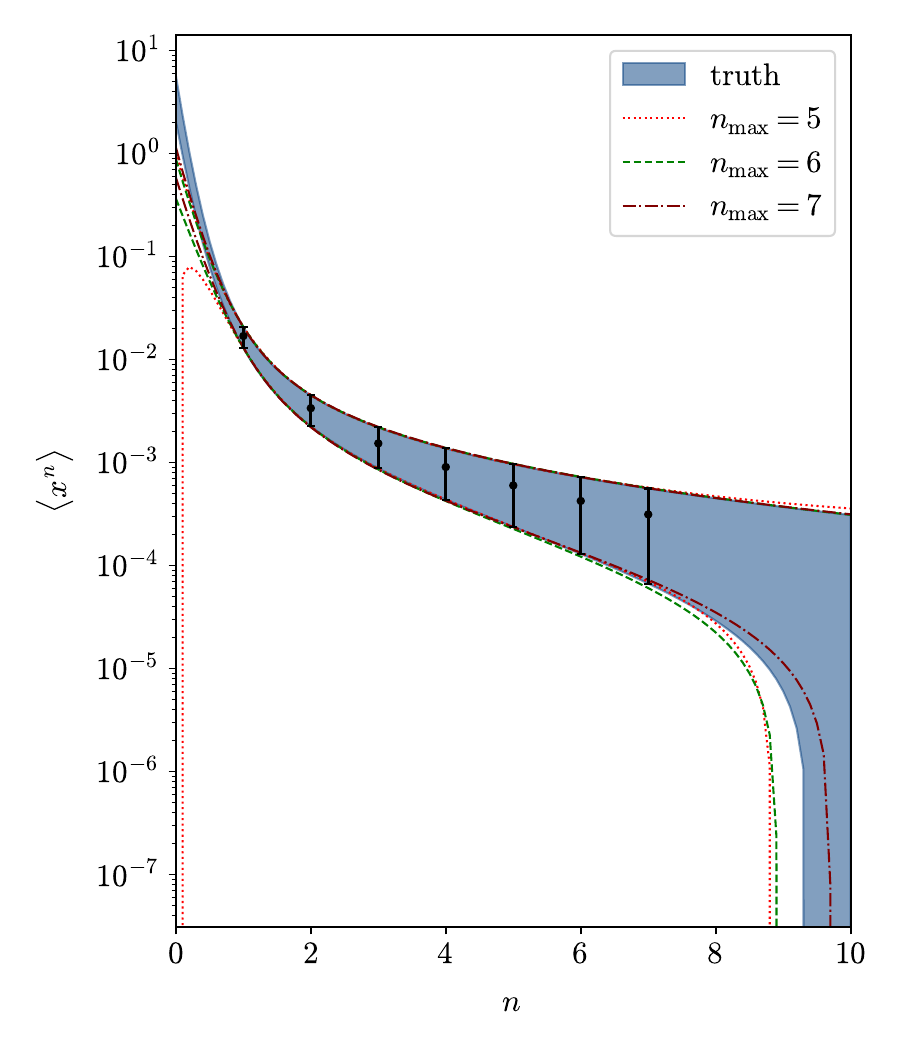}
    \end{subfigure}
    \begin{subfigure}[t]{0.30\textwidth}
        \centering
        \includegraphics[width=\textwidth]{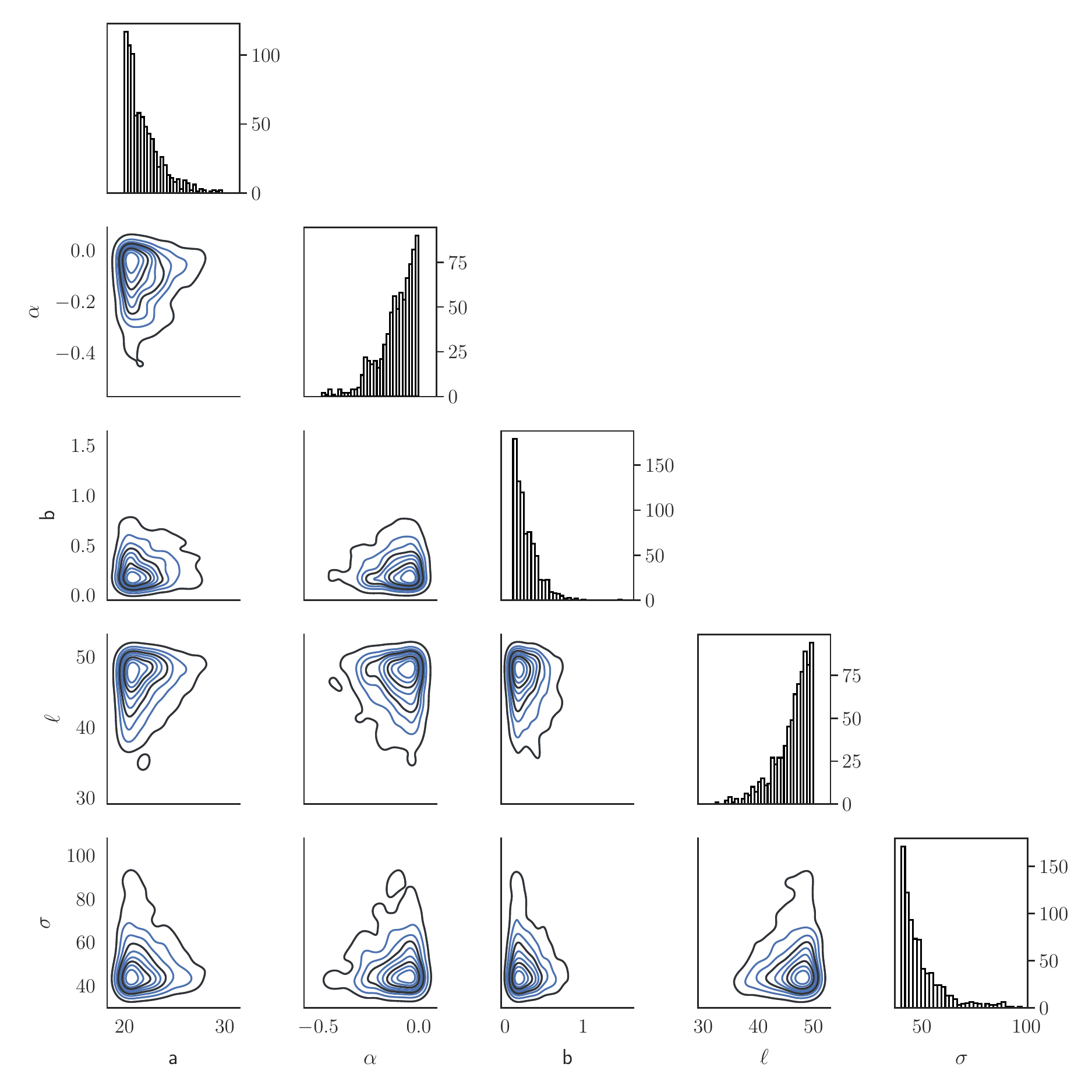}
    \end{subfigure}
    \caption{cosh dip kernel reconstruction of \NNPDF $s$ quark PDF}
    \label{fig:NNPDF_sea_GIBBS_fancy}
\end{figure}

For the \NNPDF $s$ quark PDF, if we try to retain the settings (grid scheme 2, $x^\alpha$ kernel prior, $\sigma$ truncation)\footnote{here, our $\sigma$ is now sampled in $\sigma\in[40,100]$}, we notice that we achieve good reconstruction in moment space but an $x$ space reconstruction that does not quite match (and shows no indication of converging to) truth at low $x$. Results can be seen in Figs. \ref{fig:NNPDF_sea_MATERN} and \ref{fig:NNPDF_sea_LSE}. Fortunately, the Gibbs-style cosh dip kernel can successfully address issues of multi-modality, whereas for the previous gluon dataset its performance matched that of OOTB kernels. We place an extra restriction of $a\in[20,30]$ to enforce a minimum depth in $\ell(x)$ for low $x$. Results are shown in \ref{fig:NNPDF_sea_GIBBS_fancy} and demonstrate the cosh dip kernel's promise in reconstructing other similarly unconstrained gluon/sea PDFs.

\subsection{Strategies to fit ``severely" unconstrained gluon/sea quark PDFs}

\begin{figure*}[h!]
    \centering
    \begin{subfigure}[t]{0.40\textwidth}
        \centering
        \includegraphics[width=\textwidth]{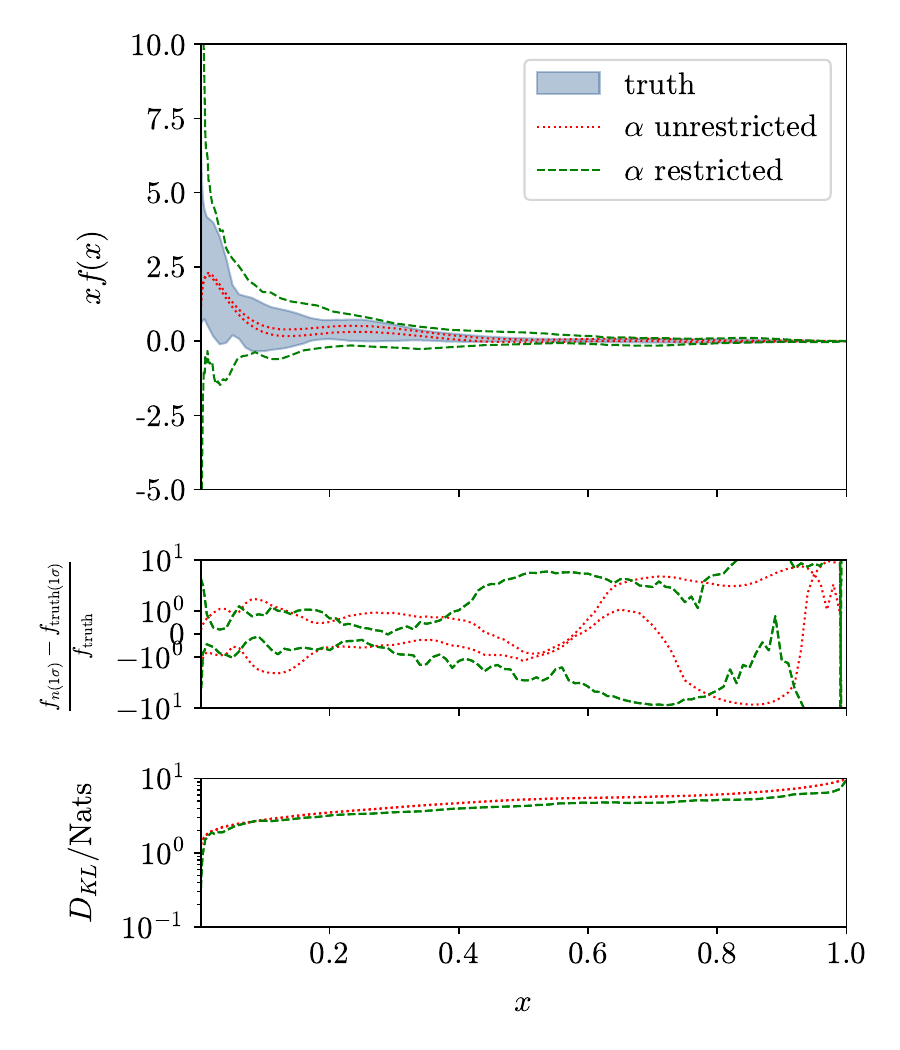}
    \end{subfigure}
    \begin{subfigure}[t]{0.40\textwidth}
        \centering
        \includegraphics[width=\textwidth]{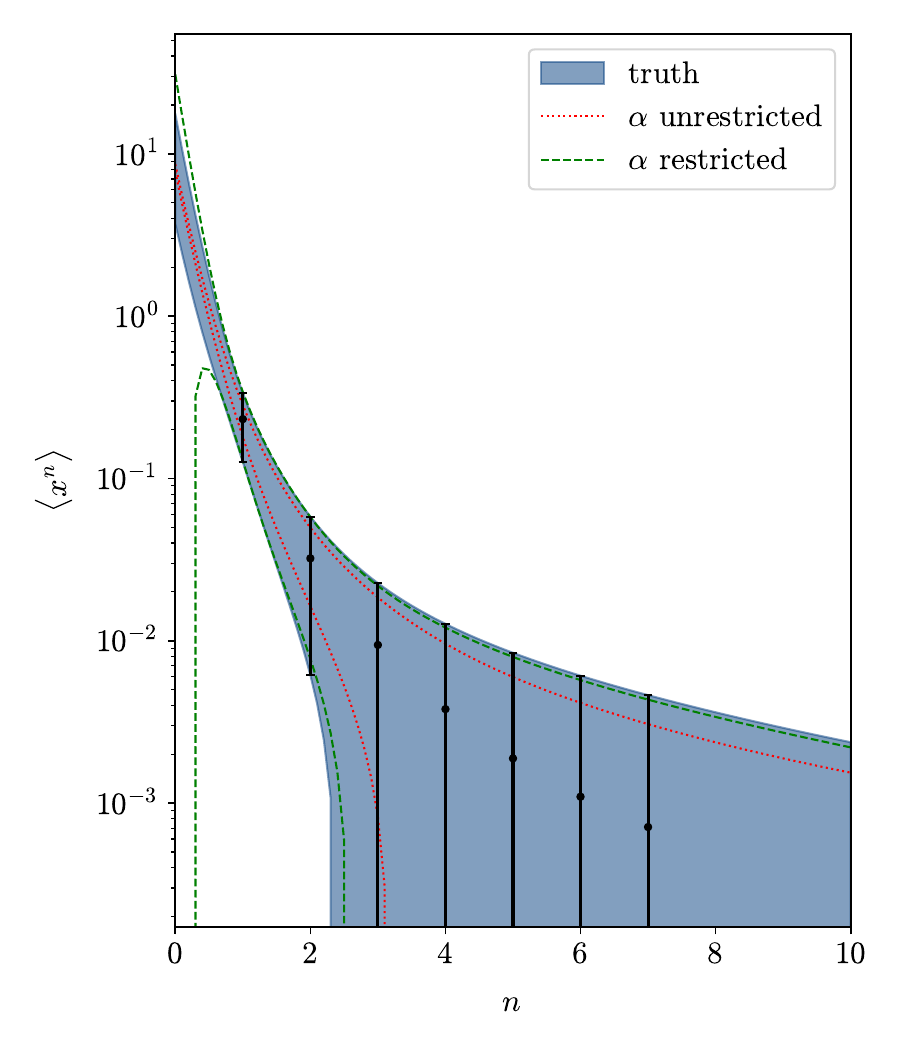}
    \end{subfigure}
    \begin{subfigure}[t]{0.40\textwidth}
        \centering
        \includegraphics[width=\textwidth]{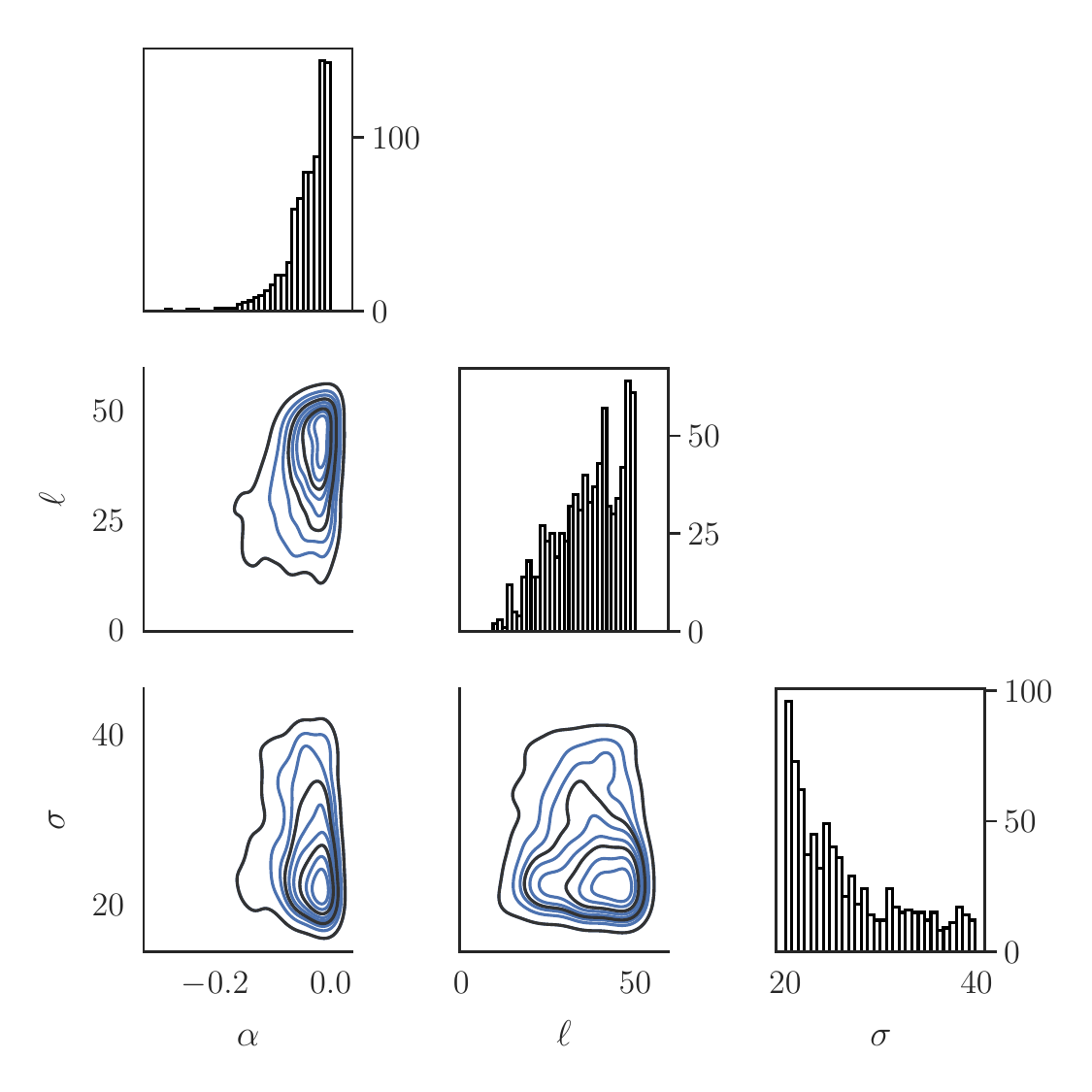}
        \caption{$\alpha$ unrestricted}
    \end{subfigure}
    \begin{subfigure}[t]{0.40\textwidth}
        \centering
        \includegraphics[width=\textwidth]{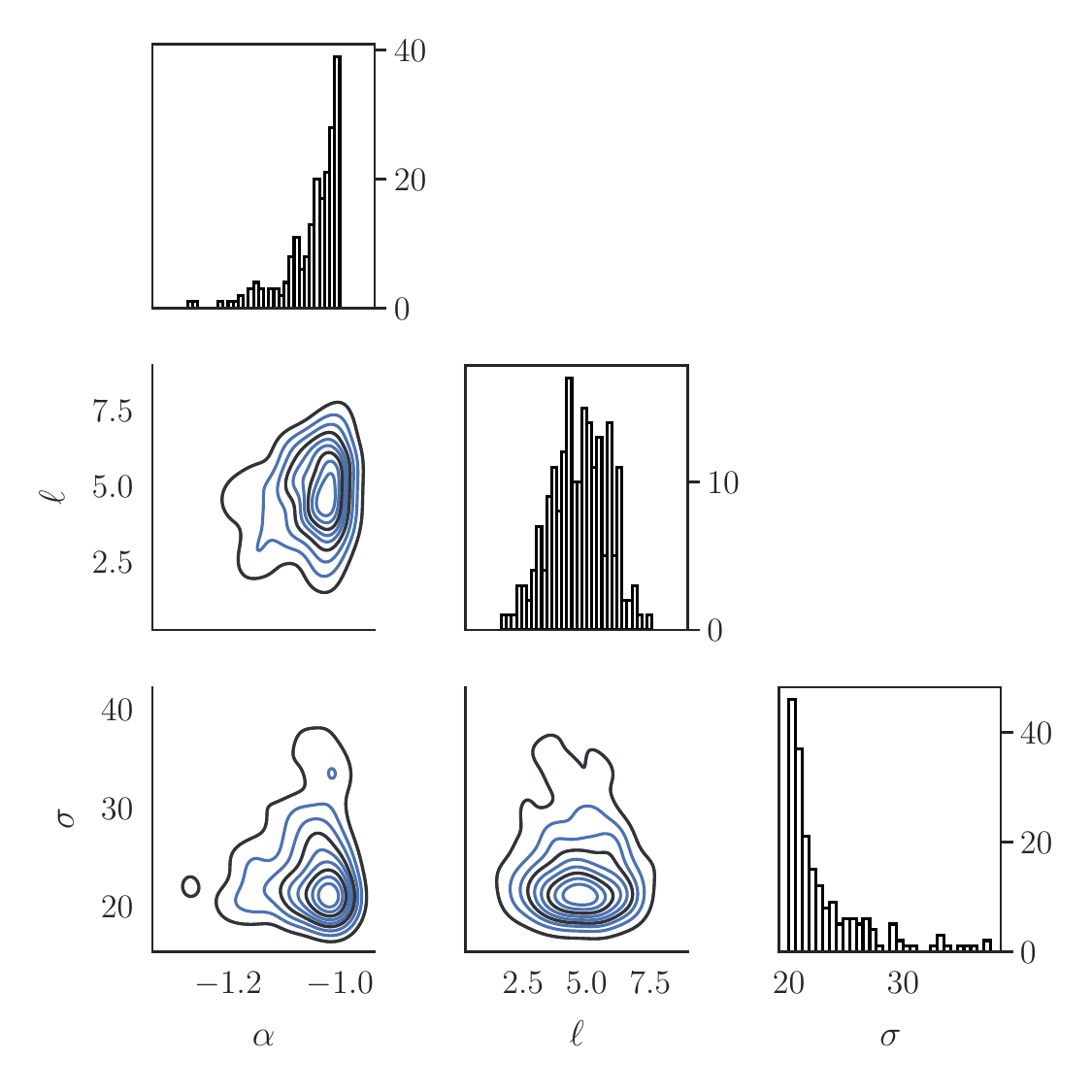}
        \caption{$\alpha$ restricted}
    \end{subfigure}
    
    \caption{\FANTO gluon PDF $\alpha$ truncation study with \MATERN{3/2} kernel at $\nmax=7$. \biascorrnote}
    \label{fig:FANTO_gluon_ohaScan_MATERN32}
\end{figure*}

``Severely" unconstrained gluon/sea quark PDFs are here represented by the \FANTO gluon dataset, and would be the hardest class of datasets to faithfully reconstruct. For the purposes of brevity, let us retain all the prior information we added to tackle the (un)constrained gluon datasets in the previous section (for both the \MATERN{3/2} and cosh dip kernels). We notice a wide sort of degrees of divergences in our PDF, uncertainties effectively spanning all real numbers (on the $y$ axis) for $x=0$. As we will see with this dataset, if the sampled value of $\alpha$ peaks at a high number (i.e. 0), we will not be able to obtain posteriors with wide uncertainties. For this reason, just as we restricted the sampling of $\sigma$ earlier, we will restrict $\alpha$ to a range that excludes $\alpha\sim 0$. This range ideally should be determined through tuning, but for our case, we will pick $\alpha\in[-2,-1]$. Results demonstrating the effectiveness of $\alpha$ restriction are shown in Fig \ref{fig:FANTO_gluon_ohaScan_MATERN32}.\footnote{Note that the $y$ axis has been truncated and will continue to be done so for this dataset for easier viewing.}

\begin{figure*}[h!]
    \centering
    \begin{subfigure}[t]{0.30\textwidth}
        \centering
        \includegraphics[width=\textwidth]{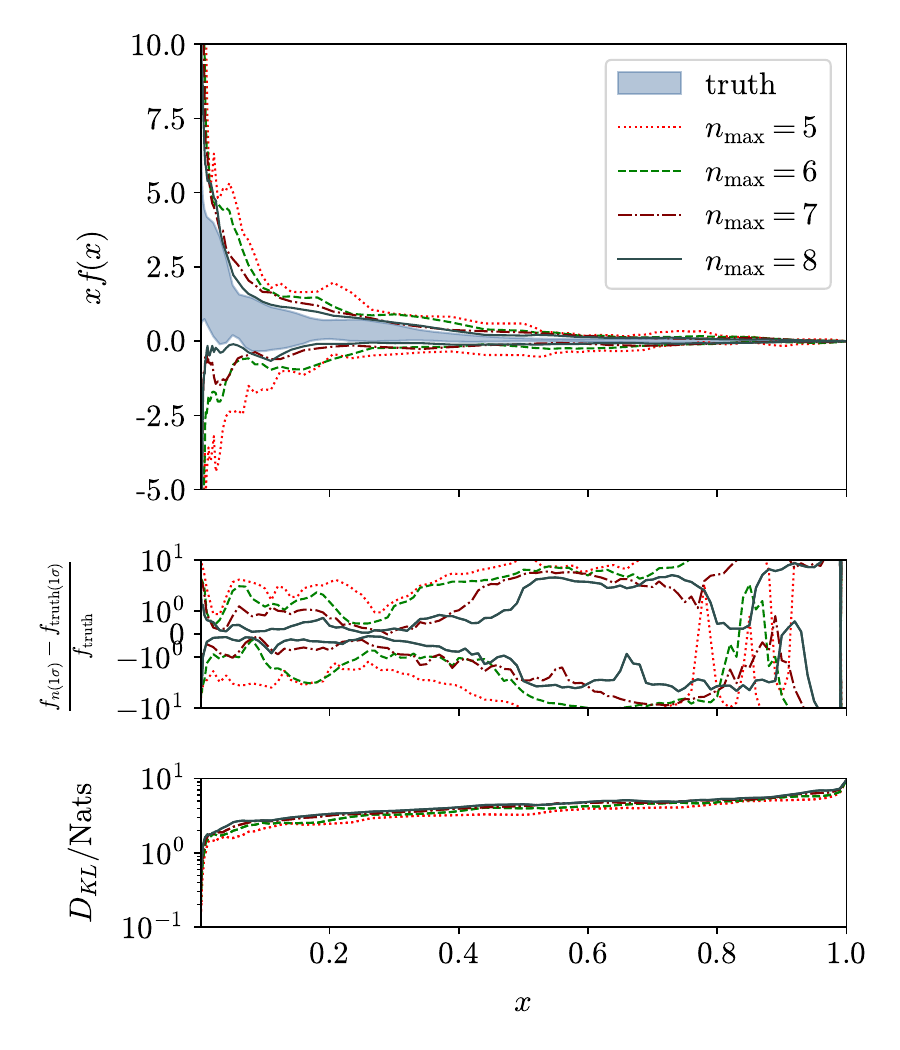}
    \end{subfigure}
    \begin{subfigure}[t]{0.30\textwidth}
        \centering
        \includegraphics[width=\textwidth]{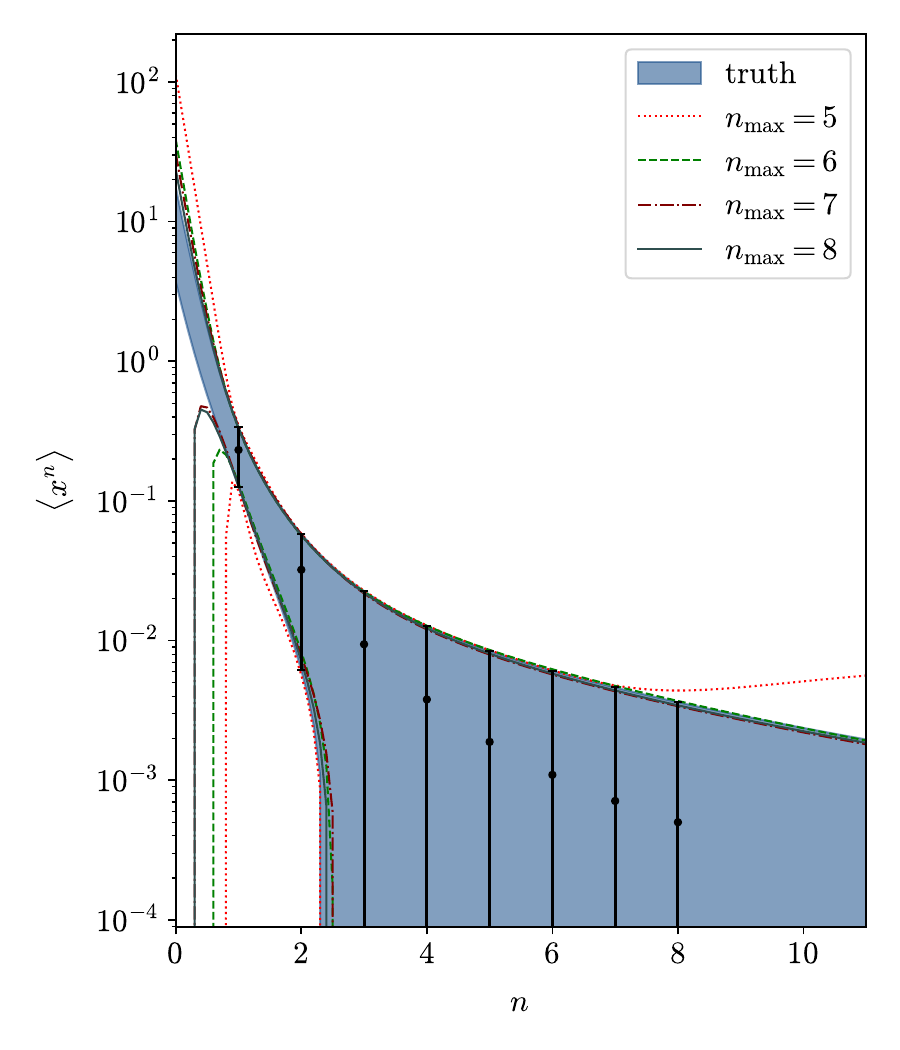}
    \end{subfigure}
    \caption{\FANTO gluon PDF study with \MATERN{3/2} kernel. \biascorrnote}
    \label{fig:FANTO_gluon_MATERN32}
\end{figure*}

\begin{figure*}[h!]
    \centering
    \begin{subfigure}[t]{0.30\textwidth}
        \centering
        \includegraphics[width=\textwidth]{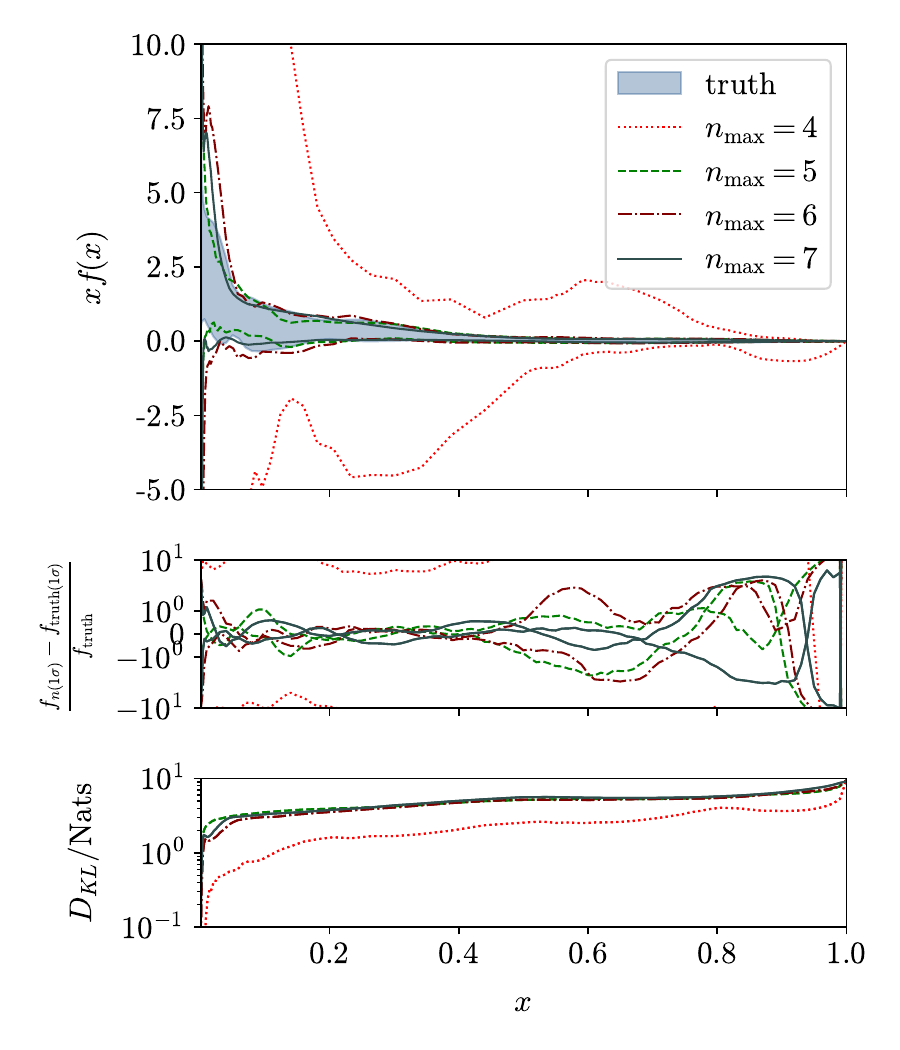}
    \end{subfigure}
    \begin{subfigure}[t]{0.30\textwidth}
        \centering
        \includegraphics[width=\textwidth]{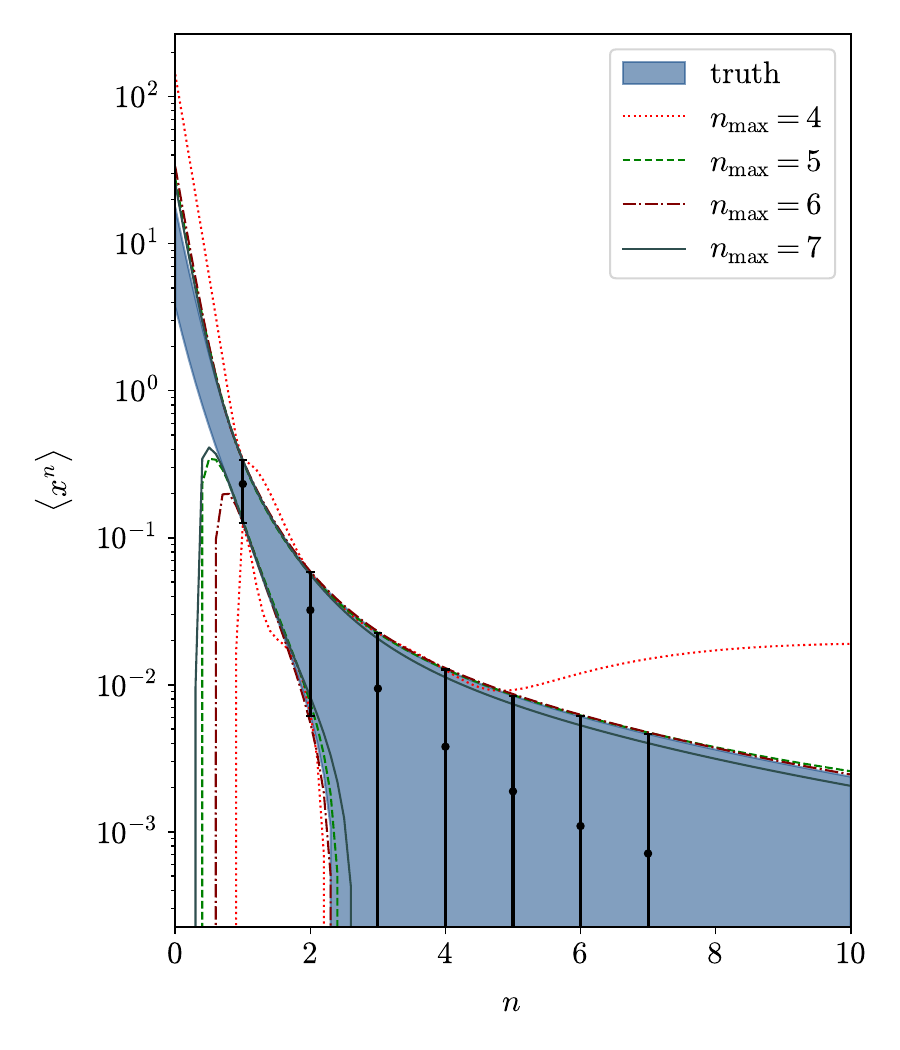}
    \end{subfigure}
    \caption{\FANTO gluon PDF study with cosh dip kernel. \biascorrnote}
    \label{fig:FANTO_gluon_GIBBS_fancy}
\end{figure*}

With this extra restriction in place, we can show reconstruction results for both the \MATERN{3/2} (\ref{fig:FANTO_gluon_MATERN32}) and the cosh dip (\ref{fig:FANTO_gluon_GIBBS_fancy}) kernels. We see fair reconstructions from both kernels, with the cosh dip kernel converging to truth in $x$ space much faster than the \MATERN{3/2} kernel. It is also worth noticing how with the cosh dip kernel the reconstruction for higher moments does not seem to matter much, indicating that in a lattice calculation perhaps only the first few moments need to be calculated with high precision to be fed into a GP framework, with the inherently ill posed reconstruction problem we are faced with rendering any higher moments to be only marginally useful.

\section{Conclusion and further directions}

Our GPR framework uses a combination of novel and traditional Bayesian tools to tackle the Hausdorff finite moment problem. In the application of our GPR framework to valence PDF datasets, we see that while there are some tradeoffs and optimal parameter choices to consider for constrained and unconstrained valence PDFs, there is a general robustness in reconstruction across kernels/priors. Moreover, spline-based Gibbs-style kernels address systematic issues present in OOTB kernels for unconstrained valence datasets that would be very challenging to perform with standard parametric means. When turning to gluon/sea PDFs, we can re-purpose these Gibbs style kernels to add inductive $x$ correlation information to perform a reconstruction. All in all, our GPR framework shows great promise in tackling the Hausdorff finite moment problem given a limited number of PDF moments for various parton classes. The number and quality of LQCD moment data will only improve, but at beginning stages, the sparsity of data may call for only OOTB kernels to be used as opposed to more flexible spline based Gibbs kernels. Nevertheless our framework can contribute to LQCD's first principles determinations of parton distribution functions that can complement other theoretical and experimental determinations of PDFs/related observables. Finally, we can re-purpose our findings to other areas of science where the finite moment problem is relevant, making our study potentially impactful in two distinct ways.

\section*{Acknowledgments}

R.K. acknowledges D. Pefkou, A. Walker-Loud, and F. Yuan for reading the manuscript and offering valuable suggestions and insights. R.K. is supported in part by the DOE NNSA LRGF Fellowship under cooperative agreement DE-NA0003960 and by the U.S. Department of Energy, Office of Science, Office of Nuclear Physics, under contract number DE-AC02-05CH11231. This research used resources of the National Energy Research Scientific Computing Center (NERSC), a Department of Energy User Facility using NERSC award NP-ERCAP0027666.

\clearpage
\newpage
\appendix

\section{First pass \NNPDF moment reconstructions with grid scheme 2}
\begin{figure*}[h!]
    \centering

    \begin{subfigure}[t]{0.40\textwidth}
        \centering
        \includegraphics[width=\textwidth]{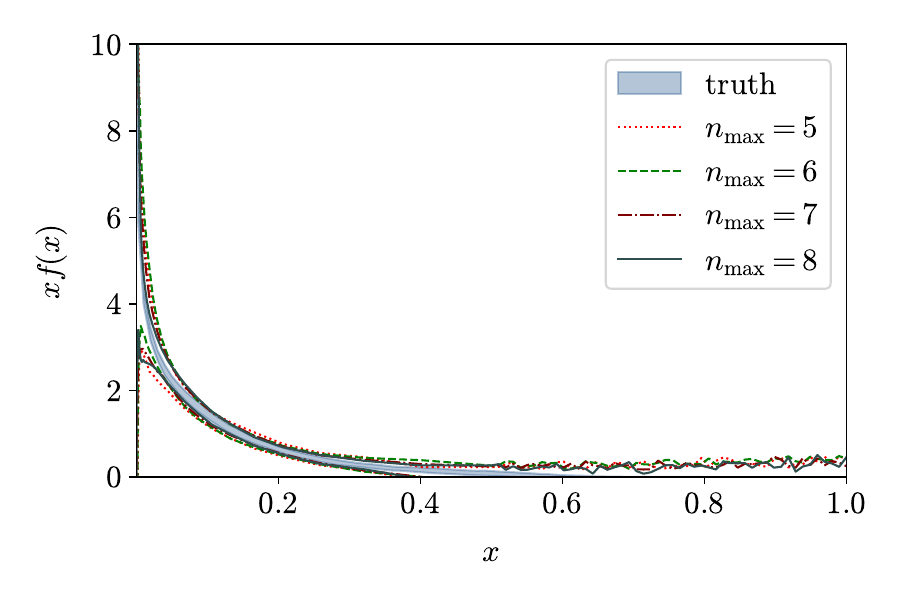}
    \end{subfigure}
    \begin{subfigure}[t]{0.40\textwidth}
        \centering
        \includegraphics[width=\textwidth]{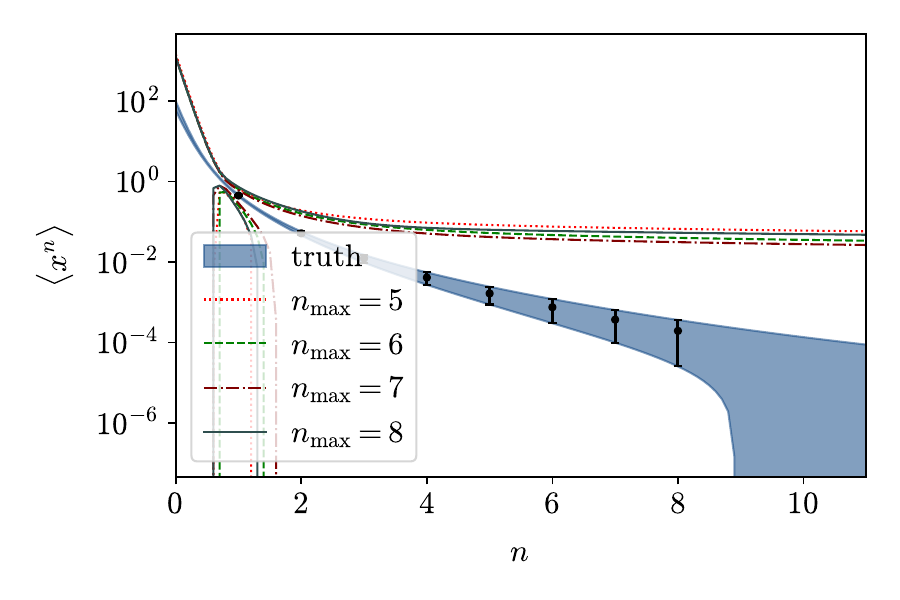}
    \end{subfigure}
    \begin{subfigure}[t]{0.35\textwidth}
        \centering
        \includegraphics[width=\textwidth]{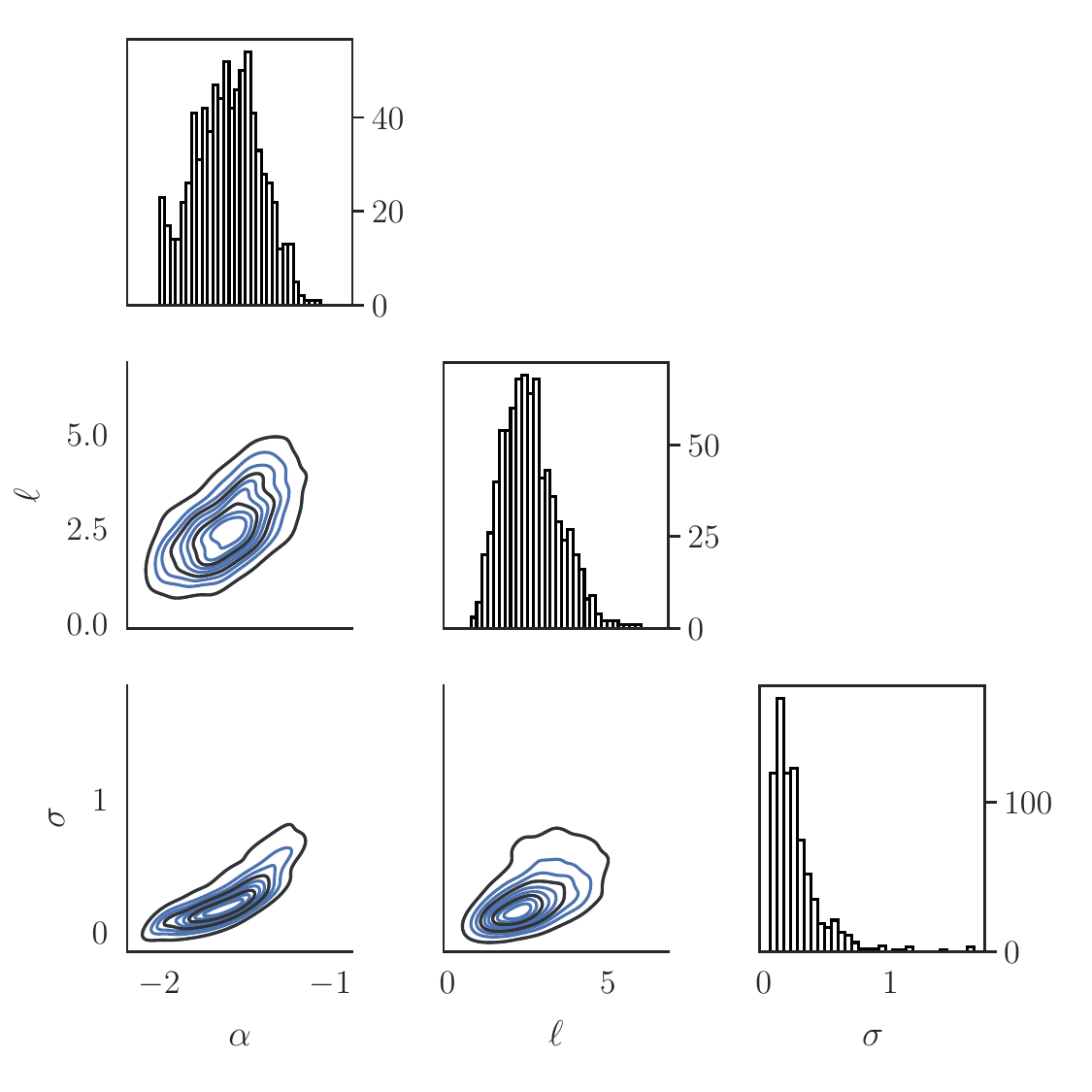}
        \caption{}
        \label{fig:NNPDF_gluon_SE_alphaScan:Triangle}
    \end{subfigure}
    \caption{A first pass at reconstructing \NNPDF gluon PDF with grid scheme 2 with \MATERN{3/2}. \biascorrnote }
    \label{fig:NNPDF_gluon_MATERN32_log}
\end{figure*}

\begin{figure*}[h!]
    \centering

    \begin{subfigure}[t]{0.30\textwidth}
        \centering
        \includegraphics[width=\textwidth]{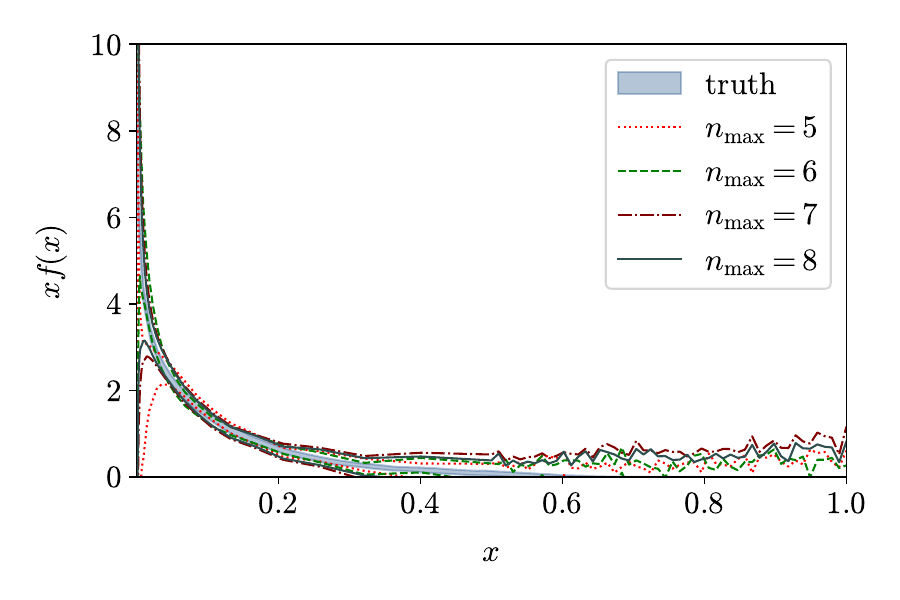}
    \end{subfigure}
    \begin{subfigure}[t]{0.30\textwidth}
        \centering
        \includegraphics[width=\textwidth]{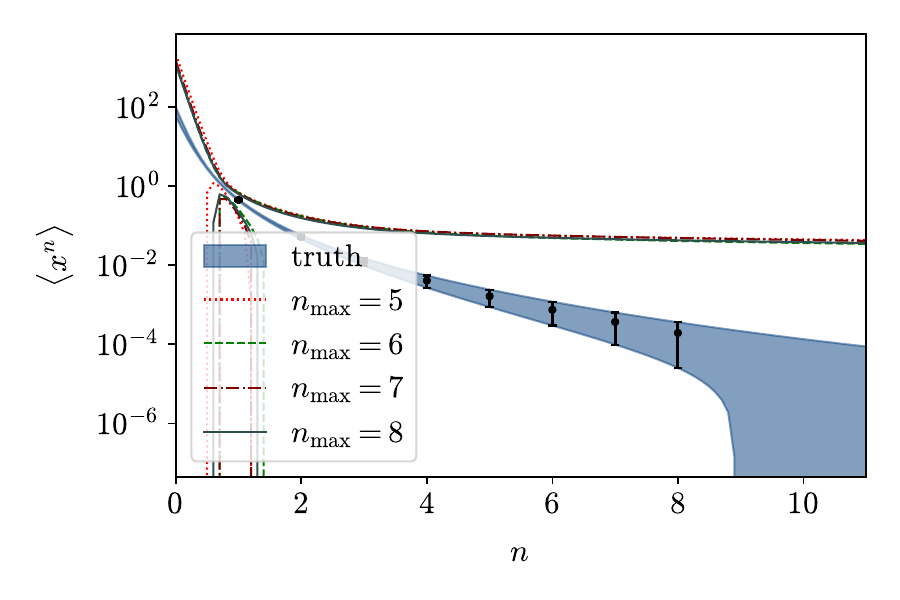}
    \end{subfigure}
    \begin{subfigure}[t]{0.25\textwidth}
        \centering
        \includegraphics[width=\textwidth]{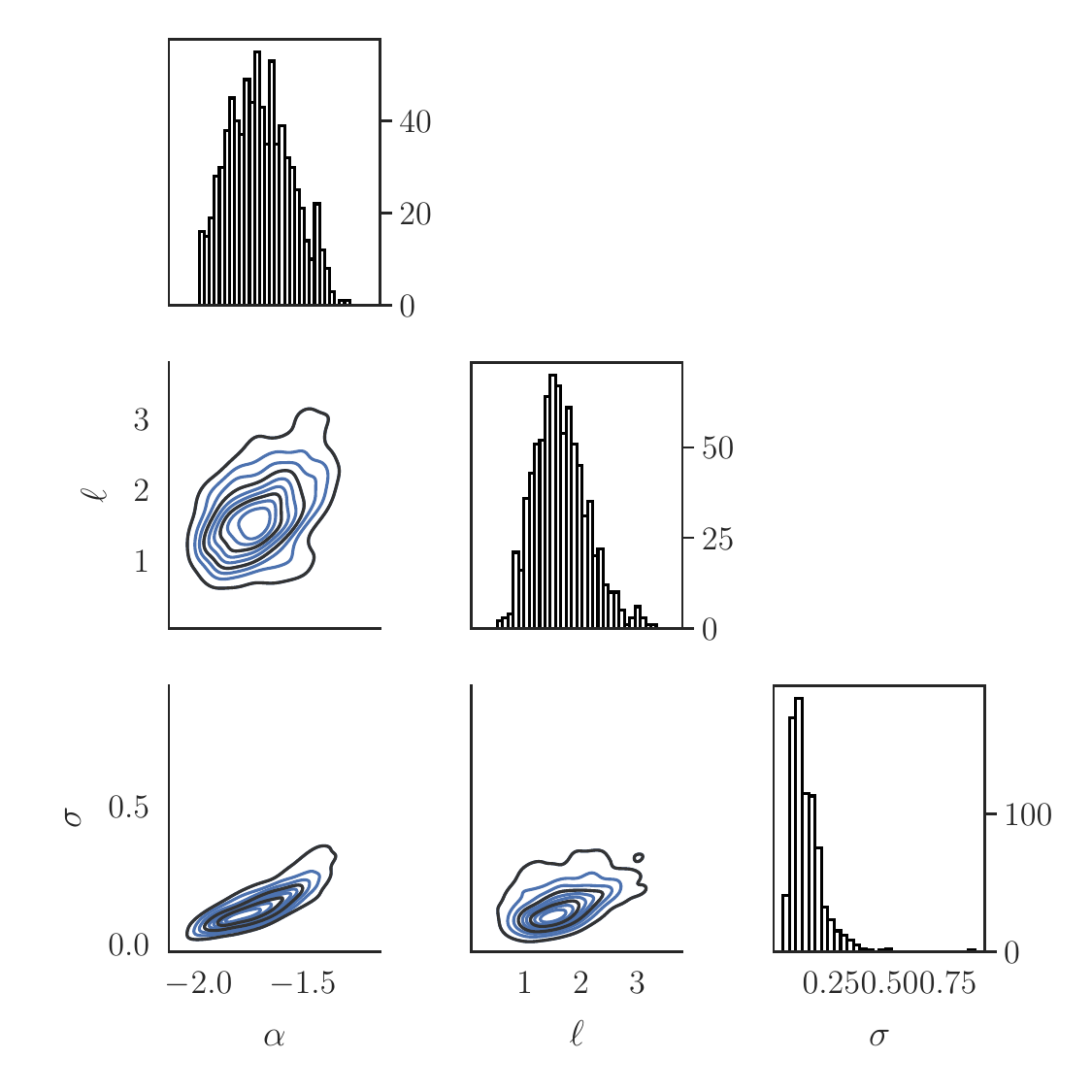}
        \caption{}
        \label{fig:NNPDF_gluon_SE_alphaScan:Triangle}
    \end{subfigure}
    \caption{A first pass at reconstructing \NNPDF gluon PDF with grid scheme 2 with LSE. \biascorrnote }
    \label{fig:NNPDF_gluon_LSE_log}
\end{figure*}

\begin{figure*}[h!]
    \centering

    \begin{subfigure}[t]{0.30\textwidth}
        \centering
        \includegraphics[width=\textwidth]{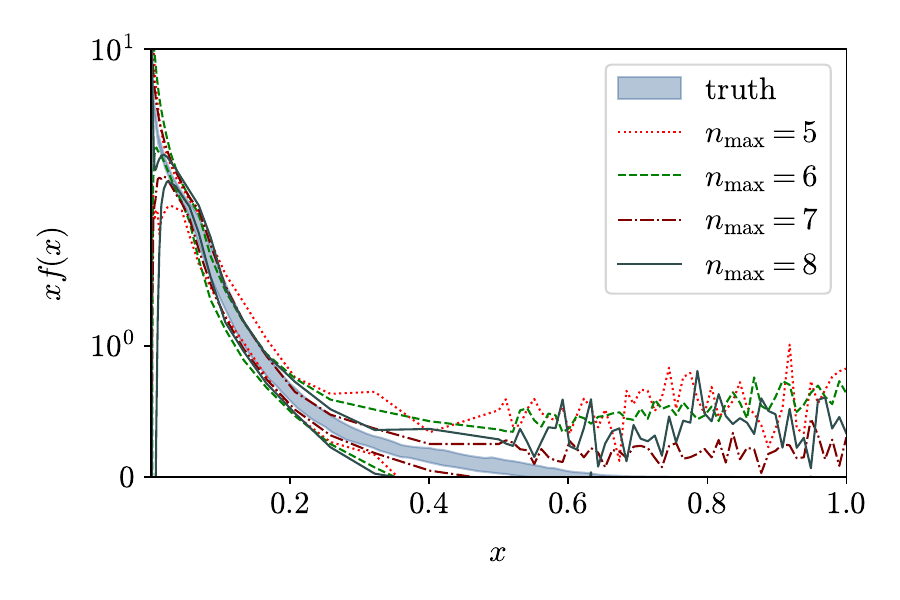}
    \end{subfigure}
    \begin{subfigure}[t]{0.30\textwidth}
        \centering
        \includegraphics[width=\textwidth]{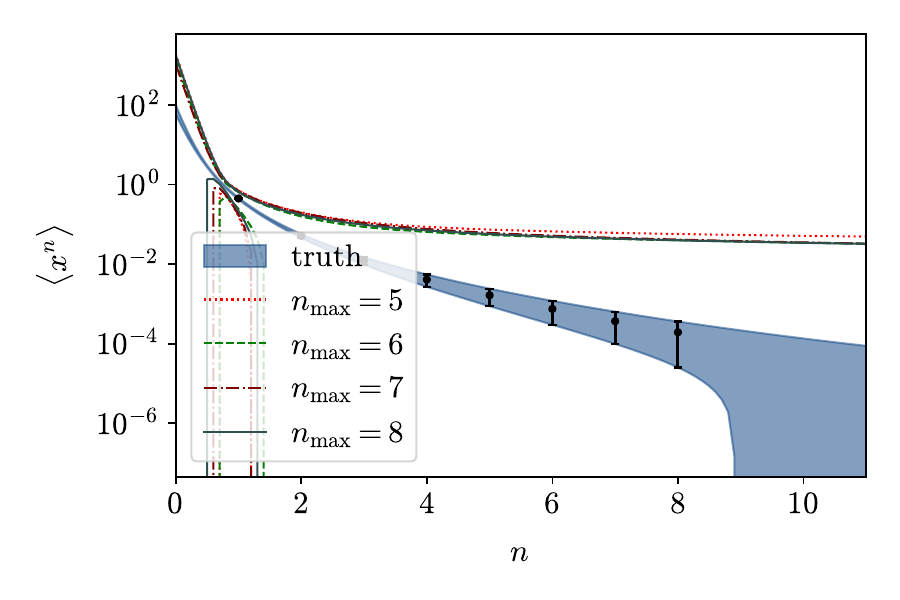}
    \end{subfigure}
    \begin{subfigure}[t]{0.25\textwidth}
        \centering
        \includegraphics[width=\textwidth]{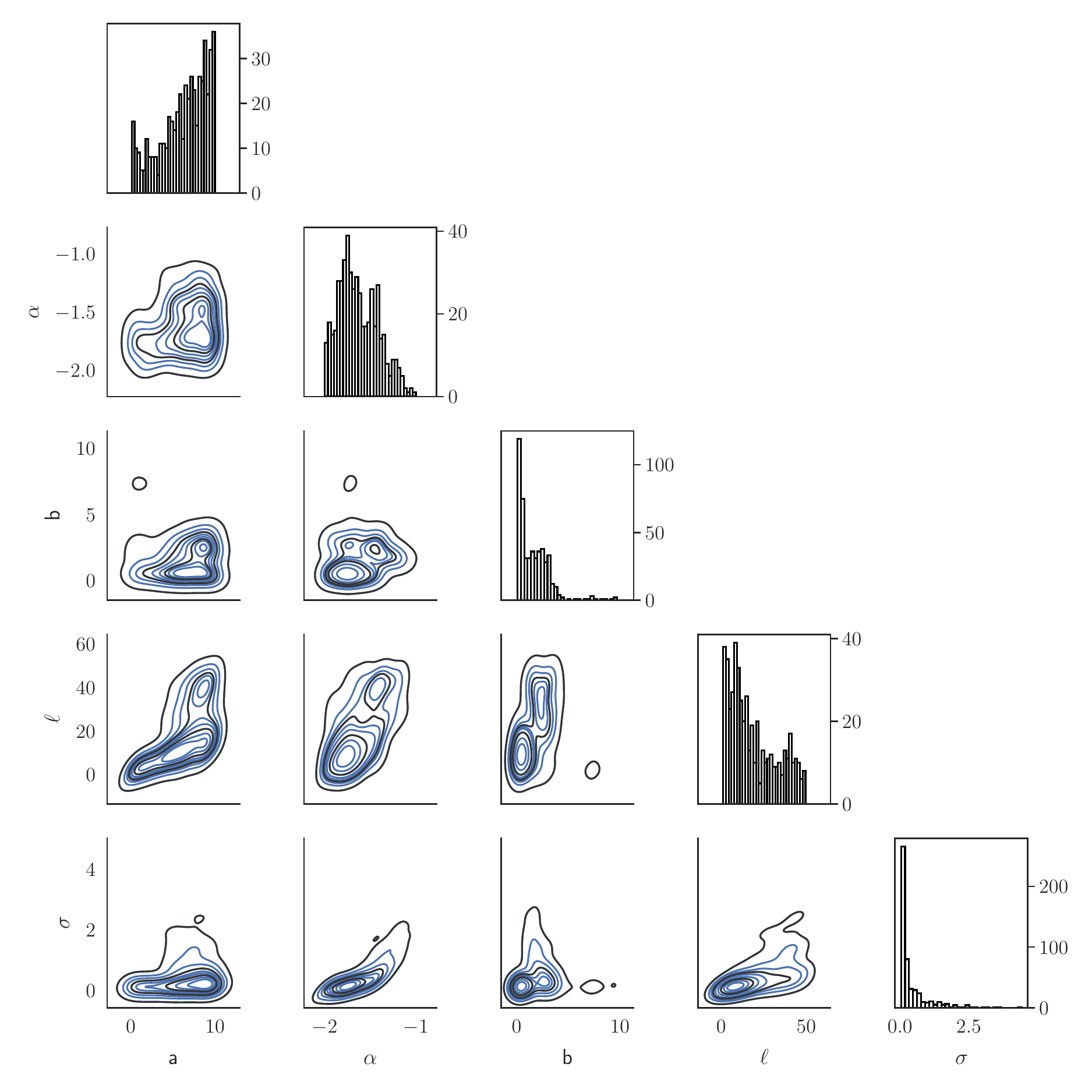}
        \caption{}
    \end{subfigure}
    \caption{A first pass at reconstructing \NNPDF gluon PDF with grid scheme 2 with cosh dip. \biascorrnote }
    \label{fig:NNPDF_gluon_GIBBS_fancy_log}
\end{figure*}

\section{Further studies with \FANTO}
\label{sec:overflow_FANTO}
\begin{figure}
    \centering
    \begin{subfigure}[t]{0.40\textwidth}
        \centering
        \includegraphics[width=\textwidth]{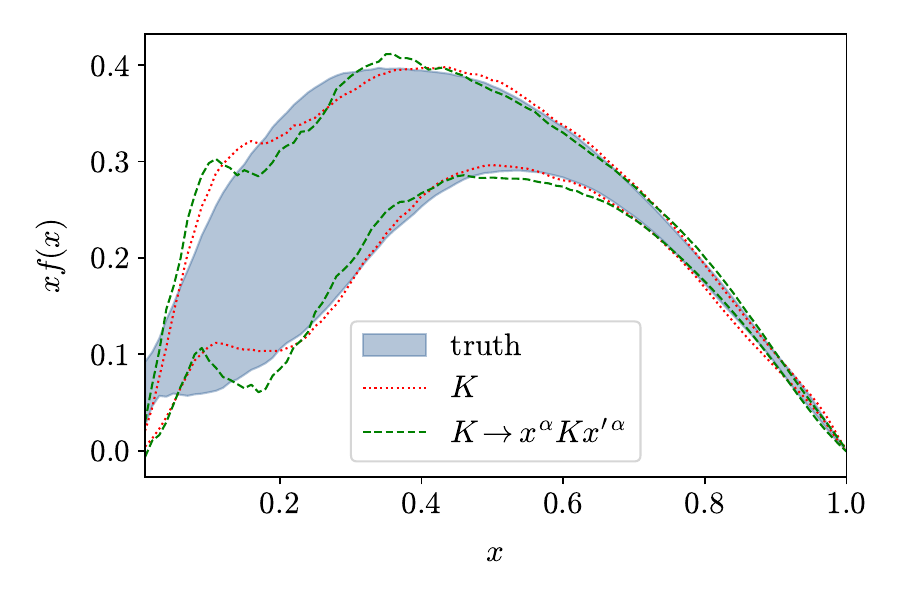}
    \end{subfigure}
    \begin{subfigure}[t]{0.40\textwidth}
        \centering
        \includegraphics[width=\textwidth]{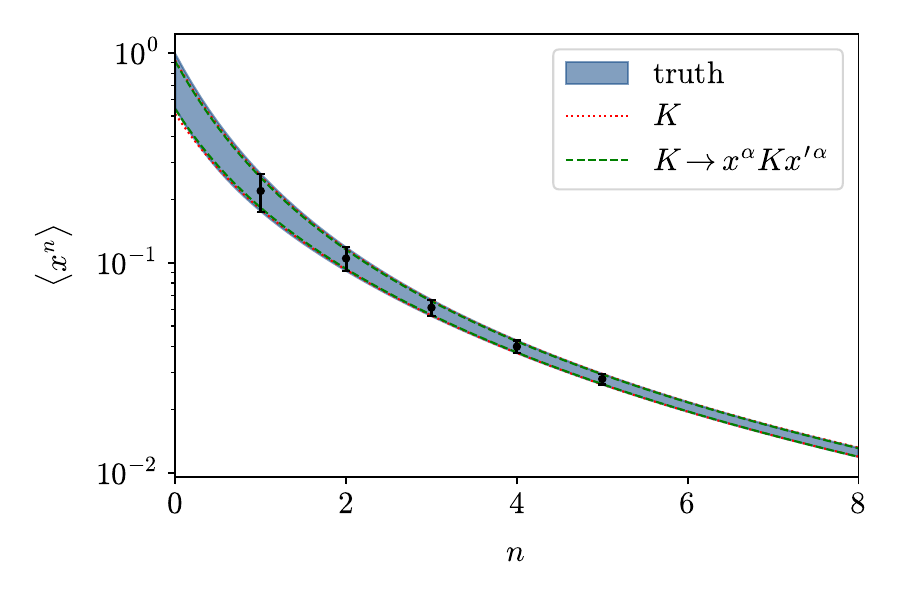}
    \end{subfigure}
    \begin{subfigure}[t]{0.30\textwidth}
        \centering
        \includegraphics[width=\textwidth]{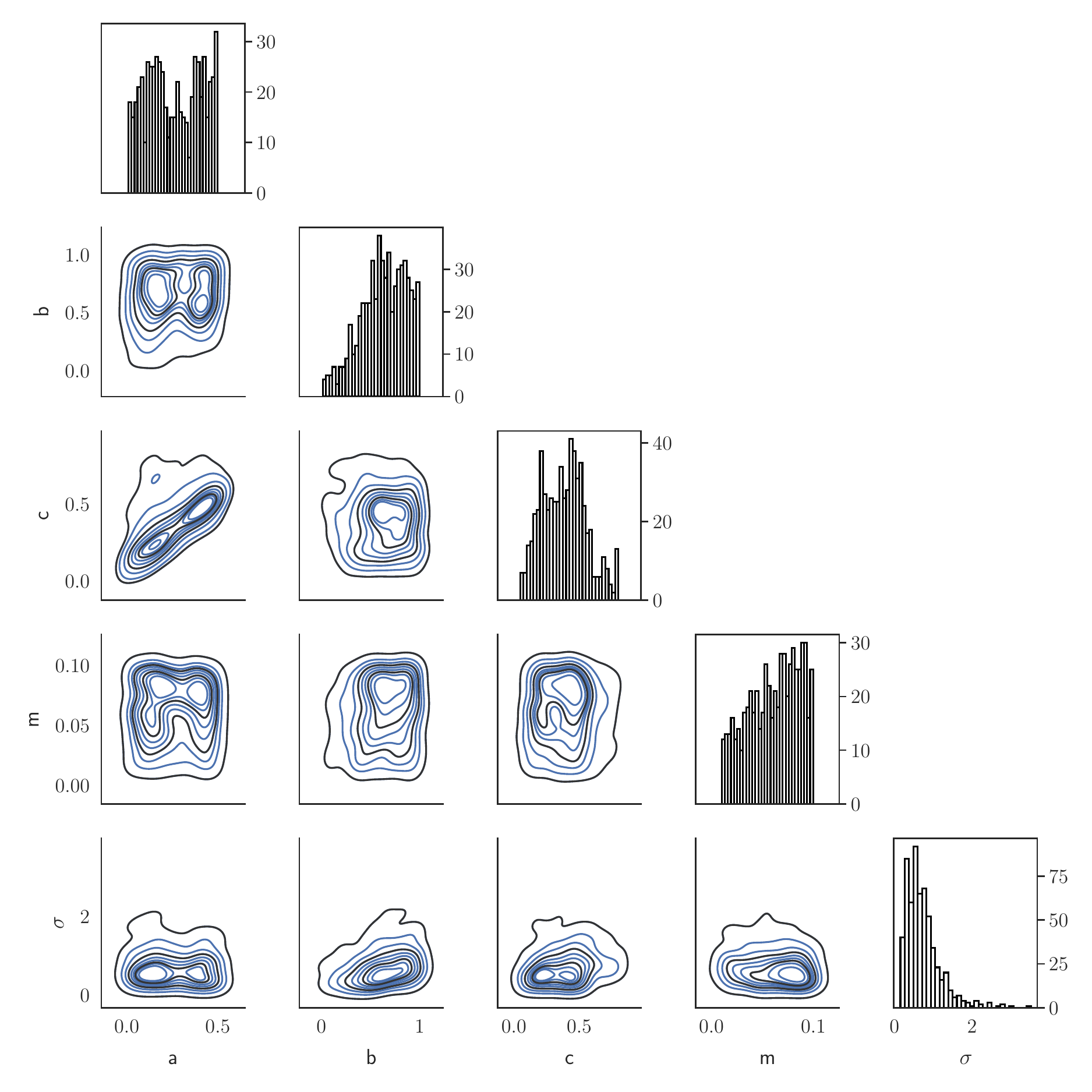}
    \end{subfigure}
    \begin{subfigure}[t]{0.30\textwidth}
        \centering
        \includegraphics[width=\textwidth]{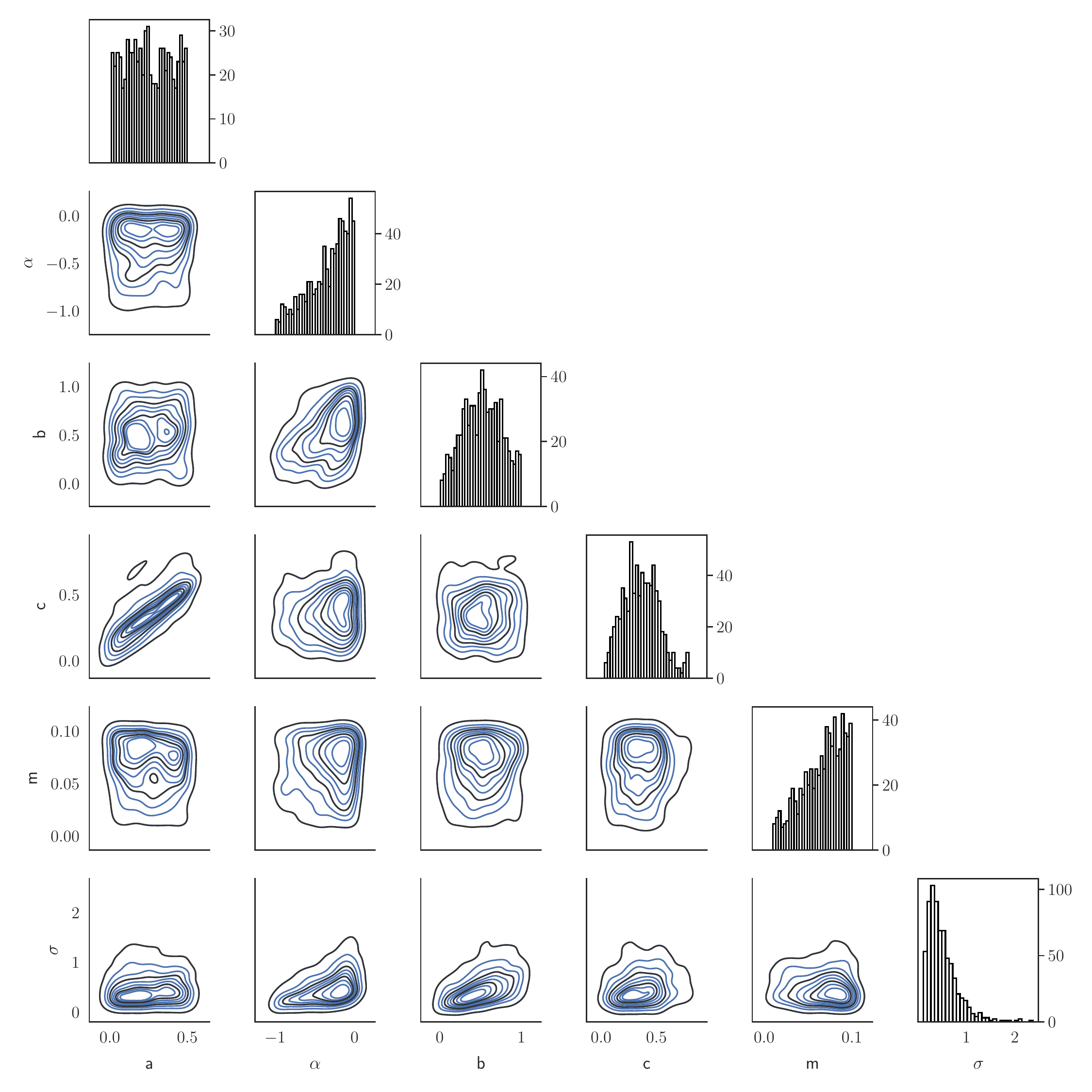}
    \end{subfigure}
    \caption{investigating $x^\alpha$ kernel prior effects with \FANTO valence  dataset}
    \label{fig:FANTO_alphaScan}
\end{figure}

\bibliography{main}

\end{document}